%% file: rpithes.tex
\documentclass{report}

\usepackage[hang]{caption2}      

\begin{document}

\include{rpititle-phd}   
\include{rpiack}

\include{rpiabs}
\include{rpichap1}

\include{rpichap2}
\include{rpichap3}
\include{rpichap4}
\include{rpichap5}

\include{rpichap6}

\include{rpichap7}

\include{rpichap8}

\include{rpiapp}

\include{rpibib}

\end{document}

%% file: rpititle-phd.tex
    
%
\title{\bf A General Symbolic Method\\with Physical Applications}        
\author{Greg M. Smith}        

\date{\today}        

%
\maketitle        

\tableofcontents        

%% file: rpiack.tex
 
\section{ACKNOWLEDGMENT}
 
While certainly taking full responsibility for the contents of this thesis, it
would yet be absurd of me to take full credit, for many people have contributed,
however indirectly, to its completion. I'd like to at least take the trouble to
thank a few of them. 

I'd first like to note my obvious debt to those scientists of various
disciplines whose ideas have influenced mine. In this connection the
bibliography is intended to be fairly thorough, though not exhaustive, in
indicating the works that, with respect to this paper, I've come into valuable
contact with. I'm satisfied to leave it to the reader to arrive at his own
conclusions as to the degree of originality of, or influences on, the thoughts
expressed herein.

My exploration of the above works has been facilitated by the suggestions and
criticisms of my committee: Professors Bringsjord, Drew, McLaughlin, and Zenzen.
These criticisms have decidedly increased the cogency of the presentation which
follows and, not only the content, but also the form of this work has thus
benefited. I'm especially indebted to my advisor, Professor Drew, for his
patience and encouragement during the many revisions this work has undergone. 

Finally, I'd like to thank my parents for the innumerable efforts they've made
on my behalf over the years. I hope they too find some measure of satisfaction
in the completion of this thesis.

%% file: rpiabs.tex
 
\section{ABSTRACT}

   This thesis derives general physical results by an entirely formal process.
   Beginning with a brief examination of the notion of language itself, it next
   explores Physics in a schematic fashion in order to arrive at conclusions on
   the relationship between experience and language. This investigation leads to
   the hypothesis that there is no separate reality to which language refers,
   and therefor to the test of constructing physical theory without reference to
   experiment: If experience is not to direct the interpretation of language
   then language must yield its own interpretation.

   To make such an idea acceptable it is next shown how references to such a
   presumably fictional entity such as an exterior reality may arise within
   language itself, and how such references may, and must, be retained. From
   this starting-point an entirely formal language is developed, along with an
   associated algebra and a Calculus, neither of which are restricted to finite
   quantities. 

   With the completion of the general symbolic system the derivation of both
   Relativity Theory and Quantum Theory, as well as the formal structures to
   which they apply, including space-time and sums-over-histories, follows from
   a purely non-empirical and finitary basis. The dynamical and thermodynamical
   laws yielding the phenomenological aspects of experience, such as are
   described by variables for pressure, volume, and temperature, as well as the
   divisions comprising phases of matter, are also argued to naturally follow on
   this basis. It is therefor plausibly claimed that the formal approach has
   succeeded in yielding its own interpretation and in thus reproducing what has
   previously been asserted to be of necessarily empirical origin. 

   It is, however, then found that this system is comprised of formally
   incompatible parts. It is thus apparently necessary to either reject the
   restriction to finite quantities or else accept the necessity of augmenting
   the formal system with a properly exterior reality, by which it is meant that
   ``experience" must ``inform" the system. Development of formal non-finitary
   theory is then argued to provide a plausible means of unifying the formalisms
   of Relativity and Quantum Theory. 

%% file: rpichap1.tex
 
\chapter{INTRODUCTION AND HISTORICAL REVIEW}

\section{Introduction: On the Notion of Language}

   Every sentence ever written carries with it an implicit assertion of a power
of that sentence to convey an idea and a more general affirmation of written
 language itself. How indeed could anyone speak of something to which language
 cannot refer? And language is pervasive; many kinds of languages have been
 created by many different peoples to serve varied ends.

   Besides obvious examples, one may also speak of music as being a kind of
   language, albeit one with a very flexible grammar. Music, in fact, considered
   as a language, highlights the active role played by interpretation in
   determining the ideas conveyed; any given musical score may be performed in a
   variety of ways by different artists and received variously by distinct
   audiences. Along with interpretation often comes an assertion of what a
   language ought to mean and what its author meant. It's often difficult for a
   consensus to be achieved on this last issue, especially when the author is
   dead !
 
   Mathematics is said to be a language; the language of the sciences. While
   music often allows for a great deal of flexibility in its interpretation,
   Mathematics strives to be a precise language so that its utilization may have
   only one result - a uniquely correct interpretation. Mathematics is also
   often pure; It need not refer to anything outside of itself, in contrast to a
   musical score where each written note corresponds to some sound to be made.

   Consideration of these examples motivates a preliminary stipulation. Language
   will be said to be in use whenever a system of written symbols has some
   relevance. This is a deliberately unrestrictive convention, but it has its uses.

   First of all, in connection with mathematics, attention is shifted away from
   any so-called ``mathematical objects" which might be supposed to lead an
   ``existence" outside of math itself. Thus sets need not have any Platonic
   ``reality", and there need not even be mention of an outside real world to
   which an applied mathematics might refer. Such considerations are obviated
   and one may ask merely whether or not a given string of symbols is to be used. 

   Secondly, as illustrated by musical scores, language is then a very general
   notion. There need not be a uniquely determined relationship between the
   symbols and what the language refers to. It is merely the possibility of some
   connection being adopted that justifies asserting that a language is in use.
 
   This last observation might bring to mind the search for alien intelligent
   life in the universe. Clearly one cannot presume, beforehand, to know the
   language used by an undiscovered race of beings. Nevertheless, if a radio
   source, for example, were to exhibit an unexpected and physically unexplained
   regularity it would not be surprising for there to be conjectures of an
   intelligent source for the ``signal" and attempts to decipher the language
   utilized in its construction.

   It seems, then, that the above definition of language is the only one which
   adequately addresses the way in which the term is commonly used, although it
   doesn't give a constructive idea of what a language may be like or what it
   may be about: Language is defined by those who use it. 

   It is the intention of this thesis to explore language in as general a way as
   possible by specifying the least restrictive and most natural system of
   writing the author has conceived of. It is hoped that a convincing treatment
   of this symbolic system itself will shed some light on the ways in which one
   understands those ``things" to which language refers.

\section{Motivation and Historical Review}   
 
   Before entering upon the developments particular to this thesis, an overview
   of some relevant prior work in the areas of Linguistics and Cognitive
   Science, of Metamathematics, and of Physics will be given. The results in
   these fields will indicate some plausible expectations for the work to be
   presented here, but this will also be an opportunity to indicate the points
   of divergence between this thesis and the perspectives advocated in other work.

\subsection{Linguistics and Cognitive Science}
 
   The form of modern Linguistics and Cognitive Science has been strongly
   influenced by the work of Chomsky\footnote{ An introduction to this work, as
   well as references to more technical expositions, can be found in
   \cite{Pinker1}.}. This work is foundational in the sense that, even for those
   that disagree with it, it is ultimately necessary to justify any serious
   theory in this area by how it relates to Chomsky's approach and results.
   What, then, are some of Chomsky's conclusions about human language?

   Language, according to Chomsky, is a discrete combinatorial system; ``A
   finite number of discrete elements (in this case, words) are sampled,
   combined, and permuted to create larger structures (in this case, sentences)
   with properties that are quite distinct from those of their elements. For
   example, the meaning of \textit{Man Bites Dog} is different from the meaning
   of any of the three words inside it, and different from the meaning of the
   same words combined in reverse order."\footnote{ See pg. 84 of
   \cite{Pinker1}.} This may be contrasted with so-called ``blending systems"
   wherein ``the properties of the combination lie \textit{between} the
   properties of its elements, and the properties of the elements are lost in
   the average or mixture. For example, combining red paint and white paint
   results in pink paint."\footnote{ See pg. 85 of \cite{Pinker1}.} Given that
   the properties of the elements of a blending system are lost in a
   combination, it can hardly be sensible to attempt to formulate general
   statements about a mixture in terms of its elements. Conversely, in a
   discrete combinatorial system the elements may be recovered from their
   combination and general rules may be given for the formation of such combinations. 

   The rules for the combination of the elements of a discrete combinatorial
   system are called its ``grammar". ``A grammar specifies how words may combine
   to express meanings; that specification is independent of the particular
   meanings we typically convey or expect others to convey to us."\footnote{ See
   pg. 87 of \cite{Pinker1}.} The utilization of a grammar is dependent upon the
   categorization of words into parts of speech: ``A part of speech, then, is
   not a kind of meaning; it is a kind of token that obeys certain formal rules,
   like a chess piece or a poker chip. ... When we construe an aspect of the
   world as something that can be identified and counted or measured and that
   can play a role in events, language often allows us to express that aspect as
   a noun, whether or not it is a physical object. For example, when we say
   \textit{I have three reasons for leaving}, we are counting reasons as if they
   were objects. Similarly, when we construe some aspect of the world as an
   event or state involving several participants that affect one another,
   language often allows us to express that aspect as a verb. For example, when
   we say \textit{The situation justified drastic measures}, we are talking
   about justification as if it were something the situation did, though again
   we know that justification is not something we can watch happening at some
   particular time and place."\footnote{ See pg. 106 of \cite{Pinker1}.} 

   Because grammar operates on categories of words it gives sentences a modular
   structure which may be described, as is familiar from Computer
   Science\footnote{ See \cite{Papadimitriou}.}, by ``production rules". For
   example, a noun phrase consists of an optional determiner, followed by any
   number of adjectives, followed by a noun. A sentence consists of a noun
   phrase followed by a verb phrase. A verb phrase consists of a verb followed
   by a noun phrase.\footnote{ See pg. 98 of \cite{Pinker1}.} It will be noticed
   that in these rules the order of the parts of speech within the (written)
   sentence is crucial to the correct application of the grammar. Languages in
   which the grammar may really be given in this form are called ``isolating
   languages". Such a scheme is not always carried out, however. In so-called
   ``inflecting" languages, such as Latin, the spelling of the words themselves
   is modified in order to reflect the role they play within the sentence. as is
   done in the conjugation of verbs. When such cases are provided it is possible
   to rearrange the words within a sentence and still determine the role played
   by each word. English is, in fact, partially isolating and inflecting. It is
   clear, however, that any inflected language may be rewritten in a particular
   order so that it may be interpreted as an isolating language, and the
   generality of the notion of grammar given above is thus retained in all cases.    

   The notion of grammar indicated so far allows a different form of production
   rule for each kind of phrase or part of speech within a given language, and
   also allows for variation in the grammars of distinct languages. It has,
   however, been found\footnote{ See pg. 111 of \cite{Pinker1}.} that all
   grammatical rules within a given language, as well as all grammatical rules
   within all languages, may be put in a standard form. This constitutes
   Chomsky's theory of a Universal Grammar.  This empirically discovered
   independence of grammar from the particular culture in which it is used has
   been asserted to justify the claim that language ability is an inherited and
   intrinsic property of individuals. The inheritance of grammatical language
   ability has further been argued to explain the rapidity and particular way in
   which children learn language. Thus one aspect of linguistic analysis
   indicates an inherent component of language and calls into question the
   degree to which one can discern the properties of a truly external world.

   More generally, the inherent aspects of language may be asserted to hinge, in
   part, on the notion of similarity. The identification of words as being of
   particular parts of speech constitutes recognizing a kind of similarity, and,
   moreover, the generalizing of specific examples to general rules which is
   necessary for learning also requires the identification of a similarity. The
   problem is that similarity is apparently in the mind of the beholder:
   ``Suppose we have three glasses, the first two filled with colorless liquid,
   the third with a bright red liquid. I might be likely to say the first two
   are more like each other than either is like the third. But it happens that
   the first glass is filled with water and the third with water colored with by
   a drop of vegetable dye, while the second is filled with hydrochloric acid-
   and I am thirsty."\footnote{ See pg. 416 of \cite{Pinker1}.} This points to a
   boot-strap problem if one is to be able to learn and not have some inherent
   sense of similarity. It may be observed, then, that pre-existing notions of
   similarity are posited by the cognitive scientists in order to solve the old
   philosophical problem of induction.

   Another observation in Linguistics and Cognitive Science points to a
   difficulty in speaking of a real external world. Here the observation has to
   do with the metaphors that people use in language. It has been found that
   there are two basic metaphors in language\footnote{ See pg. 354 of
   \cite{Pinker2}.}: that of location in space and that of force, or cause. The
   first may be illustrated by the sentences: ``Minnie told Mary a story. Alex
   asked Annie a question. Carol wrote Connie a letter." Thus ``Ideas are gifts,
   communication is giving, the speaker is the sender, the audience is the
   recipient, knowing is having."\footnote{ See pp. 353-354 of \cite{Pinker2}.}
   Force as a metaphor may be illustrated with the following pairs of sentences:
   ``The ball was rolling along the grass." and ``The ball kept on rolling along
   the grass". ``John doesn't go out of the house." and ``John can't go out of
   the house." ``Larry didn't close the door." and ``Larry refrained from
   closing the door." ``The difference is that the second sentence makes us
   think of an agent exerting some force to overcome resistance or overpower
   some other force. With the second ball-in-the-grass sentence, the force is
   literally a physical force. But with John, the force is a \textit{desire}: a
   desire to go out which has been restrained. Similarly, the second Larry seems
   to house one psychic force impelling him to close the door and another that
   overpowers it."\footnote{ See pg. 354 of \cite{Pinker2}.} 

   The sentences above can hardly seem unusual, and the pervasiveness of such
   metaphors is readily discovered. Indeed, ``space and force are so basic to
   language that they are hardly metaphors at all."\footnote{ See pg. 357 of
   \cite{Pinker2}.} Space and force are, however, integral parts of the
   description of physical reality. If, as the above quotations suggest, these
   concepts are building blocks of language itself, then what is the necessity
   of an external reality anyway? How may references to reality be distinguished
   from any other kind of reference if such physical terminology is (perhaps)
   always in use? 

   This discussion of Linguistics and Cognitive Science may be summed up as
   follows: There is reason to doubt that Cognitive Science has arrived at a
   clear distinction between language and the external reality which it,
   presumably, describes. This would be difficult to do, of course, but it is
   still somewhat unsatisfactory that it seems to be no help on this point.
   Nevertheless, it takes a realist position in speaking of a real world and
   reasoning that humans must have an inherent ability to recognize similarities
   in order to be able to deal with ambiguous data, as in the example with the
   glasses,  which isn't just in the mind. This pre-existing faculty for
   classification extends to the common possession of a Universal Grammar which
   operates on ordered strings of symbols, and which is the foundation for
   communication. This grammar may be given in terms of production rules, in a
   canonical form, for the manipulations of a finite discrete combinatorial
   system. In any event, it is clear that Linguistics and Cognitive Science have
   led to the conclusion that language may be understood in terms of a formal
   symbolic system. The consideration of formal systems in and of themselves,
   apart from consideration of any external physical ``reality", is the domain
   of mathematics, so this will be the next subject surveyed.
 
\subsection{Metamathematics}
 
   The modern study of formal systems within mathematics is referred to as
   metamathematics, the study of which originated with David Hilbert\footnote{
   For a leisurely exposition of metamathematics see \cite{Hofstadter}.}.
   ``Metamathematics includes the description or definition of formal systems as
   well as the investigation of properties of formal systems. In dealing with a
   particular formal system, we may call the system the \textit{object theory},
   and the metamathematics relating to it the its \textit{metatheory}.

   From the standpoint of the metatheory, the object theory is not properly a
   theory at all ... but a system of meaningless objects like the positions in a
   game of chess, subject to mechanical manipulations like the moves in chess.
   The object theory is described and studied as a system of symbols and of
   objects built up out of symbols. The symbols are regarded simply as various
   kinds of recognizable objects. To fix our ideas we may think of them
   concretely as marks on paper; or more accurately as abstracted from our
   experience with symbols as marks on paper. ... The other objects of the
   system are only analyzed with regard to the manner of their composition out
   of the symbols. By definition, this is all that a formal system shall be as
   an object of study for metamathematics.

   The metatheory belongs to intuitive and informal mathematics (unless the
   metatheory is itself formalized from a metatheory, which here we leave out of
   account). The metatheory will be expressed in ordinary language, with
   mathematical symbols, such as metamathematical variables, introduced
   according to need. The assertions of the metatheory must be understood. The
   deductions must carry conviction. They must proceed by intuitive inferences,
   and not, as the deductions in the formal theory, by applications of stated
   rules. Rules have been stated to formalize the object theory, but now we must
   understand without rules how these rules work. An intuitive mathematics is
   necessary even to define the formal mathematics.

   ... The methods used in the metatheory shall be restricted to methods, called
   \textit{finitary} by the formalists, which employ only intuitively
   conceivable objects and performable processes. No infinite class may be
   regarded as a complete whole. Proofs of existence shall give, at least
   implicitly, a method for constructing the object which is being proved to exist.  

   This restriction is requisite for the purpose for which Hilbert introduces
   metamathematics. Propositions of a given mathematical theory may fail to have
   a clear meaning, and inferences in it may not carry indubitable evidence. By
   formalizing the theory, the development of the theory is reduced to form and
   rule. There is no longer ambiguity about what constitutes a statement of the
   theory, or what constitutes a proof in the theory. Then the question whether
   the methods which have been formalized in it lead to contradiction, and other
   questions about the effect of those methods, are to be investigated in the
   metatheory, by methods not subject to the same doubts as the methods of the
   original theory."\footnote{ This excellent and authoritative quote may be
   found on pp. 62-63 of \cite{Kleene}.}

   According to the above characterization, metamathematics achieves clarity of
   meaning and definiteness of implication for the \textit{object theory}, but
   the object theory is not all of mathematics. The undefined
   \textit{metatheory} forever remains as an exterior criterion of the
   acceptability of the formalized object theory, and it is thus implicitly
   impossible, according to this standpoint, to ever arrive at a complete notion
   of mathematics which is both clear in its meaning and definite in its
   implications. It is clear, however, that this scheme allows a progressive
   clarification and development of the formal scheme. What, then, has been
   achieved along these lines?

   It has been indicated that the object theory is judged from the perspective
   of the metatheory; Thus the metatheory must, if the meaning of statements in
   the object theory is to be definite, definitely classify each statement in
   the object theory as being either acceptable or not. This classification is
   achieved by a mapping of each statement in the object theory onto either
   ``True" or ``False"; This mapping is called the ``interpretation" of the
   object theory. It is clear, then, that from the metamathematical perspective
   the notion of truth is contingent upon there being a separate metatheory
   which determines, via an interpretation, what is to be true. 

   Metamathematics is also supposed to make the acceptability of proofs of an
   assertion manifest, where a proof is a formal derivation of a true statement.
   Formal operations, being rules for the combination of meaningless symbols,
   can, in themselves,  only lead from one combination of meaningless symbols to
   another and thus make one such combination depend upon another. Once an
   interpretation is provided, however, there is hope that certain formal
   operations may preserve the initial truth value in a progression of formal
   combinations leading to the statement it is desired to prove. Such a proof is
   not absolute, but rather expresses the truth of the conclusion as being
   contingent upon the truth of the initial statement in the derivation. This
   method cannot remove the ultimately contingent nature of such proofs, so that
   certain statements cannot be proven but may only be assigned the value
   ``true" according to the interpretation. Such statements are called
   ``axioms." The truth-preserving formal rules would then be termed ``correct".
   Those statements which may be proven are termed ``theorems."

   The axioms and the particular proof theory applied to an object theory are
   determined by the metatheory. Thus, from the metamathematical perspective,
   once a formal system is chosen, formal mathematics boils down to the
   selection of the axioms and the formal rules for carrying out proofs.  Now it
   will be noticed that, once axioms and the formal rules for proofs are given,
   the formal system inherits an order leading from axioms to theorems. The
   direction and extent of this order will be determined by the axioms and proof
   system. There is thus a correspondence between the rules for proof accepted
   by the metatheory and some ordinal number. Such a selection cannot, in the
   above sense, be proven correct but may only be a matter of preference. The
   choice made in modern mathematics may, for the most part, be said to be for
   first order logic with  proof based on the acceptance of arguments by
   induction, in the usual sense, on the integers, the integers having ordinal
   number $\omega$\footnote{ For a critical discussion of this standpoint see
   \cite{Cohen}.}. 

   What is the result of this chosen perspective? Hilbert's original motivation
   in proposing the metamathematical approach was to show, at least, that
   mathematics was both ``consistent", so that no false statement may be proven,
   and ``complete", so that any true statement may be proven. As is well known,
   G\"{o}del\footnote{ See either of \cite{Hofstadter} or \cite{Kleene}.} dashed
   the hopes for the fulfillment of this program when he showed that, for any
   object theory which may model arithmetic and for which induction is defined
   to be over the integers, it may be proven that the object theory is both
   incomplete and cannot be proven to be consistent. It will be illuminating to
   proceed at this point to a discussion of the role of judgements based on the
   metatheory in the proof of the first of these assertions.

   The key to the proof of these results lies in the notion of G\"{o}del
   numbering. G\"{o}del numbering encodes statements in first order logic as
   numbers in the arithmetic system which is formalized. Such numbers may then
   be arguments of predicates in the formal system so that it becomes possible,
   from the metamathematical perspective, to interpret some such statements as
   being self-referential. In particular, G\"{o}del was able to construct a
   statement G such that from the metamathematical perspective G asserts that G
   is not a theorem in the object theory. It then is found to be impossible to
   prove G. This may be explained as follows: It is asked whether or not G is a
   theorem. If it is, then G, being a theorem, must assert a truth. G, however,
   according to the metatheory, asserts that G is not a theorem, so this is a
   contradiction. If the object theory is to be consistent then this alternative
   must be rejected. On the other hand, supposing G is not a theorem, it then
   follows, again according to the metatheory, that G asserts a truth. Then,
   since G is both true and not a theorem of the object theory, it follows that
   the object theory is incomplete. 

   Now G being true entails its negation being false, and, false statements
   being underivable, it follows that neither G nor its negation may be a
   theorem of the object theory. G is thus an ``undecidable" proposition; It's a
   statement the object theory can make no assertion about, one way or the other.
   In order to remedy this ``hole" in the object theory it is necessary that the
   object theory be amended so that it adopts either G or its negation as a new
   axiom. The first option, usually referred to as being the standard one, is
   straight-forward enough. The second alternative, referred to as the
   non-standard one, it turns out, requires the augmenting of the object theory
   with a new class of symbols and a corresponding reinterpretation of
   predicates to admit the new class of symbols. This option points directly to
   an extension of the form of the object theory, as is illustrated by the
   reformulation of the Calculus within non-standard analysis\footnote{ See
   \cite{Robinson}.}. The interesting thing about non-standard analysis is that
   it allows rigorous definition and manipulation of infinite and infinitesimal
   quantities. It is clear, in any event then, that G\"{o}del's Incompleteness
   Theorem, rather than merely being prohibitive, may be fruitful as well.   

    It should be carefully noted that the metamathematical basis for G\"{o}del's
    theorems was, in part, the acceptance of induction over the integers. While
    this is the standard approach, it is worth noting that G\"{o}del's theorem
    asserting the unprovability of consistency may be refuted if one is willing
    to accept transfinite induction over a larger ordinal. It has been shown
    that induction over an ordinal known as $\epsilon_0$ allows a consistency
    proof to be carried through\footnote{ See \cite{Gentzen}.}. 

   In conclusion, the discussion of metamathematics has revealed a number of
   things. Metamathematics sets out, first of all, on the dual basis of an
   object theory and a metatheory. While the object theory is formalized so that
   acceptable statements may be definitely recognized, the metatheory, which
   justifies and interprets the object theory, is not necessarily formalized at
   all. The metatheory operates on an intuitive basis, which, in part, is simply
   a way of saying that the way that the metatheory operates is undefined. There
   is a parallel here between human language and physical reality on the one
   hand, and the object theory and the metatheory on the other. Proceeding
   nonetheless to consider the implications of this approach, it is found that
   the interpretive role played by the metatheory leads to the usual notions of
   truth and proof in an axiomatic mathematical system. Proof theory introduces
   a notion of order in the object theory. There is also a consequent parallel
   between the kinds of inductions accepted for proofs and the size of ordinals
   that must be admitted in the object theory. G\"{o}del's theorems then show
   that despite the definiteness of the notions of truth and proof which
   metamathematics utilizes, the presumably categorical power of these notions
   doesn't obtain. It does, nevertheless, point to a limitless flexibility in
   the formation of the formal system. This flexibility is manifest in the
   extensibilty of the system of axioms and in the potential variety of variable
   types that may be constructed in non-standard theories. Infinities and
   infinitesimals, for example, naturally find their way into the formal system
   in this way. This fluid nature of the formal language of metamathematics
   suggests that if any comprehensive conclusions about the relationship between
   reality and language are to be arrived at, they might not derive from the
   given form of a language, but may be justified by a systematic referal to an
   exterior reality instead. For this reason attention now turns to an overview
   of Physics.
   
\subsection{Physics}    

   Science is distinguished from other mental pursuits by its recourse to
   experimentation, this, it may be said, being a systematic procedure for
   determining how to use language to refer to ``experience". While ``reality"
   is a scientifically indeterminate word, it is supposed that, via
   experimentation, Science refers to \textit{something} other than a formal
   symbolic game\footnote{ The notion of reality, from a quantum-mechanical
   perspective, is discussed in \cite{Espagnat}.}. Thus the exploration of
   Physics will begin with a discussion of experimentation, the scientific method. 

   The scientific method may be described in a number of steps. First there is
   Observation. In this phase of the process it is simply asserted that
   something is noticed in the ``outside world" and this something is
   \textit{described}. It is important to note, right away, that science
   therefor deals with that which, though taken to come from outside of the
   linguistic apparatus, is, nevertheless, describable within this system. 

   After Observation comes a stage in which Questions may be formulated. Here
   the essential activity is that of trying to guess at any possible connections
   between different aspects of the physical situation and the particular
   phenomenon noticed in the previous step. In other words, an attempt is made
   to imagine ways in which the occurrence of the noticed phenomenon might depend
   on other physical events which then would precondition it. Loosely stated,
   this amounts to seeking causes for the observed effect. It should be noted
   that such questions, like the observed phenomenon, arise in an undefined way,
   as saying that such ideas are found by using one's imagination does not
   define imagination itself. It is also worth noting that any presumed cause,
   like any other condition which may be distinguished, cannot always obtain; a
   term which always applies is not descriptive. Under these conditions it is at
   least conceivable that a supposed cause and the observed effect may be related.

   The third stage of the scientific method is the formulation of a Hypothesis.
   The Hypothesis is a clearly stated trial explanation for the observed
   phenomenon chosen, according to best judgement, from the possibilities
   considered in the previous stage. This step is also undefined; it is simply a
   judgement call. It is clear that, from the scientific perspective, it cannot
   matter how questions or hypothesis are formed. The only thing that can matter
   is the usefulness of these questions and hypotheses once they are ``found."  

   The fourth step is to carry out a controlled Experiment. This means that a
   well defined series of pairs of physical situations is created such that in
   the first member of each pair the supposed cause obtains while the other half
   of the pair, called the control, is a situation which is identical to the
   first except that the supposed cause does not obtain. Having created these
   initial conditions, the experiment consists in waiting to see, and noting in
   each case, whether or not the effect to be explained occurs. It is presumed
   that any "real" cause would be in effect during these trials and would show
   itself, at least sometimes, in the effect occurring in the non-controls and
   not occurring in the controls. Any hypothetical cause is accepted or rejected
   accordingly.

   The final stage in this progression of the scientific method is that of
   Theorizing. Those relationships which pass the experimental step are taken,
   at least provisionally, to be factual. These relationships are then thought
   about (another undefined step) and an attempt is made to create a
   mathematical theory which comprises all of these relationships. This theory
   may make predictions as to what will happen in experiments not yet performed,
   and this becomes a test for the theory itself to pass. In any event, having
   cycled through these five stages, Science always returns to the first to look
   for new observations.  
 
   A number of questions may be raised concerning the adequacy of the scientific
   method outlined above. The world being tremendously complex, which phenomena
   are worth observing? Since phenomena, in order to be observed in the above
   sense, must be described, doesn't this precondition what may be observed?
   Might it not be a consequence of this that only phenomena which are
   sufficiently simple may be observed? If the number of hypotheses that might
   explain an effect is not finite, doesn't it follow that the experimental
   investigations might run themselves into a small corner in the sense of
   restricting the way the world is viewed by scientists? And might it not
   happen that experiment never checks all hypotheses which might explain a
   given effect? The scientific response to these issues may be inferred by
   examining Newton's Rules of Reasoning in Philosophy\footnote{ See Part III of
   \cite{Newton}.}. While these rules do not constitute a formal oath of office
   for all scientists, it may be fairly asserted that they do reasonably reflect
   the practical attitude of most scientists.   

   Newton's Rules of Reasoning in Philosophy are:

   Rule I: We are to admit no more causes of natural things than such as are
   both true and sufficient to explain their appearances.

   Rule II: Therefor to the same natural effects we must, as far as possible,
   assign the same causes.

   Rule III: The qualities of bodies, which admit neither intensification nor
   remission of degrees, and which are found to belong to all bodies within the
   reach of our experiments, are to be esteemed the universal qualities of all
   bodies whatsoever.

   Rule IV: In experimental philosophy we are to look upon propositions inferred
   by natural induction from phenomena as accurately or very nearly true,
   notwithstanding any contrary hypotheses that may be imagined, till such time
   as other phenomena occur, by which they may either be made more accurate, or
   liable to exceptions.

   Rule IV refers to ``natural induction", and it should be noted that this term
   has been the subject of much debate. For the purposes of the argument to
   follow, this phrase will be replaced by ``experimentation". This seems
   reasonable as, after all, Newton's rules are being interpreted and applied,
   rather than merely cited.

   Newton's first rule begins by subordinating causes to effects: Particular
   causes may be admitted only if they explain observed effects. However, it
   should be clearly noted that causes are otherwise unrestricted by Newton's
   rules and it is nowhere indicated how one is to choose among many particular
   causes which might explain a particular effect. It is only asserted that if
   one of these causes is acceptable according to experiment, and is therefor
   ``true", then it may be adopted. This is in agreement with a minimal
   interpretation of the experimental procedure outlined above. 

   It is clear then that both the experimental method and Newton's commentary on
   it leave the acts of observation, questioning, and hypothesizing without
   limits defined prior to experimentation. It follows from this that no limits,
   in principle, may be asserted to apply to the effective range of experimental
   method save any which may restrict the \textit{form} of statements and
   hypotheses so that they are subject to controlled experiment. Thus the
   scientific method is not restricted to investigate only particular
   observations, and there is no bar to future effects necessitating the
   abandonment of previous causes, as is indicated explicitly in Rule IV, so
   that experimental investigation need not lead to a narrow outlook. It is not
   necessary that all hypothetical causes be checked.
 
   What may be said about the form of statements about experimental observations
   and causes? It may be observed that the experimental method would be
   fruitless if it were merely the case that all observed effects were in
   one-to-one correspondence with causes, for then there would be no conceptual
   gain in speaking of causes: Note here that even in a deterministic physics,
   where prediction may in principle be perfect and invertible, an additional
   idea of the passage of time is involved. Thus, in general, effects are not
   equivalent to causes though they proceed from them. Furthermore, if the
   description of effects is not to be superfluous, then their number must be
   finite; In a description of an infinity of elements terms could be omitted
   and still one would retain a description which may correspond to the
   original. Appealing to Rule I then, it is clear that both causes and effects
   may and must be representable by a finite number of finite numbers, and this
   may be taken to be the form that experiment requires statements adopt. This
   conclusion is implicit in Newton's Rules III and IV where the ``qualities of
   bodies" are spoken of as being ``accurately or very nearly true" and it is
   said that the ``universal qualities" of bodies ``admit neither
   intensification nor remission of degrees". That physical quantities are to be
   represented by finite numbers is not explicit in the previous definition of
   the experimental method, though it is an unspoken canon of scientists. 

   Consider now Rule II: It is, as Newton asserts, a consequence of Rule I. It
   goes beyond Rule I in pointing out that physical explanation shall therefor
   be universal in that all phenomena must be viewed within a single framework.
   In fact, Rule II has the effect of imposing the requirement that no part of
   the universe, nor any aspect of its activity, may be considered forever
   isolated from the rest, as in that case the isolated part would be considered
   physically irrelevant through not having effects which experiments need
   address. This then also justifies Newton's third rule, and Rule II
   furthermore serves as a constraint on theory, pushing physicists towards
   unification.  

   The discussion thus far has focused on experiment as a method of analyzing
   experience but it hasn't yet been shown that this method, as opposed to the
   methods of linguistics and metamathematics, results in anything other than a
   formal symbolic game. Does the scientific method really access something
   separate from language itself, some external reality? In order to address
   this question the final aspect of the scientific method, that of theory, will
   be considered next.

   There is a well-known understanding of the relationship between theory and
   experience: ``Every theory can be divided into two separate parts, the formal
   part, and the interpretive part. The formal part consists of a purely
   logico-mathematical structure, i.e., a collection of symbols together with
   rules for their manipulation, while the interpretive part consists of a set
   of ``associations", which are rules which put some of the elements of the
   formal part into correspondence with the perceived world. The essential point
   of a theory, then, is that it is a \textit{mathematical model}, together with
   an \textit{isomorphism} between the model and the world of experience (i.e.,
   the sense perceptions of the individual, or the ``real world" - depending
   upon one's choice of epistemology)."\footnote{ See pg. 133 of
   \cite{Everett}.}. Here again, as was the case with Linguistics and
   Metamathematics, there is a separation between experience and language,
   and where that separation is to be found, is, ultimately, intuited rather
   than defined. 

   If the separation between the real and the merely formal is not defined a
   priori, then perhaps, as a minimum, that separation is at least definitely
   stated in contemporary physics. Contemporary physics is, for the most part,
   quantum theory, and the mathematical model of quantum theory was given,
   essentially, by von Neumann\footnote{ See \cite{Neumann1}.}. Rather than
   discuss the intricacies of this model it is sufficient for current purposes
   to note here that calculations in this formalism are of two types: ``Process
   1" calculations which apply whenever new experimental data are acquired in
   ``measurements", and ``Process 2" calculations which apply for those times in
   the interim between measurements\footnote{See pg. 3 of \cite{Everett}.}. This
   distinction in the formal operations of the theory clearly and formally
   distinguishes ``real" data from merely formal variables. This formal
   separation would unquestionably draw a definite line between the ``real" and
   the formal if it weren't also the case that the interpretation of the model
   leads to difficulties\footnote{ For a quick description of these difficulties
   see pp. 3-10 of \cite{Everett}. For a selection of some different ways of
   viewing the quantum formalism also consult pp. 181-195 of \cite{Bell}.}.
   These difficulties, known as the ``measurement problem", have lead to a
   variety of elaborate attempts at resolution\footnote{ See \cite{Wheeler}.},
   but, as yet, none are uniformly accepted. 

   Contemporary physics, having failed at clearly separating formalism and
   ``reality", may even be doubted to have adopted any such separation in the
   first place, instead almost reducing physics to formalism. To begin with, it
   has been shown, by those studying ``Quantum Logic", that the formal apparatus
   of quantum theory need not be motivated by experiment, but is, rather, a
   formal embodiment of the propositional calculus\footnote{ For an overview of
   this formalism see \cite{Baer}. For an in depth treatment see
   \cite{Varadarajan}.}. The reduction of physics to formalism, it may be
   argued, is further supported by a founder of quantum theory, Niels Bohr, in
   the following comparison of quantum and classical physics: ``In the case of
   quantum phenomena, the unlimited divisibility of events implied in such an
   (classical) account is, in principle, excluded by the requirement to
   \textit{specify} the experimental conditions. Indeed, the feature of
   wholeness typical of proper quantum phenomena finds its logical expression in
   the circumstance that any attempt at a \textit{well-defined} subdivision
   would demand a change in the experimental arrangement incompatible with the
   \textit{definition} of the phenomena under investigation."\footnote{ See pg.
   4 of \cite{Bohr}.}( italics mine). As to the concept of a physical condition,
   Bohr seems to be identifying it with, if not replacing it by, its formal
   description. Some may yet feel that these points are exaggerated and merely
   ``philosophical", and do not really apply to practical physics. In answer to
   this it is sufficient to turn consideration to the modern theory of quarks,
   for quarks are supposedly fundamental constituents of physical description
   motivated by experiment, and, at the same time, are subject to
   ``confinement". Thus, while quarks are believed to ``exist" they cannot, in
   principle, ever be seen! Quarks can impact experimental results only
   indirectly, and are thus primarily products of theory and not observation. 

   Beyond formalization of the ``existence" of external reality, science has
   even gone so far as to attempt the formalization of the individuals
   themselves who 
   have experiences, so that none could speak of anything which is \textit{not}
   formal. This program may be seen to be a consequence of Science's previously
   mentioned drive towards universality. According to von Neumann: ``... it is a
   fundamental requirement of the scientific viewpoint - the so-called principle
   of the psycho-physical parallelism - that it must be possible so to describe
   the extra-physical process of the subjective perception as if it were in
   reality in the physical world - i.e., to assign to its parts equivalent
   physical processes in the objective environment, in ordinary
   space."\footnote{ See pg. 418 of \cite{Neumann1}.} The physical world being
   formalized, as indicated above, it then follows that individual experience
   ought to be as well. Along the lines of deterministic theory, work in
   Computer Science has produced Turing machines as well as other apparatus of
   Artificial Intelligence\footnote{See \cite{Hofstadter} for a basic coverage.
   For detailed theory consult \cite{Hermes}.}. von Neumann went so far, in his
   theory of self-reproducing automata, as to formalize deterministic machines
   that would carry on many of the formal operations of life-forms\footnote{See
   \cite{Neumann2}.}. von Neumann also addressed non-deterministic theory by
   showing that nearly deterministic behavior may be synthesized by the parallel
   organization of non-deterministic elements\footnote{ See \cite{Neumann3}.}.
   Consequently it is now common practice in physics to, in effect, replace
   observers by Turing machines\footnote{See pg. 64 of \cite{Everett}.}. 

  The above discussion has fairly well summarized the results of contemporary
  physics. A review of physics, then, indicates that Science attempts to answer
  the need for specifying a definite interpretation of its formalism through
  experimentation. The nature of the experimental procedure itself indicates
  that there are no necessary restrictions to the range of the interpretations
  thus derived, though such investigations are formally restricted to
  measurements expressible by finitely many finite numbers. Science is concerned
  with having \textit{an} explanation of observations, but by no means need
  there be a unique explanation or theory. In fact, theories need not arise
  solely or uniquely from experiment either, as is illustrated in the case of
  quantum theory. This freeing of theory from strict correspondence with
  observation has resulted in modern physical theory being almost totally formal
  in nature. It has also left in question the degree to which anything besides a
  formalism is required or even specifiable. 

\section{A Departure from Prior Approaches} 	

   Having surveyed Linguistics and Cognitive Science, Metamathematics, and
   contemporary Physics, an attempt will now be made to glean any wisdom common
   to these disciplines. Some of those things which these disciplines have in
   common will be found to hold in the symbolic system developed here.
   Particular aspects specific to some of these areas will also feature in the
   pages to follow. And, as it will also be observed, there is a similarity in
   the difficulties facing these theories which will be an important indicator
   of a point of departure which distinguishes the theory developed here from
   previous developments. 

   It may be recalled that each of the noted areas does make use of written
   discrete combinatorial systems. Manipulations of these systems were, in all
   cases, specified by clearly stated formal grammars. In applying grammars
   rules were in effect according to the category of symbols being manipulated,
   there being no concern for any particular additional meaning attached to the
   symbols. This lends a certain rigor to these manipulations, but, in all
   cases, there was also an appeal made to an intuited level of interpretation
   and justification which, in the case of Physics at least, yields confusion if
   not contradiction. In no case was there an a priori clear division between
   what is to be merely formal and what is to be intuited.

   The study of Linguistics additionally pointed to the particular relevance of
   the order of written symbols while also noting that the Universal Grammar may
   be taken to derive from an intuited notion of similarity or
   categorization. Consideration of Metamathematics taught that the notions
   Truth and Proof aren't as categorical as might be formally desired, but that
   they are instead grounded, ultimately, in a choice of interpretation of the
   symbolism. Surprisingly, metamathematics, instead of demanding rigid and
   ultimate truths, requires an infinite flexibility in an allowed step-wise
   extension of any sufficiently complex formal system. This extension results
   in, among other things, a rigorous definition of infinities and
   infinitesimals. Experimentation, which was discussed in order to curtail
   this embarrassment of riches, demands, instead, operating with finite
   quantities, but still hasn't categorically solved the problem of the
   interpretation of the formalism.
   
   A concern common to all of the above considerations is, given that language
   is to play a role at all, the need for precisely determining the proper role
   and extent of the respective formalisms. In this connection Wittgenstein's
   dictum\footnote{For an overview of Wittgenstein's philosophy see
   \cite{Pears}.}, that that which cannot be said ought to be passed over in
   silence, seems pointedly relevant. It may, however, be pointed out that he who
   would wholely reject the relevance of language can hardly \textit{state} his
   case! Furthermore, it is hardly possible to \textit{describe} that which is
   outside of language. It seems, besides, that language ought to play some part
   in our experience as without such interpretation that experience 
   is, arguably, a chaotic jumble. None of this, however, resolves the
   difficulties encountered above wherein language referred to something in
   addition to itself. 

  The notion that language might be able to represent and refer to something
  \textit{outside} of itself might seem questionable. Perhaps such a conception
  is even self-contradictory, yet one might also conclude that empty formalism
  is the only alternative to such an approach and feel thereby driven to infer
  some external ``physical" basis for, or \textit{content} of, language.
  Re-examination of the nature of the axiomatic approach will suggest, however,
  that this is a false dichotomy and that one might instead adopt a third
  viewpoint which embraces none of the arbitrary formalisms such as have arisen
  in pure mathematics and yet also avoids such problematic combinations of
  intuition and formalism as have arisen in Physics. The considerations to
  follow outline such a re-examination and give an overview of the argument to
  be presented at length in the remainder of this thesis.
  
  Recall that one lesson of G\"{o}del's Incompleteness Theorem is that any
  sufficiently powerful axiomatic system will have ``holes" corresponding to
  undecidable propositions. There are no stated constraints on how such ``holes"
  may be filled so that, in general, no axiom system is \textit{formally}
  preferable to another. Usually one is thus led to recognize the limitations
  and arbitrary nature of the axioms of any \textit{particular} metamathematical
  system. Not withstanding this result the axiomatic method is still employed
  by both mathematicians and physicists alike in order to draw definite
  conclusions from definite premises. 
  
  The formal symmetry in the admissibility of distinct axiomatic theories is
  usually broken either as a matter of taste, as in the case of pure
  mathematics, or via reference to experiment, as in the case of applied
  mathematics and Physics. That the relationship between empirical data and
  formal structures is indefinite has perhaps already been plausibly indicated
  and will be further advocated in Chapter 2. Any plausibility in these
  observations argues against the second means of identifying relevant
  formalisms though it cannot, by any means, \textit{prove} that such a
  rejection is mandatory. If such a course is adopted, it yet remains to
  \textit{formally} characterize the ``physically relevant" structures within
  the formal theory. In this connection it may be anticipated that the presumed
  indefiniteness in the empirical relevance of \textit{particular} formal
  structures, together with the previously discussed physical requirement that
  ``measurements" be expressible as finite collections of finite (real) numbers,
  leads to the expectation that the finiteness of data might be the
  \textit{only} possible formal requirement that \textit{could} characterize an
  empirically relevant symbolism. This observation identifies an interesting
  class of formalisms to be investigated and gives some hope of success in
  replacing the empirical method with a strictly formal one. 
  
  The role of experiment being at least questionable, the axiomatic method yet
  remains arbitrary in that distinct and incompatible axiomatic systems are, in
  themselves, of equal standing. The selection of \textit{any} particular
  collection of axioms being arbitrary, it would seem that any acceptable
  non-empirical symbolic method must be non-axiomatic. In light of this it may
  be asked how one is to determine a self-sufficient and \textit{constructive}
  symbolic method. An answer: Rather than thinking of G\"{o}del's Incompleteness
  Theorem as being a constraint on \textit{particular} axiom systems it is
  instead possible to interpret it as being a critique of the axiomatic method
  itself. Such a critique is a statement of what the formalisms of all such
  distinct axiomatic systems have \textit{in common}. 
  
  The acceptability of such a statement suggests that the formal manipulations
  of a non-axiomatic system be taken to be those compatible with the adoption of
  \textit{any} particular axiomatic system. This may also be thought of as
  requiring that the formal manipulations demanded by any given axiomatic system
  may be thought of as special restricted cases of the application of the formal
  rules of the non-axiomatic system. Such restrictions in the applicability of
  formal manipulations are determined by the axioms of the given axiomatic
  system so that such systems must be embeddable in the non-axiomatic system.
  
  The axioms of such embedded formal systems may be thought of as summarizing
  the  ``truths" of the experiences of particular individuals so that, if this
  identification is made, it may be seen that the formal rules of the
  non-axiomatic system then ought to be consistent with the testimony of
  ``observers". If such an identification is to be made then it remains to
  at least plausibly illustrate that the ``realistic" structure of everyday
  language may nevertheless be thought of in terms of, and justified by, the
  strictly formal considerations of a non-axiomatic theory. In this sense the
  formalism may be said to generate its own interpretation. Such a development,
  which coincides with an intuitive categorization of symbols such as Linguists
  infer in the operation of the Universal Grammar, is given in Chapter 3.
  Accepting such an identification then allows the derivation of the formal
  manipulations of the non-axiomatic system to be carried out, beginning with
  Chapter 4, in an intuitively plausible manner. 
  
  It should be noted that the non-axiomatic formal approach, by definition, does
  not presume that axioms may not be adopted, but, rather, attempts to
  investigate what is common to all such axiom systems. Definite conclusions
  correspond, as always, to definite axioms. While a non-axiomatic theory may
  make note of such connections it cannot endorse any conclusions as being
  categorically ``true". Conversely, any definite results must be those that may
  be derived on the basis of some axiomatic theory. If this is kept in mind it
  should be clear that this thesis can never make claims that one must
  necessarily accept, nor can it include formal manipulations that amount to
  proofs in the absense of reference to any axioms. In this sense a
  non-axiomatic theory may be said to be based upon the notion of contingency
  rather than truth. It may also be noted that, in accordance with the notion of
  a non-axiomatic theory, all non-axiomatic theories may be considered to be
  identical. Thus the formal structure of \textit{the} non-axiomatic formal
  system will be identified in this thesis as being that of the General Symbolic
  System.
  
  In order to make the above ideas clearer it may be helpful to make a detour
  in order to compare these ideas with some other historically important ideas
  not already discussed. It seems plausible, first of all, that if the
  non-axiomatic method is to be self-sufficient, then it might profitably be
  compared to the approach of various mathematical schools of thought. This
  comparison will then lead to a consideration of some of the philosophical
  ideas of Wittgenstein.
  
  Historically, there have been three main schools of mathematical
  thought, namely: Formalism, Logicism, and Intuitionism. It is easy to
  quickly point out differences between the non-axiomatic method
  and the approach of each of these schools. Formalism has already been
  discredited by G\"{o}del, but the symbolic method to be developed here
  manifestly differs from Formalism in the sense that there is to be no object
  theory separate from any intuitive metatheory. Logicism attempts to construct
  all of formal mathematics starting from a propositional system of assertions,
  assertions such as that a particular mathematical object has a particular
  property. The point of divergence between the approach developed here and that
  of Logicism is, as is explained in section 5 of Chapter 3, that the general
  symbolic system doesn't adopt any particular kinds of mathematical
  objects as being such final categories. Finally, the Intuitionists are known
  for their rejection of the notions of mathematical infinities and of the Law
  of the Excluded Middle, which mandates that all statements are either true or
  false. The above approach, while evidently sharing the Intuitionistic attitude
  to truth, will distinguish itself from Intuitionism in fully embracing formal
  infinities.
  
  Having distinguished the desired formalism from that of all standard
  mathematical schools of thought it would seem that the above ideas are on very
  philosophically shaky ground. In this vein it will be interesting to consider
  some of Wittgenstein's ideas. 
  
  One of the foundational goals of Wittgenstein's philosophy\footnote{Refer
  again to \cite{Pears} for a discussion of Wittgenstein's philosophy.} is to
  prevent the imprecise use of language. Wittgenstein felt, in fact, that with a
  precise understanding of language there could be no questions, properly
  speaking, that could not be answered. In such a case the long-standing
  paradoxes of philosophy would have to therefor be rooted in an improper use of
  language. While his later philosophy attempted to explore language
  empirically\footnote{These ideas were published posthumously in
  \cite{Wittgenstein1}.}, and is therefor apparently to be rejected here, his
  earlier philosophy, like the program to be pursued here, attempted to
  understand the nature of experience through an analysis of pure
  formalism\footnote{Wittgenstein's earlier philosophy is given by him in
  \cite{Wittgenstein2}.} and therefor might be akin to the ideas to be developed
  here. 
  
  Although Wittgenstein's early ideas do proceed from a Logicist viewpoint, they
  will nevertheless be discussed. Wittgenstein espoused the idea that
  language gave a ``picture" of reality, wherein the structure of language
  \textit{exactly} mirrored the structure of an external reality. Wittgenstein
  thus postulates a neumenal world of ideas, similar to Plato's idea of forms,
  parallel to a physical reality. Such an approach differs from the theory to be
  developed here in necessarily speaking of an external reality, and it seems
  suspicious that language and physical reality might coincidentally have
  identical structures, but the fact that the nature of language is to exactly
  determine the nature of experience is nevertheless quite similar to the idea
  that, in the general symbolic system, the formalism generates its own
  interpretation.
  
  In further elaborating his idea of the ``picture" Wittgenstein came to the
  conclusion that physical reality is, like ordinary written language, composed
  of atomic elements corresponding to those symbols representing ideas which are
  not further analyzed. In accordance with his Logicist views he also, therefor,
  came to the conclusion that language, and therefor physical reality also, is
  ultimately tautological in nature, there being no a priori way to distinguish
  atomic elements of reality. This led him to later reject the results of these
  investigations because he could not reconcile the complete symmetry between
  the atoms of his neumenal world and the apparent diversity of the physical
  world. 
  
  Wittgenstein's early ideas may be compared to those of this thesis in the
  following way. While the structure of language is the only structure there
  will be that \textit{may} be discussed in the general symbolic system, it is
  not presumed in the developments to follow that there will be atoms in the
  symbolic system into which all other expressions may be analyzed. Such an
  analysis amounts to a proof in a formal system, while such atoms correspond to
  axioms. The general symbolic system rejects, by definition, such a conclusion.
  Instead of this, as is explained in section 9 of Chapter 7, the considerations
  of the general symbolic system will lead to finite \textit{levels} of
  description which may be further elaborated to an arbitrary degree of
  complexity. Additionally, acceptance of the methods of this thesis leads,
  unlike Wittgenstein's early work, to the development, in Chapter 7, of a
  purely formal symbolic system to a point where the structure of language
  apparently \textit{does} accurately reflect the diversity of experience. This
  is all that will be said of Wittgenstein's ideas here, and discussion now
  returns to an overview of the mode of development of the general symbolic
  system.
  
  Beyond the synthetic definition of, and presumed uniqueness of, the general
  symbolic system, it remains to indicate the particular formal manipulations
  which it should adopt. Here it suffices to generalize the formal rules of a
  particular simple axiomatic system as, after all, such structure must be
  embedded in the general symbolic system itself. Such a requirement is, as is
  indicated in Chapter 4 beginning with section 4, not merely restrictive, but
  decisive: The general symbolic system embeds the formal structure of Lattices
  and is thereby identified to be a division ring with a partial order.
  
  The notions of conventional mathematics are developed with a bias in favor of
  finite quantities, this perhaps being associated with the atomism which
  corresponds to proofs in axiomatic theories. The methods of the Calculus, in
  their reliance on the $\epsilon$-$\delta$ definition of a limit, are a case in
  point. Though finite quantities are perhaps favored by empiricists, no such
  preference is to be found in the basis for the development of the general
  symbolic system. Both for the sake of its own development, as well as for the
  sake of maintaining its comparability to the usual mathematics, Chapter 5 is
  devoted to the task of generalizing, within the general symbolic system, the
  formal operations of the Calculus. This generalization takes particular
  advantage of the non-commuting ``multiplication" of the general symbolic
  system and thus echoes the stress which Linguistic analysis laid on the
  importance of the notion of order in language. Furthermore, Chapter 6 takes
  advantage of the generalized Calculus in order to arrive at some final
  conclusions as to the algebraic nature of the formal rules of the general
  symbolic system.
  
  With all that has been said above, it should still be recalled that the
  purpose of the development of the general symbolic system is the investigation
  of the \textit{possibility} that a purely formal, yet non-axiomatic, theory
  might nevertheless have a structure which yields predictions which agree with
  experiment. Such a program may be thought of as follows: If it fails, then
  this might be taken as a ``proof" that experiment is needed in order to
  ``cause" physical theory to correspond to experience. If it succeeds, however,
  then it is difficult to see how such a cause may be asserted to be necessary:
  Occam's Razor argues against such a supposition. 
  
  As already noted above, the \textit{presumed} formal characterization of
  empirically relevant symbolic structure is that of the finiteness of such
  symbolic structures. As the formal rules of the general symbolic system are
  taken to have been completed in Chapter 6, the first six sections of Chapter 7
  are dedicated to exploring such special structures within the general symbolic
  system. This investigation leads to novel derivations of the formalisms of
  both General Relativity and Quantum Theory, as well as unsurprising
  characterizations of each as corresponding, respectively, to invertible and
  non-invertible manipulations of symbols. In so far as the empirical results of
  all of modern physics is conventionally taken to be encompassed within these
  formalisms it might perhaps also be concluded that the general symbolic
  system, in conjunction with the condition of structural finiteness, has
  succeeded to the same extent as empiricism in yielding an accurate
  determination of the nature of experience. 
  
  In the seventh section of Chapter 7 it is argued that the nature of the
  derivations of the Relativistic and Quantum-theoretical formalisms indicates
  that, if such is possible, all finite structure may be represented within
  either of these complementary schemes, though, as is characteristic of the
  non-axiomatic method, such a conclusion should not be thought to have been
  \textit{proven} to be mandatory. The next section of Chapter 7 inclines, in
  fact, to the conclusion that such \textit{finite} structures are formally
  incompatible. Resolving this conflict presumably requires either the
  consideration of non-finite formal structures or resorting to empiricism. 
  
  As the non-axiomatic method is conceptually consistent with adopting
  non-finite formal structures, the feasibility of adopting this first
  resolution of the above difficulty is investigated in the last section of
  Chapter 7. These considerations lead to the plausible conclusion that the
  Relativistic and Quantum-theoretical formalisms may be extended, in a mutually
  consistent manner, by the methods of Non-Standard Analysis. In particular, the
  formal manipulations of the general symbolic system are to be governed by
  \textit{both} of the formalisms of Relativity and Quantum Theory, the only
  difference between this and the finitary case being that the symbols are no
  longer required to be finite in general. Established non-standard methods are
  then used to illustrate the relationship between this extended formal system
  and the usual finitary formalism which is thus embedded within it. 
  
  As a final observation, the role which Non-Standard Analysis plays in
  presumably reconciling Relativity and Quantum Theory as parts of the general
  symbolic system suggests that an identification may be made between the notion
  of the externality of the ``physical reality" which finite empiricism forever
  appeals to and the presumed conceptual inadequacy of any finite structures
  within the general symbolic system. If this identification is made then, it is
  proposed, these two approaches may be thought of as being indistinct. While it
  would be inconsistent with the very notion of a non-axiomatic theory to demand
  such an identification, nevertheless it is hoped that such an identification
  might make rejecting serious consideration of such systems less likely.

%% file: rpichap2.tex
 
\chapter{A SCHEMATIC OVERVIEW OF PHYSICS}

\section{Introduction}

   The introductory chapter has provided a general orientation and motivation
   for the approach to be taken in this paper. The primary question raised by
   the discussion there is the necessity of language referring to an external
   ``reality" even in such an objective discipline as Physics. Before proceeding
   to the formal development of the theory it will be helpful to re-examine, in
   a practical and more detailed way, the functioning of the scientific method,
   including the actual style of activity of both individual kinds of physicists
   and some of the historical development of Physics itself. It is hoped that
   then, after having more comprehensively illustrated the dubiousness of there
   having been a clear distinction between formalism and ``realism" or
   empiricism, it will be possible to proceed more convincingly with the formal development to follow. 
 
\section{The Ideal Relationship Between Theory and Experiment}
 
   The growth and functioning of modern Physics can be understood in terms of a symbiotic relationship between two general activities of physicists: That of theory construction on the one hand, and that of observation on the other.

   Theoretical physicists make predictions as to what will happen next in a given situation, and thus explain such events whenever they're successful. Theoreticians also attempt to encompass as many conceivable situations as possible within as few distinct predictive schemes, or theories, as possible. Physics thus progresses whenever theoretical physicists are able to construct a theory to explain phenomena not previously understood, to predict new phenomena, or to ``unify"\footnote{ Hereafter the convention is adopted that the quotation of a word or phrase which is used, in context, in a definite sense will, if applicable, be presumed afterwards to indicate the definition of that word or phrase.} old theories by creating a single theory which explains what was previously explained by two or more old ones. This activity yields a kind of unified understanding of the world.

   Theoreticians can't claim to have ever explained everything, nor can it reasonably be claimed that all of their predictions have been tested or confirmed. Apparently, the world is too uncooperative for that. There has thus been a definite role to be played by physicists who don't merely predict, but who also observe.

   Some of the terms used to describe these observations repeatedly occur in varying combinations. The recurrence of such ``events", which usually have relatively simple descriptions, apparently gives an opportunity for an exhaustive and methodical exploration of a part of experience. This exploration is undertaken by the experimental method.

   By adopting conventions for the terms to be utilized in description it becomes possible to compare observations and thus to engage in ``measurement". The recurring conditions which define events also identify parts of description which refer to ``measuring devices". Measurement, by organizing descriptions in a conventional form, serves to help determine the extent to which theory succeeds, and even the failure of theory to explain a measurement still represents an advance for Physics, for it increases knowledge of the world and gives new challenges to the theoreticians.

   It should be clear, then, that theory and observation reciprocally benefit the understanding which physicists aim for. Theory, as confirmed by experiment, serves as a basis for prediction and specifies and clarifies what is known, while observation identifies what is yet to be understood whenever unexplained measurements arise. In any event, theory cannot fruitfully deny the observations made in experimentation.

\section{A Realistic Examination of the Behavior of Scientists}

   While, as just indicated, there are certainly mutually beneficial aspects to
   the relationship between theory and experiment which, in a manner of
   speaking, drive the engine of Science forward, there are also conflicts which
   arise in their interplay. This is, in part, because it is often difficult to
   recognize and accept when a previously successful system of concepts can no longer be relied upon when attempting to understand new experiences. It may even be said that theory preconditions what is noticed. There is a consequent on-going tension between groups of physicists, each comprising a school of thought, when their theories are faced with the extensive variety of experimental data. This tension may be seen in the debates between such schools of thought. 

   For this reason the sketch of the activities of physicists made in the last section, in stressing the mutually beneficial aspects of the relationship between theory and experiment, bears some further development. To this end some further general remarks about theory will be made, and it will be shown that physical debate has taken a predictable form. 
   
   In a physical theory prediction is a strictly mathematical operation: Given certain mathematical objects representing what is known, in the form of initial and boundary conditions, parameters, and physical constants, the theory then generates, in a well-defined way, other mathematical objects which are supposed to tell something of what is to happen later, in the future. The predicted future, in being derived from what is currently known, cannot be any more precisely defined than what is already known. Now the transformation by which the known thus yields predictions may or may not be invertible. If it is invertible, then prediction may be said to be a sort of relabeling, and physicists would, in this sense, know exactly as much about the future as they know about the present. It is usually otherwise, the future being somewhat less accessible than the present. When a theory's predictive transformation is not invertible, when the future is somewhat indefinite, then prediction may be said to be ``statistical", by definition.

   This consideration helps to explain the way in which experiments are, in fact, interpreted. If a theory yields, apparently, more information about the future than the past, then it may be concluded that some of the description of what is known is superfluous or physically irrelevant. If, alternately, prediction is less accurate than statements of the known, then, it being desirable to know the future as completely as possible, this deficiency may be attributed to the world being essentially statistical in nature, to a defect in the physical theory, or to an inadequacy in the determination of what is known by measurement. Exactly one of these factors must be the culprit, for it would be fruitless and ambiguous to maintain more than one independent cause for the same effect. 

   Physicists who don't believe the world to be essentially statistical and who have faith in the physical theory would then assert that there ``really" are more comprehensive initial data from which one could make better predictions. They would then advocate attempts to measure such data or assert that statistical methods are resorted to for practical reasons. Such has been the case for Classical Thermodynamics and Statistical Mechanics.  

   Physicists who accept such a physical theory and the initial data must assert that the world is essentially statistical in nature. This has been the position of adherents to Quantum Theory as exemplified by Niels Bohr's Copenhagen Interpretation. Bohr, while not denigrating all of the results of the non-statistical General Relativity Theory, nevertheless rejected its methods and maintained the adequacy, or ``completeness", of the statistical Quantum Theory.

   If a physicist accepts the adequacy of the initial data but rejects there being any essential indefiniteness in the world, then they are compelled to fault a physical theory which makes statistical predictions. This was Einstein's reaction to Quantum Theory. Einstein couldn't easily accept the idea that ``God plays with dice", and so he spent his last years searching for a generalization of his relativistic theory which could account for quantum phenomena without necessitating an essential randomness in the world. Such a theory would then replace Quantum Theory\footnote{ For Einstein's presentation of his physical beliefs see \cite{Schilpp}. For an independent and in-depth appraisal of Einstein's realism, see \cite{Fine}.}.

   Most physicists today fall into one of these last two groups, but neither camp can claim a definitive victory so long as Quantum Theory and General Relativity can each claim unique successes and still resist unification. Each position was advocated in the famous Bohr - Einstein Debate. It is commonly believed that Bohr won this debate, and Quantum Theory enjoys a corresponding popular approval among physicists, but this is a position maintained by practical success and not, as indicated above, a position proven to be unassailable. In fact there still remains controversy over the interpretation of Quantum Theory, and so Relativists may remain ``convicted, but not convinced".

  The notion of a statistical prediction has led to an overview of physical procedure and been useful in comparing the canonical schools of thought in contemporary physics. It has done even more: It has convincingly shown that each school of thought is offering a different solution to the same problem of the meaning of physical theory, and that the debate between these schools has taken a predictable form.

\section{A Criticism of the Role of Experiment}

   The preceding analysis has highlighted the difficulty in uniquely
   interpreting the relationship between theory and experiment, and the
   consequent disputes that have arisen. Now experiment is merely a particular
   realm of experience to which language may address itself. In fact, according
   to Niels Bohr \footnote{ See pg. 3 of reference \cite{Bohr}.}; ``...by the
   word `experiment' we can only mean a procedure regarding which we are able to
   communicate to others what we have done and what we have learnt." Thus
   language must, in general, possess the same ambiguity of interpretation as
   Physics if experimentation is still uncritically referred to. Before carelessly accepting such an implication it would be best to critically examine the role of experiment in physics. This will be done in the following and it will lead to some interesting conclusions.

   The defining virtue and test of Science is its ability to predict, where
   prediction, properly speaking, is a matter of telling what will happen in an
   experiment that has never been done before. Such a prediction can only be
   based on something which is presumed independently of the experiment actually
   being performed. Such prior knowledge comprises physical theory and may also
   be taken to \textit{define} ``reality" itself.

   In order that Science may claim permanent relevance it is necessary that the supply of new experiments be inexhaustible, and thus infinite in number. It follows that reality may, in this sense, be said to be infinite as well. It would contradict the definition of reality, however, if observations were to be said to create reality: ``Acts of observation", in so far as such a term is used at all, must correspond to entirely unpredictable events.

   Because a physical theory and the corresponding notion of reality must be presumed in order to make predictions, it is impossible, especially by experiment, to prove its correctness. It may, however, be ``falsified" \footnote{ For a detailed discussion of this notion see reference \cite{Popper}.} by its inability to make acceptable predictions.

  Consider, for example, the experimental results that are in accordance with
  Newton's Theory of Gravitation. These same results are in accord with
  Einstein's Theory of Gravitation as well, so that the same experimental data
  have received two quite distinct theoretical explanations. From an empirical
  standpoint, Einstein's theory is distinguished from Newton's in that it is
  able to naturally account for additional phenomena which Newton's theory can
  address only with great difficulty if at all. Einstein's theory is in this
  sense scientifically preferable to Newton's, although there can be, as
  indicated above, no empirical guarantee that future observation won't favor another theory over Einstein's.

   There cannot, therefor, be a direct path from experiment to theory, and consequently no way to empirically determine which individual theoretical assumptions are to be made or rejected in the construction of a theory: It is necessarily the theoretical system as a whole which must confront experiment. This indicates a kind of miracle, for any physical theory, including those that agree with experiment, must then be constructed in a way that has no well-defined connection with experimental results.

   Considering the nebulous relationship between theory construction and the results of experimentation, it may well be wondered whether or not an attempt to construct physical theory without reference to experiment could be successful. In fact, if experiment is to be relied upon then it, in being inexhaustibly novel, can never allow for a final physical theory. In other words, a permanent role for experiment implies that physical theory forever be provisionally acceptable at best, and that there can be no place for necessity in Physics.

   Conversely, if definite conclusions are desired, then it is ultimately necessary to ignore experiment in the construction of physical theory. This may seem like a heretical idea, but it is the only choice to make if, instead of rejecting the idea out of hand, it is desired to determine if physical theory can be constructed in this way. This is a worthwhile question to resolve, so this presumption will be made at this point.

   Summing up then, it is usually assumed that there are two separate realms to
   be recognized: Physical theory and the reality to which it refers. Experience
   is then supposed to guide the construction of the physical theories which
   describe it, but it has been argued that there cannot be any defensible
   systematic procedure for incorporating the results of experiment into a
   physical theory. Accepting this conclusion leads to the realization that
   physical theories must therefor be miracles of independent creation which,
   nevertheless, do make predictions that agree with experiment, at least
   provisionally\footnote{ For an extensive argumentation as to the freedom
   allowed in theory construction in Science, consult \cite{Poincare}.}. This
   being the case, it must then be possible to create physical theory without
   utilizing any experimental results whatsoever, and thus to abolish
   consideration of experimentation as a realm separate from physical theory
   itself.

%% file: rpichap3.tex

\chapter{THE UNIFICATION OF LANGUAGE AND EXPERIENCE}

\section{Introduction}

   In order to follow the course set out in the last chapter it will be
   necessary, first of all, to indicate how a language which makes reference to
   experiences need not, in fact, be referring to anything outside itself. This
   may be done by indicating how the ``realistic" way in which language is used
   may be interpreted in a strictly formal sense. If this is to be done
   convincingly then a number of intuitively understood terms must be given
   explicit formal roles within language and the structure which is normally
   thought to be found within a reality external to language, as indicated in
   language itself, must be shown to have a strictly formal derivation. The
   discussion in this chapter will be carried out in conventional language and
   will be, in this sense, metamathematical. This should not, however, lead to
   the same difficulties indicated in the first chapter. This is because it is
   proposed that the formal system to be developed is not an ``object language"
   which is separate from and subordinate to conventional language, but is,
   rather, equivalent to it. In other words, it is not presumed that the
   language developed is a formal language which may be intuitively judged in
   relation to a metalanguage, but instead the language itself is designed to
   incorporate and symbolize the intuitions of individuals. This is the
   condition which, as has been indicated, is required by the identification of
   the axiomatic systems embedded within the general symbolic system with the
   testimony of individuals.

\section{Individuals and Language}

   Whatever applies to the relationship between theory and experiment will also
   affect the more general relationship between language and experience. In
   particular, the choice just made in the last chapter requires that experience
   not be considered separate from language itself. Nevertheless, language still
   contains statements referring to that which is experienced, and so such
   statements must remain intelligible. Thus statements about experience must be
   both accommodated and yet denied to refer to a separate category outside of
   language. This may seem paradoxical.

   The unique solution to this difficulty which is adopted here is to identify
   the recognized significance of any statements made by individuals about their
   experiences with these statements themselves, so that such statements may be
   taken be a natural development within the language itself. Then all
   statements may be made in the form of references to experiences and yet no
   such separate realm need be recognized. Any ``separation" will be a feature
   of the language itself. This identification has a number of consequences. 

   As all statements in language might be the assertions of some individual at
   some particular instant, it would seem to follow that all statements must
   have realizations in experience. However, because such statements, being
   merely strings of symbols, can only be judged according to their formal
   properties, it follows that while it may be intuitively appealing to base the
   considerations to follow on realistic explanations and examples, such an
   approach may be misleading, for it is easy to forget that such examples are,
   themselves, only to be judged according to their formal properties. The
   developments to follow will, accordingly, be kept, so far as seems
   reasonable, strictly formal and abstract.

   It is evident that great care must be taken in dealing with statements about
   experience from an individual perspective. The delicate matter of
   constructing a strictly formal symbolic system which nevertheless
   accommodates the testimony of individuals is the task taken up next.

\section{Perception and Meaning}

   The construction of the symbolic system starts, as indicated above, from the
   perspective of a single individual. It is assumed that the reader is an
   individual and therefor may understand the meaning of the terms ``reader" and
   ``individual" without a definition being provided.

   It may be said that the symbolic system is to be utilized by this individual
   for the purpose of comprehending his experiences. In keeping with the
   individual perspective it may also be said that symbols may be utilized in
   the description of experiences at a particular instant. The developments to
   follow will be based on the consideration of the description of experiences
   at a particular instant, and the symbols will be said to identify what is
   experienced. The sense, if any, of these terms beyond their purely formal
   utilization is left to the reader's insight.

   Description will be provided by the combination of written symbols. The
   assignment, by an individual, of certain symbols to a particular instant will
   be said to comprise an act of ``perception." Such symbols will, in virtue of
   this assignment, be said to have been made ``meaningful" to this individual.
   Because a given symbol either definitely does, or does not, appear in a given
   statement, it follows that a given symbol is, or is not, meaningful to an
   individual at a particular instant. Moreover, not all symbols need be meaningful.

   Because written symbols may themselves be said to be experienced, it
   necessarily follows that perception may result in symbols being assigned, by
   an individual, to other symbols. This ``meta-labeling" provides a means of
   inter-relating symbols because any symbols assigned to the same symbol are,
   in virtue of this very fact, grouped together. Such symbols may be said to be
   ``associated." 

   The association of symbols just mentioned raises the question of whether or
   not the development of the symbolic system can give anything distinct from
   set or class theory. In these theories the ``existence" of sets or classes is
   required so long as they don't violate any of the formal axioms of the
   theory. Such ``existence" has not been mandated here, for the written strings
   of symbols are thus far presumed to correspond to testimony, and be
   merely given, and are not required by some formally stated rules to be
   formed. So long as this remains the case such an identification would be
   unjustified. This point will be returned to later, in the derivation of
   physical theories, where a condition for the existence of associations
   between symbols will be given.

\section{An Important Illustrative Example}

   The meaning of a particular symbol is, at best, a necessarily private matter.
   It is not generally possible to say ``what" a particular meaningful symbol
   means. It is only possible to note that certain symbols are consistently
   assigned to others. This situation may be made intuitively as well as
   formally clear with an example involving two hypothetical individuals, Joe
   and Tom, say, who may be supposed to refer to certain symbols common to both
   of them.

   Let it be supposed that the symbols ``stop signal" and ``apple" may be
   referred to by both Joe and Tom, and that each indicates to the other that he
   takes these symbols to be meaningful. Now ``stop signal" and ``apple" may be,
   for the sake of argument, assigned in Joe's mind to an experience identified
   as ``A", while ``go signal" is assigned in his mind to the experience
   indicated by ``B". It is not generally presumed that an individual's mind may
   be read like this, but this fiction is indulged in just to make a point.

   Even supposing that Tom shares the same experiences to which ``A" and ``B"
   refer, it is perfectly admissible for Tom's corresponding experiences to be
   the reverse of Joe's. Joe and Tom could then agree in saying that the symbols
   ``stop signal" and ``apple" both refer to things which are ``Red" and other
   things, such as the ``go signal" may be agreed to be ``Green". However, Red
   would then be assigned to the experience A in Joe's mind and to the
   experience B in Tom's. The reverse would be the case for Green.

   Any system of symbols corresponding to experience will accommodate such
   examples, and will therefor share in this indefiniteness as to the meaning of
   particular symbols. If, then, a symbol is supposed to have a definite
   meaning, then this meaning must be established solely within an individual's
   mind. It is possible, however, to establish definite relations between
   symbols. In the above example, in fact, the convention was reached that the
   meta-label Red was to have the symbols ``stop signal" and ``apple" assigned
   to it.

   The categories arrived at, such as Red and Green above, can be useful
   conventions which are stipulated according to some symbols having some
   meaning in common. In this case the red things have, according to both Joe
   and Tom, common meanings. Joe takes red things to share the meaning A, while
   Tom sees red things as sharing the meaning B. Again, this does not mean that
   red has the same meaning for Joe and Tom.

   Such categorizations are always strictly provisional: there are no final
   categories and there can be no mention of necessary truths. In the above case
   it can never be asserted, for example, that there cannot be a symbol, such as
   ``Fire Truck" which is not linked in Joe's mind to the experience A. This may
   be because ``Fire Truck" isn't meaningful to Joe. It may be the case that Tom
   wouldn't assign ``Fire Truck" to the experience A either. It may also be the
   case that Tom may never have the experience A and so never be able to
   understand why Joe associates ``apple" and ``stop signal" as being ``red" but
   doesn't group ``Fire Truck" under this metalabel. The usage of a particular
   symbol, such as red, can only be justified and stipulated solely by
   individuals, so that it can never be asserted that a particular symbol need
   apply to a particular experience. Distinct individuals may merely agree or
   disagree in their assignment of symbols.

   The most precision with regard to meaning that may be expected of language,
   then, besides its including individual symbols of indefinite meaning, is that
   it also provide for the construction of provisionally relevant relations or
   groupings of symbols. The meta-labeling system discussed so far differs from
   the system of the usual formal logic in that it is not based on relations
   with a fixed and known number of ``arguments". Meta-labeling allows just such
   definite but flexibly chosen connections between symbols to be established in
   writing as to give partial exposition of the relations it represents. The
   specification of which symbols are associated by a given meta-label
   identifies a definite structure without, as is necessarily the case, making
   the meaning of any symbol evident. Language, in not being reliably able to
   convey a separate meaning, is thus seen to be strictly ``formal" in nature.

\section{The Development of Language and the Notion of Truth}

   How are languages to be judged? If the way in which symbols are combined in
   writing, and which combinations are to be allowed, is established once and
   for all, then it may be said that a particular well-defined language is in
   use. It otherwise is possible to arbitrarily amend the relations expressed by
   the language and, in this sense, the language will have been changed. Given
   various symbols which are connected, by meta-labeling, in definite ways, it
   is generally possible to give definite rules, called the ``grammar", for how
   such symbols may be combined in written expressions. Expressions which
   satisfy the rules may be said to be ``correct." It is thus necessary that a
   well-defined language have a grammar.

   If given more than one language, how may they be compared? Certainly the
   comprehensiveness of a language is a primary consideration. Can a language be
   shown to be comprehensive?

   If a language is developed to the point where several symbols may be
   consistently associated by a meta-label, then this meta-label may be said to
   identify a particular ``thing", and those symbols which it usually associates
   may be taken to be ``properties" of that thing. Once this may be done
   multiple times, it may then be said, in a strictly formal sense, that these
   certain things ``exist" to which the language refers.

   Properties may be assigned to different things, although, as indicated
   before, such categorizations need not be permanent nor are they necessary.
   Assertions, relative to these adopted conventions, may then be made in the
   language, these assertions taking the form of indicating that a thing does or
   does not have a particular property assigned to it. The assertion, by an
   individual, that a thing has a particular property is, like all perceptions,
   an action which can have no necessary external justification. Even the
   individual himself may not provide a reason for a thing to have a property.
   The property is simply a part of the perception of the thing itself.
   
   The discussion above, in showing that the notion of an assertion or
   ``proposition" may be
   derived from the metalabeling which perception results in, should help to make clear
   the relationship between the theory herein developed and the approach of the
   school of thought known as ``Logicism". For, though the formal propositional
   apparatus from which Logicism proceeds has been shown to arise within the
   theory, nevertheless it must not be thought that the theory is equivalent to such an
   approach. 
   
   A difference between the two approaches may, in fact, be discerned
   by considering the old philosophical issue of the discussion of a
   particular object such as, for example, a chair. Whereas a logicist would, by
   the very nature of his approach, be constrained to assert that a chair does,
   ``objectively" speaking, have certain properties, the approach espoused here
   starts from describing the perceptions of a \textit{given} individual, and
   therefor does not dismiss, at the outset, the possibility that individual
   perceptions may differ. Considering a case of discussion of the color of the
   chair in which some individuals ``are" color blind should make clear the
   primary role of \textit{convention}, in which individuals ``agree to agree",
   as opposed to objective truth, and
   therefor also indicate an advantage of the present theory over Logicism.
   While, failing such agreement, effective discussion might be impossible,
   nevertheless undercutting the \textit{possibility} of such disagreement can only be
   justified by arbitrarily dismissing some perceptions in favor of others. 

   Despite the above observations, if the descriptive system is greatly
   developed then it may contain many things and many assertions about them.
   Such a sufficiently developed system might exhaust all of the written
   statements any individual is likely to make in a lifetime. Such a system, in
   virtue of its complexity, might be deemed comprehensive, but, it may be
   asked, does it ``tell the truth"?

   Because experience, as a realm separate from language, has been abolished
   from consideration, there can be no way to judge a language apart from its
   formal properties. Thus there can be no external criterion on which to base
   an assertion of the truth-value of a statement. The only possible assertion
   of the adequacy of a language that may be sought, then, is in arguing that
   the language is, in its own nature, capable of representing any written
   relation. It is also worth noting that in rejecting the notion of truth there
   can consequently be no worry about consistency, nor can G\"{o}del's
   Incompleteness Theorem necessarily apply. 

   The way in which statements ``about experience" can be made in a language
   without the necessity of an external realm of experience has been discussed.
   It has been found that the notions of things, properties, and assertions
   about things can be given purely formal characterizations. The notion of
   intrinsically true statements must, in the context of a non-axiomatic theory,
   be abandoned, but it is still reasonable to talk about the descriptive
   adequacy of a language.

%% file: rpichap4.tex

\chapter{BEGINNING CONSTRUCTION OF THE SYMBOLIC SYSTEM}

\section{Introduction}

   The last chapter addressed the way in which experiential statements may be
   given purely formal interpretations. There certain terms were introduced and
   certain arguments made, but it was not addressed how various formal
   manipulations were to be decided upon. Such a formal development of the general symbolic system is begun in this chapter.

\section{Reserved Symbols and the Formal Nature of Language}

   It is not possible, as has been demonstrated, to in any way prejudge which
   symbols may play a role in individual perception in the general symbolic
   system. If the identification of meaningful symbols is left entirely to the
   individual, then the symbolic system must be such that, for any correctly
   written expression, each potentially meaningful symbol may be replaced by
   another to yield a new expression which may take on, according to individual
   perception, the same meaning as the last. This may be thought of as a
   recoding of the previous expression. This recoding will be the first aspect
   of the symbolic system to be specified.

   In the written symbolic system symbols will be distinguished simply by their appearances as written; 'a' is, for example, distinct from 'b' in this sense. This will be indicated by writing a$\not=$b. It is important to note that no ``equality" has been defined here.

   The convention will be adopted that writing a$\mapsto$b will indicate that
   the symbol 'a' is to be replaced by 'b' whenever it appears in a written
   expression. Suppose, for example, that a$\mapsto$c and b$\mapsto$d. If this
   is done in such a way that if any pair of images, such as 'c' and 'd', are
   such that c$\not=$d then the corresponding pre-images, such as 'a' and 'b',
   are also such that a$\not=$b, then the replacement effected will be said to be a ``relabeling", it being presumed that relabelings are defined on all symbols.

   It will never be the case that x$\not=$x for any symbol 'x', and there is no requirement that an image and its corresponding preimage be distinct. Thus a given symbol may be retained under a given relabeling and, in fact, under all relabelings considered. The symbolic system thus allows that certain symbols may be used in the same way in all expressions in different relabelings. In this way it is possible to always retain particular symbols to play fixed roles in the symbolic system, but it has not been necessary to stipulate the ``existence" of symbols as necessarily given: Such ``reserved symbols" may be conventionally set aside for such use. Both '$\not=$' and '$\mapsto$' will be taken to be reserved symbols. Consequently, given a relabeling in which a$\mapsto$b and c$\mapsto$d, an expression such as a$\not=$c will be replaced by b$\not=$d, it being presumed that $\not=$ $ \mapsto$ $ \not=$ by convention.

   It can be seen that having reserved symbols is the key to the adoption of a conventional language in which symbols are always used in the same way. Ironically, then, it is the very indefiniteness of the meaning of any symbol which will allow a language with rigid rules of combination, with a definite grammar, that is, to be constructed. In other words, the meaninglessness of language will be what allows a definite ``formalism" to be used.

\section{Partial Order and Formal Structure}

   In the last section it was found that the indefiniteness of the meaning of symbols provided a key to introducing reserved symbols which are adopted conventionally and allow for the possibility of a definite language being used. It will now be shown that the meta-labeling between symbols engenders a certain structure which may be explicitly indicated in the language through the use of reserved symbols indicating an order relation between symbols. The meta-labeling also will provide a definite means of combination of individual symbols, thus leading to the notion of an ``expression".

   Consider the assignment of one symbol to another. In the example involving Joe and Tom both 'apple' and 'stop signal' were assigned to 'Red'. It may also have been the case that some symbol had been assigned to 'apple', and some other symbol assigned to that, and so on. This chaining of the connections between symbols motivates a definition.

   It will be said that 'x' is a descendent of 'y' iff

      \hspace{1in}either 'x' is assigned to 'y'

      \hspace{1in}or     'x' is assigned to a descendent of 'y'.

   Obviously this is a recursive definition which is not constructive in the sense that a definite end to the recursion is not to be expected for any reason formally indicated. The assignment of symbols is not, however, something to be questioned: It may be said that it is contradictory to presume that individuals could stipulate such connections in an act of perception and not be able to identify any chain of assignments which would identify descendents of a given meta-label.

   With this definition of descendents it is then possible to construct an order relation between meaningful symbols. It will be said that 'x' is ``less than or equal to" 'y', written x$\le$y, if 'x' is a descendent of 'y' or if it is 'y' itself. 'x' will be said to be ``equal to" 'y', written x = y, if both x$\le$y and y$\le$x hold. Clearly, if x = y then y = x. If x$\le$y but x = y doesn't hold, then x$<$y will be written. It will also be convenient to adopt the following conventions: x$\le$y may also be written as y$\ge$x, and x$<$y may be written as y$>$x. The symbols $\le$, $\ge$, $<$, and $>$ are taken to be reserved symbols.

   Only the structural relationships indicated by meta-labelings can have any significance within the formal system, and these relationships may be conveyed by the reserved symbol $\le$. Suppose, for example, that a$\le$b, a$\mapsto$c, and b$\mapsto$d. Then (a$\le$b)$\mapsto$(c$\le$d), so that the meta-labeling structure between the pre-images 'a' and 'b' is inherited by the images 'c' and 'd'. For this reason it may be said that the images are used ``in the same way" as the corresponding pre-images.

   Because, again, only the meta-labeling structure can be of formal significance, and because the symbol '=' indicates an ``equality" based on meta-labeling, it follows that any two symbols 'x' and 'y' such that x = y may be substituted for one another anywhere within an expression. Thus meta-labeling is the source of ``the principle of substitution."

   With these definitions it is clear that $\le$ satisfies the standard definition of a ``partial order":
\begin{equation} \mbox{x}\le \mbox{x for every x.} \label{eq:part1}\end{equation}
\begin{equation} \mbox{x}\le \mbox{y and y} \le \mbox{z implies that x} \le \mbox{z.} \label{eq:part2}\end{equation}
\begin{equation} \mbox{x} \le \mbox{y and y} \le \mbox{x implies that x = y.} \label{eq:part3}\end{equation}

   If, for meaningful x and y, either x$\le$y or y$\le$x, then x and y will be
   said to be ``comparable". Any system of meaningful symbols, all of which are
   comparable, will be said to form a ``chain" and it may also be said that the
   order $\le$ is then ``total" for these symbols.

   Along with the notion of a partial order there will also be an associated
   notion of a ``bound". Given two meaningful symbols x and y, any symbol w such
   that w$\le$x and w$\le$y will be said to be a ``lower bound" of x and y.
   There need not be any lower bounds of x and y, but there may also be many.
   Among such w there may be at most one unique ``greatest" one, denoted by x$\wedge$y, such that e$\le$x$\wedge$ y for any lower bound e.

   It will be noticed that the string of symbols x$\wedge$y  stands for a single symbol determined by meta-labeling with respect to the two symbols x and y. Thus x$\wedge$y is to be considered as a kind of unit. In order to make this clear it will be convenient to introduce the reserved symbols '(' and ')' and write (x$\wedge$y).

   There will be notions of an ``upper bound" and for a ``least upper bound" for meaningful symbols which are defined in correspondence with the analogous notions above. A unique least upper bound of x and y will be denoted by (x$\vee$y), where $\vee$ is taken to be a reserved symbol. 

   Both of $\wedge$ and $\vee$ act as a means of combining, according to meta-labeling, several individual symbols into one unit. Because, for example, (a$\wedge$b) is considered as a unit, it makes perfect sense to consider a compound such as (a$\wedge$b)$\wedge$c. Such a compound, combining several individual symbols, will be called an ``expression". Expressions will be primary constituents of the written symbolic system.

   The above considerations have introduced $\wedge$ and $\vee$ and found their significance as a means of the formal combination of individual symbols into expressions. It has also been shown how the meta-labeling involved in perception may be explicitly indicated by using the reserved symbol $\le$. These are fundamental steps in the development of the formal system, but there remain further developments to consider.

\section{A Preliminary Consideration of the Formal System}

   Prior discussion has indicated that the general symbolic system to be
   developed here cannot admit of the notion of truth nor proceed on an
   axiomatic basis: Its most concrete realizations of relationships can only be
   expressions of contingency. While no axioms or truths may be embraced, it
   would come to the same thing if any were explicitly rejected either. Thus
   this system must be formally compatible with the \textit{structure} of \textit{any}
   axiomatic system. This observation justifies summarily citing and applying the results
   of a particular axiomatic theory in so far as the general symbolic system
   cannot have formal rules for the manipulation of symbols which are
   incompatible with it. This state of affairs will be taken advantage of in what follows.
   
   A symbolic system with an associated partial order $\le$ and the operations $\wedge$ and $\vee$ may be viewed on the basis of the usual Set theory, and then it comprises what is commonly known as a LATTICE \cite{Birkhoff}. If attention is restricted to only meaningful symbols then the standard developments of Lattice Theory will provide necessary specifications of the symbolic system. For such a structure it can be shown\footnote{For the Lattice-Theoretical results cited in this section see pp.1-17 of reference \cite{Birkhoff}.} that the following results, which, for convenience, are cited as they appear in the above reference, must hold:

   L2. x$\wedge$y=y$\wedge$x and x$\vee$y=y$\vee$x. (COMMUTATIVITY)

   L3. x$\wedge$(y$\wedge$z)=(x$\wedge$y)$\wedge$z and x$\vee$(y$\vee$z)=(x$\vee$y)$\vee$z. (ASSOCIATIVITY)

   CONSISTENCY: x$\le$y is equivalent to imposing either of x$\wedge$y=x and x$\vee$y=y.

   THEOREM 8: Any system with two binary operations which satisfies L2, L3, and CONSISTENCY is a Lattice, and conversely.

   Thus L2, L3, and CONSISTENCY completely characterize Lattices. It must be the
   case, then, that at least one of these conditions must be violated in order that a symbolic system not comprise a Lattice.     

   Further results for Lattices:

   Lemma 3: If y$\le$z then (x$\wedge$y)$\le$(x$\wedge$z) and (x$\vee$y)$\le$(x$\vee$z).

   Lemma 4: (x$\wedge$y)$\vee$(x$\wedge$z)$\le$x$\wedge$(y$\vee$z).

   Suppose that O is such that, for every meaningful symbol x, O$\le$x holds, and that O$^\prime$ is defined identically. It follows that O$\le$O$^\prime\le$O, so that O=O$^\prime$. Thus any such symbol O, and likewise any symbol I such that x$\le$I for any meaningful x, must be unique. Such O and I will be universal bounds for meaningful symbols.

   O and I have been introduced, in part, in order to show that the $\le$ in
   Lemma 4 cannot be strengthened to an equality. This can be shown by
   considering the system of inequalities O$\le$x, O$\le$y, O$\le$z, x$\le$I,
   y$\le$I, and z$\le$I. Then x$\wedge$(y$\vee$z)=x$\wedge$I=x and
   (x$\wedge$y)$\vee$(x$\wedge$z)=O$\vee$O=I, where it's not the case that x=I.
   The equality in Lemma 4 may hold in some Lattices, however. Such Lattices will be called ``DISTRIBUTIVE".

   THEOREM 10: In a distributive lattice, if c$\wedge$x=c$\wedge$y and c$\vee$x=c$\vee$y, then x=y.

   Theorem 10 provides something like a cancellation law for distributive lattices.

   Consider a distributive lattice with O and I. From the given lattice a new
   distributive one may be constructed which extends the original lattice by
   including, for each x in the original lattice, a corresponding new symbol
   x$^\prime$. This new symbol is to be defined solely by the requirement that
   O$\le$x$\le$I and O$\le$x$^\prime\le$I. In such a lattice it will be the case
   that x$\wedge$x$^\prime$=O and x$\vee$x$^\prime$=I. Such an x$^\prime$ will
   be called the ``complement" of x. According to Theorem 10, a symbol x can
   have at most one complement in a distributive lattice. A lattice in which
   each symbol has a complement will be called a ``complemented" lattice. A complemented distributive lattice will be said to be ``BOOLEAN".

   These considerations have been preparation for  

   THEOREM 16: In a Boolean Lattice: 

       L9. (x$^\prime$)$^\prime$=x, and

       L10. (x$\wedge$y)$^\prime$=x$^\prime\vee$y$^\prime$.

   A final lattice theoretical result will be of central importance here:

   THEOREM 7: Any chain is a distributive lattice.
   
   The above survey of Lattice Theory describes a formal structure which may be
   generated according to an individuals metalabeling, so the general symbolic
   system must therefor be able to somehow embody such a structure. It must be
   stressed, however, that the general symbolic system has \textit{not} been
   shown to be a lattice. The reason for this distinction will be given in the
   next section.

\section{The Necessity of Generalizing Lattices}

   With the last result cited in the last section it is appropriate to ask
   whether or not lattices, generated according to the order $\le$ and the
   operations $\wedge$ and $\vee$, are suitable structures for the general
   symbolic system. It is to be expected that they are not, for, in addition to
   being based upon metalabeling structure, lattices stipulate that such
   metalabeling structures must have least upper and greatest lower bounds while
   such requirements cannot be part of a non-axiomatic theory.

   The selection of meaningful symbols, and of the meta-labeling between them,
   is almost entirely unrestricted and left up to the individual. From this it
   follows that any formal rules for symbols which may be adopted for the
   symbolic system must apply whatever the meaningfulness or order relations of
   the symbols may be. Now it is possible for any three particular symbols to
   form, according to individual perception, a chain, so that any acceptable
   formal rule must apply as if the whole symbolic system is a lattice and, in
   fact, a chain. Theorem 7 then indicates that such a lattice must be distributive. 

   The lattice structure is the minimal embodiment, as demonstrated by Theorem
   8, of the partial order $\le$ and the associated operations $\wedge$ and
   $\vee$ which are generated by unrestricted meta-labeling with least upper and
   greatest
   lower bounds. Because, as has
   been shown, free meta-labeling is precisely what may and must be
   conveyed by language, it is unacceptable to have a further restriction placed
   upon such a lattice structure. Thus the general symbolic system must be able to
   accommodate lattices in general, and non-distributive lattices in particular.
   It may be noted, in passing, that the necessity of not restricting the
   general symbolic system to Boolean lattices decisively distinguishes it from
   the mathematical methods acceptable to the Intuitionists. 

   The above considerations lead to the conclusion that the general symbolic
   system must be distinct from, and yet embody, lattices formed by meaningful
   symbols. This may only be accomplished by embedding such lattices in a distinct
   formal system. It will then be necessary that this formal system itself have
   operations which are analogues of the operations $\wedge$ and $\vee$, and
   that these analogous operations satisfy a distributive law.

   In order to justify the program pursued in this thesis it is necessary that
   the formal system may consistently reflect the ``intuited" combinations of
   meaningful symbols perceived by individuals and yet also obey formal rules
   which apply generally. Thus, for any manipulations in the formal
   system, it will still be necessary to be able to identify corresponding
   meaningful expressions and, in this way, be able to assert that the system
   has yielded its own interpretation. Now there will generally be more than one
   expression in the formal system corresponding to a given meaningful one, for
   since a lattice of meaningful symbols may be \textit{embedded} within the
   formal system, it would otherwise be formally identical to such a lattice.
   It necessarily follows that the general symbolic
   system has a ``formal redundancy": The process of ``interpretation" whereby
   meaningful expressions are extracted from expressions in the formal system is
   generally a many-to-one transcription.  
   
   In conclusion, the only apparent shortcoming of the choice of the lattice
   structure as the general symbolic system lies in its inability to maintain
   both the freedom
   necessary in meta-labeling and obedience to generally applicable formal
   rules. Nevertheless, because the general symbolic system must embed the
   lattice structure, a unique and definite means of specifying the structure of
   the formal symbolic system has been indicated. This consists in generalizing
   the formal rules for a distributive lattice while explicitly retaining those
   lattice-theoretical results which don't depend on which symbols are
   meaningful and therefor fall into the order $\le$. Other results may not be
   taken over so directly, but may still suggest extensions. The task of this
   generalization will be taken up in the next section.

\section{The Symbolic System as a Division Ring}

   The general symbolic system will be constructed in a very nearly direct
   analogy with the structure of lattices found above, though it must be
   stressed that what is pursued here is merely the specification of formal
   rules for symbolic manipulation which are not inconsistent with and yet are
   sufficient for
   the embodiment of the lattice structures discussed above. In particular, the
   general symbolic system will necessarily begin with two reserved symbols,
   referred to as operations, denoted by +, called ``addition" and corresponding to $\vee$, and $\ast$, called ``multiplication" and corresponding to $\wedge$. The symbols $\wedge$ and $\vee$ will still be retained in expressions whenever all non-reserved symbols are those that would result from the interpretation of an expression in the general symbolic system, and their appearance will indicate this state of affairs. 

   In the following, certain formal operations will be defined in terms of
   allowed substitutions. Recall that the ``equality" of symbols, indicated by
   =, justified the principle of substitution. The formal symbol = will,
   correspondingly, be taken to be an extension of the equality previously
   defined by metalabeling. Thus = will always justify substitution. Another
   note: The presentation will parallel that of a typical axiomatization of a
   mathematical structure, but it is important to remember that no notion of ``existence" is to be inferred. This point will soon be stressed with an observation relating to an inverse operation for $\ast$.

   Recall that x$\wedge$y was considered to be a symbol which corresponded in a particular way to the symbols x and y. + and $\ast$ will be defined to have the same formal property, so that the ``combinations" that + and $\ast$ yield may be represented by a new symbol.

   Given the combination of symbols a and b indicated by a$\ast$b, this same combination may be written as a new symbol c, where
\begin{equation}\mbox{c = a$\ast$b.}\label{eq:div1}\end{equation}
Given the combination of symbols a and b indicated by a + b, this same combination may be written as a
new symbol c, where
\begin{equation}\mbox{a+b=c.}\label{eq:div2}\end{equation}

   The requirements (\ref{eq:div1}) and (\ref{eq:div2}) indicate that the operations + and $\ast$ are, in the usual terminology, ``closed".

   The associative property, specified above in L3, applies in all lattices and
   must therefor be taken over directly. Thus the following substitutions are permitted:

\begin{equation}(a\ast b)\ast c=a\ast (b\ast c)\label{eq:div3}\end{equation}
\begin{equation} (a+b)+c=a+(b+c)\label{eq:div4}\end{equation}

   The distributive property will, as indicated in the last section, have to be imposed. Thus the substitutions:

\begin{equation} a\ast(b+c)=(a\ast b)+(a\ast c)\label{eq:div5}\end{equation}
\begin{equation} (a+b)\ast c=(a\ast c)+(b\ast c)\label{eq:div6}\end{equation}

   It is necessary to give the second rule, rule (\ref{eq:div6}), because a$\ast$b=b$\ast$a need not hold in general.

   It will be convenient, for the sake of referring to some ideas in the usual
   mathematics, to revert to language using the notion of existence for a
   moment, this being done solely for the sake of motivation of a formal
   substitution rule. In this spirit the cancellation property for distributive
   lattices, which was cited in Theorem 10, leads, as a trial generalization, to
   the investigation of the following standard group-theoretic rules:

   For any symbols a and b there are symbols x and y such that
\begin{equation}\mbox{a$\ast$x=b and y$\ast$a=b.}\label{eq:div7}\end{equation}
   For any symbols a and b there are symbols x and y such that
\begin{equation}\mbox{a+x=b and y+a=b.}\label{eq:div8}\end{equation}

   It is easily shown that rules (\ref{eq:div1}),(\ref{eq:div3}), and (\ref{eq:div7}) lead to the availability of a unique symbol ``1"\footnote{ See pg. 201 of \cite{Hilbert}.} , the ``multiplicative identity", such that 
\begin{equation}\mbox{For every symbol a, a$\ast$1=1$\ast$a=a.}\label{eq:div9}\end{equation}

   In a similar fashion, rules (\ref{eq:div2}),(\ref{eq:div4}), and (\ref{eq:div8}) lead to a unique symbol ``0", the ``additive identity", such that 
\begin{equation}\mbox{For every symbol a, a+0=0+a=a.}\label{eq:div10}\end{equation}

   It can, furthermore, be shown that 0$\ast$a=a$\ast$0=0 for any symbol a. With this in mind it is clear that rule (\ref{eq:div7}) cannot be maintained. 

   Returning now to the consideration of substitution rules, it is clear that the symbols ``0" and ``1" may be reserved, and the rules (\ref{eq:div7}) and (\ref{eq:div8}) are to be replaced by the following rules:

   For any symbol symbol a which is not such that a=0, another symbol, a$^{-1}$, may be written such that
\begin{equation}\mbox{a$\ast$a$^{-1}$=a$^{-1}\ast$a=1.}\label{eq:div11}\end{equation}

   For any symbol a another symbol, -a, may be written such that 
\begin{equation}\mbox{a+(-a)=(-a)+a=0.}\label{eq:div12}\end{equation}

   The expression a+(-b) is conveniently written as a-b. It is easily shown that
   (-1)$\ast$a= -a. a$^{-1}$ will be referred to as the ``multiplicative
   inverse" of a, while -a will be called the ``additive inverse" of a or the
   ``negative" of a. Evidently, the inversion of either multiplication or
   addition serves to erase symbols in written expressions. The operation of
   inversion gives a good opportunity to emphasize that the concept of existence
   is not to be applied. If, in fact, a were to refer to, for example, a ``dog", then what, if anything, might a$^{-1}$ refer to ? The inverse of a dog ?

   Finally, it can be shown \footnote{ See pg. 201 of reference \cite{Hilbert}.} that a certain substitution rule for addition holds: 
\begin{equation}\mbox{a+b=b+a.}\label{eq:div13}\end{equation}

   This property of addition is referred to by saying that addition is ``commutative". It turns out that multiplication is not necessarily commutative\footnote{ See pp. 92-94 of \cite{Hilbert}.}.

   The rules (\ref{eq:div1}), (\ref{eq:div2}), (\ref{eq:div3}), (\ref{eq:div4}),
   (\ref{eq:div5}), (\ref{eq:div6}), (\ref{eq:div9}), (\ref{eq:div10}),
   (\ref{eq:div11}), (\ref{eq:div12}), and (\ref{eq:div13}), taken together,
   indicate that the general symbolic system comprises what is generally known
   as a ``division ring" or a ``skew field". It should be recalled that these
   rules merely specify the way in which symbols are to be formally manipulated,
   so that, for example, the ``existence" of an inverse of a symbol is strictly
   equivalent to erasure of that particular symbol being allowed. In this sense
   algebra is here regarded to be a branch of orthography. 

   This is the extent of the exploration of the formal algebraic properties of the symbolic system that will be undertaken at this point. It remains to investigate the role of the order $\le$ in the symbolic system, and this task will be taken up in the next section.

\section{Order in the Symbolic System}

   The order in a lattice is generated and defined by the meta-labeling
   inter-relationships between meaningful symbols, and its sole purpose is the
   conveyance of this information. The general symbolic system is to be a proper
   extension of a lattice in which not all symbols may correspond to meaningful
   ones. Since not all symbols, even after interpretation, may be meaningful, it
   follows that no order can be defined, according to meta-labeling, on all of
   the symbolic system. This has the obvious consequence that Induction, and
   with it Proof Theory, cannot, in any sense, be carried out in the general
   symbolic system. It also forecloses any paradoxes associated with the Axiom
   of Choice. It will be the case, however, that part of the symbolic system
   will have an order defined on it: This will be a partial order such as is
   induced under the embedding of a lattice of meaningful symbols. There is a
   definite necessity, then, for extending the notion of order to the formal
   symbolic system.

   Those symbols which correspond to meaningful symbols will still obey the laws found before for partial orders:

\begin{equation}\mbox{x$\le$x for every x.}\label{eq:part1a}\end{equation}
\begin{equation}\mbox{x$\le$y and y$\le$z implies that x$\le$z.}\label{eq:part2a}\end{equation}
\begin{equation}\mbox{x$\le$y and y$\le$x implies that x = y.}\label{eq:part3a}\end{equation}

   The uncritical transcription of Lemma 3 would read:

\begin{equation}\mbox{If y$\le$z then x$\ast$y$\le$x$\ast$z and x+y$\le$x+z for any x.}\label{eq:lemma3}\end{equation}

   Can this rule be accepted as it is? Consider a symbol y$\le$0. From the
   second half of (\ref{eq:lemma3}) it follows that 0$\le$ -y. Thus, for any
   meaningful symbol y, exactly one of y = 0, y $<$ 0, or -y $<$ 0 must hold. Now consider y $<$ 0 and x = -1 and apply the first half of (\ref{eq:lemma3}). Then the conclusion would be reached that both y $<$ 0 and -y $<$ 0, but this definitely cannot be accepted.  

   It is thus required that x be restricted in the first half of (\ref{eq:lemma3}), so that the condition x $>$ 0 is adopted. This results in the adoption of the two rules 

\begin{equation}\mbox{If y $<$ z and x $>$ 0 then x$\ast$y $<$ x$\ast$z.}\label{eq:part4}\end{equation}
\begin{equation}\mbox{If y $<$ z then x+y $<$ x+z for any x.}\label{eq:part5}\end{equation}

   It can now be concluded that the symbolic system is a division ring on part
   of which an order $\le$ obeying the requirements (\ref{eq:part1a}) through
   (\ref{eq:part3a}), and (\ref{eq:part4}) and (\ref{eq:part5}), is defined. The nature of the symbolic system is still very abstract, but a more concrete picture will start to emerge in the next section.

\section{Finiteness and Some Concrete Examples}  

   Now that the the abstract algebraic and ordinal properties of the general symbolic system have been somewhat developed, it is possible to exhibit some of the usual mathematical objects that should be expected to find a place within the system. These considerations will start with the fundamental notions of finiteness and infiniteness, and will then proceed to the construction of symbols that multiplicatively commute with all other symbols, as well as to the construction of symbols with tailor-made commutation properties. These developments will illustrate the non-trivial nature of the rules thus far adopted.

   Consider a symbol s to which various other symbols are assigned. Suppose that
   x is one such assign, so that x is a descendent of s but not of any other
   descendent of s. Construct a new symbol s/x such that s/x has the same
   assigns as s except it omits x. If there is not a one-to-one relabeling from the assigns of s to those of s/x, then s will be said to have ``finitely many" assigns. Otherwise, s will be said to have ``infinitely many" assigns.

   If a symbol s has finitely many assigns, then it will be possible to explicitly arrange them in a terminating list. Given such a list a corresponding sum 
\begin{equation}\mbox{m = $\underbrace{1+1+1+\cdots+1}_{\mbox{m terms}}$}\end{equation}
 may be derived, where the summands 1 correspond to the distinct assigns of s. s will then be said to have m assigns and 1 will have been added ``m times". Such m will themselves be said to be ``finite", and also to be ``Natural Numbers". Natural numbers serve to ``count" the assigns of a given finite symbol.

   Besides this elementary construction of the natural numbers, other constructions starting from 1 and using finitely many applications of the operations +, -, $\ast$, and ($\cdot$)$^{-1}$, but not requiring (0)$^{-1}$, result, in the obvious way, in the generation of the Integers and Rational numbers.

   Consider a natural number m and any other symbol q. Then 
$$\mbox{m$\ast$q = (1+1+1+$\cdots$+1)$\ast$q = (q+q+q+$\cdots$+q)}$$ and
$$\mbox{q$\ast$m = q$\ast$(1+1+1+$\cdots$+1) = (q+q+q+$\cdots$+q)}.$$
so that m$\ast$q = q$\ast$m for any q. Thus the naturals commute with all symbols. So too will the integers and rationals.

   Consider now those symbols which are generated by finitely many operations +, -, $\ast$, and ($\cdot$)$^{-1}$ and finitely many other symbols x$_{1}$, x$_{2}$,$\cdots$,x$_m$. The resulting symbols will be rational polynomials in x$_{1}$, x$_{2}$,$\cdots$,x$_m$. If, as the x$_{1}$, x$_{2}$,$\cdots$,x$_m$ are varied, the same sequence of operations is applied to the new x$_i$, then a function will have been established. Thus all of elementary algebraic mathematics is available in the symbolic system.

   Besides the development of this algebra, it is also of basic importance to investigate the constructive aspects of the commutativity of the $\ast$ operation. In this case it has already been constructively established by Hilbert \footnote{ See pp. 92-94 of reference \cite{Hilbert}.} that the rules for a division ring with order allow for symbols with tailor-made commutation properties, so that this detail need not be further addressed.

   In this section some finite constructions within the symbolic system have been investigated. It has been found that the usual algebraic objects of elementary mathematics are to be found in the symbolic system. The integers and rationals, in particular, provide such examples of symbols which, in addition, commute with every other symbol. It has also been found that non-commuting symbols may, as is necessarily the case, be formally constructed. These considerations have focused on finite manipulations of symbols, but it remains to investigate infinite manipulations. This will be done in the next chapter.

%% file: rpichap5.tex

\chapter{THE CALCULUS GENERALIZED}

\section{Introduction}

   The usual presentation of mathematics is developed in terms of finite quantities: Functions and numbers are, for the most part, finite. The Calculus is developed in terms of the concept of a limit which, again, relies on finite quantities. It will be important for the sake of later considerations that this prejudice not be maintained. Then, whenever quantities turn out to be, or are assumed to be, finite, this state of affairs can be explicitly recognized rather than tacitly assumed. 

   In the sections to follow a general theory of functions, differentiation, and
   integration will be developed within the general symbolic system which
   doesn't presume that the quantities involved are necessarily finite: The
   formal methods developed can be applied independently of this condition and
   yet extend the results given by the standard finite version of the Calculus.

\section{Functions}

   Polynomials, as has been seen, give an example of functions, in the usual
   sense, in the general symbolic system. It will be necessary to give, however,
   a general characterization of functions in the general symbolic system which
   doesn't rely on their having been finitely generated. The necessity of
   maintaining the formal structure of statements under relabeling will show
   that a particular modified notion of a function is required.

   Suppose A = a$_1\ast$a$_2\ast$a$_3\ast\cdots\ast$a$_n\equiv\prod^{n}_{j=1}$a$_j$. If the a$_j$ are allowed to vary, that is, a relabeling is applied where each a$_i$ is replaced by a new ``argument", then this product will give a simple example of a function of many variables. Whatever general definition of a function is adopted, it should recognize this product as giving a function.

   Let the a$_j$ vary through, in particular, a relabeling a$_j\mapsto$q$_j$. The resulting product must be formally identical to the original and have the same interpretation. Thus each a$_j$ must vary in the same way, as given by q$_k$=w(a$_k$), where w is some well-defined rule for relabeling individual variables. It must then be the case that w(A) = w($\prod^n_{k=1}$a$_k$) = $\prod^n_{k=1}$w(a$_k$) = $\prod^n_{k=1}$q$_k$. As a special case this gives w(x$^2$) = w(x)$^2$; This will determine w.

   Choose any v$\not=$0. Then an equation may be written involving a symbol u, u not corresponding solely to x, such that w(x)$\ast$v = u$\ast$x. Now w(x$^2$) = u$\ast$x$^2\ast$v$^{-1}$ = w(x)$^2$ = u$\ast$x$\ast$v$^{-1}\ast$u$\ast$x$\ast$v$^{-1}$, so that u = v. Thus relabelings w$_u$ are given, in general by w$_u$(x) = u$\ast$x$\ast$u$^{-1}$ for u$\not=$0. In order that these relabelings may be non-trivial it is only necessary that u not commute with all of the arguments of the function to be defined. This indicates the important role to be played by the availability of symbols which don't commute with given symbols, for if no such symbols are available then even a simple product would be incompatible with relabeling, let alone a natural definition of a function motivated by it.

   It will be noticed, in the case of the product A = $\prod^n_{j=1}$a$_j$, that any symbol which commutes with all of the a$_j$ also commutes with A. This will serve as a defining characteristic of a function. 

   The considerations to follow in the development of the theory of functions
   and differentiation, while proceeding from different motivation, nevertheless
   formally parallel those already given by Dirac \footnote{ See reference
   \cite{Dirac}.} in a paper addressing algebra in quantum theory. In
   particular, Dirac formalized operations already found in Quantum Theory and
   suggested formal axioms in order to carry out these desired operations. He
   also worked within the framework of standard mathematics with finite
   quantities and accepted the notion of mathematical truth. His proposals for
   the definition of functions and differentiation were made for application to
   the Quantum Theory in particular, but were not proposed as replacements for
   the corresponding notions in mathematics itself. Here the developments
   proceed from a non-physical motivation and basis, and mathematical quantities
   are not necessarily finite. With this difference between Dirac's approach and
   the approach of this thesis in mind, the results for these subjects
   to be given below will be merely cited, but not formally derived here, unless
   they represent developments not formally parallel to Dirac's presentation. 

   DEFINITION: Given certain symbols q$_\alpha$ ordered by their index $\alpha$, a rule, called $\phi$, for generating values $\phi$(q$_\alpha$) from the "arguments" q$_\alpha$, will be called a FUNCTION iff

   a) Any $\beta$ which commutes with all of the q$_\alpha$ also commutes with $\phi$(q$_\alpha$), and

   b) Whenever q$_\alpha^\prime$= u$\ast$q$_\alpha\ast$u$^{-1}$ for any u $\not=$0, it follows that $\phi$(q$_\alpha^\prime$) = u$\ast\phi$(q$_\alpha$)$\ast$u$^{-1}$.

   Some results for functions follow.

\begin{enumerate}\item Functions are well-defined, which means that any u, such as that above, will give the same value $\phi$(q$_\alpha$).\item $\phi$(q$_\alpha^\prime$) commutes with whatever commutes with q$_\alpha^\prime$.\item If $\phi$ and $\psi$ are functions, then so are $\phi\pm\psi$ and $\phi\ast\psi$.\item V(x) $\equiv$ x$^{-1}$ is a function.\item Composition of functions gives a function.\item Functions of the same variables commute.\end{enumerate}

   Thus all arguments of a function commute.

   It is to be noted that polynomials are still, according to the new definition, functions.

   It will be noticed that a function whose arguments commute with all symbols
   will have a function value that is invariant under relabelings. Such a
   function will be called a ``constant". It is also the case that if the symbol
   u which defines the relabeling commutes with an argument q$_\alpha$ then it
   will not change and the function will be constant in q$_\alpha$. If it is
   arranged that only one argument varies and yet the function value does not
   change, then the function is constant with respect to that variable. In fact, the function will not depend upon that variable, and so u will commute with all of the arguments that the function does depend upon, and then the function value won't change under this relabeling.

\section{Differentiation}

   In this section a generalization of the usual notion of differentiation will be given. The notion of differentiation given here will not depend upon the notion of a limit, but will still give the usual result for the differentiation of sums and products of functions, and will therefor agree with the results of the usual differentiation on analytic functions. It will also turn out that, for the new notion of differentiation, every function is differentiable.

   Consider the function $\phi$(q$_\alpha$). Then [q$_\alpha$, q$_\beta$]$\equiv$ q$_\alpha\ast$q$_\beta$-q$_\beta\ast$q$_\alpha$ $\equiv$ 0.  If q$_\beta$ doesn't commute with everything, then another symbol p$_\beta$ may be found such that 
$$\mbox{[q$_\alpha$,p$_\beta$] =}\left\{\begin{array}{rl} 0 & \mbox{if $\alpha\not=\beta$} \\1 & \mbox{if $\alpha$=$\beta$}\end{array}\right.$$

   Suppose Q=$\phi$(q$_\sigma$). Define 
$\partial_\alpha\phi$(q$_\sigma$)= Q$\ast$p$_\alpha$-p$_\alpha\ast$Q.

   This is the ``partial derivative of $\phi$ with respect to q$_\alpha$". Note that if $\phi$ is constant with respect to q$_\alpha$ then $\partial_\alpha\phi$=0. Other properties of differentiation will now be given:

\begin{enumerate}\item All p$_\alpha^\prime$ satisfying [q$_\alpha$,p$_\alpha^\prime$]=1 give the same value for $\partial_\alpha\phi$.\item $\partial_\alpha\phi$ is also a function.\item $\partial_\alpha$($\phi$+$\psi$)= $\partial_\alpha\phi$+$\partial_\alpha\psi$ and $\partial_\alpha$($\phi\psi$)=($\partial_\alpha\phi$)$\ast\psi$+$\phi\ast$($\partial_\alpha\psi$).\item The Chain Rule holds: $\partial_s\phi$(Q$_\gamma$((s))=$\sum^n_{k=1}\partial_k\phi$(Q$_\gamma$(s))$\partial_s$Q$_k$(s).\item $\partial_\alpha$[$\partial_\beta\phi$(Q$_\gamma$)]$\equiv\partial_{\alpha\beta}\phi$(Q$_\gamma$) is such that $\partial_{\alpha\beta}$=$\partial_{\beta\alpha}$. In other words, the order of partial differentiation doesn't matter.\end{enumerate}
 
   Item number 3 indicates that $\partial$ is a ''derivation", and so differentiation, in the generalized sense, treats sums and products of functions in the same way as ordinary differentiation does.

\section{Integration}

   Differentiation begins with one function and returns another. It may be asked whether or not this process is invertible, so that a converse process of determining a function whose derivative is a given function can be performed. It will be shown that this problem can be solved in general and, with the condition that only constants have derivatives equal to zero, for a function which is unique up to a constant. Then some results about integration will be shown and some particular integrations, and functions arrived at through integration, will be defined. These results will be useful in the further development of the theory.

   Suppose $\Psi$ is a function of the variable s and other variables q$_\alpha$. Suppose also that a function $\Phi$ is such that $\partial_s\Phi$(s,q$_\alpha$) = $\Psi$(s,q$_\alpha$).

   Then, for p defined as above, 
\begin{equation}\mbox{$\partial_s\Phi$(s,q$_\alpha$)=$\Phi$(s,q$_\alpha$)$\ast$p-p$\ast\Phi$(s,q$_\alpha$)=$\Psi$(s,q$_\alpha$)}\label{eq:int1}\end{equation}
\begin{equation}\mbox{where s$\ast$p-p$\ast$s=1 and
q$_\alpha\ast$p-p$\ast$q$_\alpha$=0 for each q$_\alpha$,}\label{eq:int2}\end{equation}
\begin{equation}\mbox{while s$\ast$q$_\alpha$-q$_\alpha\ast$s = q$_\alpha\ast$q$_\beta$-q$_\beta\ast$q$_\alpha$=0 also holds.}\label{eq:int3}\end{equation}
   It can be seen that, for a symbol p defined to satisfy the conditions
   ($\ref{eq:int2}$), and given $\Psi$(s,q$_\alpha$), the symbol
   $\Phi$(s,q$_\alpha$) may be defined to satisfy ($\ref{eq:int1}$). This may be
   done because all that is under consideration here is a symbolic method that
   needs to be consistent with algebra, but is not such that each symbol needs
   to have an algebraic \textit{derivation}. There is thus always at least one
   such symbol $\Phi$, so all functions $\Psi$ have ``antiderivatives".

   Given such a $\Phi$, consider now the function $\Phi$+c, where c is a
   constant with respect to s. Then
   $\partial_s$($\Phi$+c)=$\partial_s\Phi$=$\Psi$. Thus antiderivatives can, at
   best, be defined up to a constant in this way. This is as close as
   antiderivatives can be to being unique. It will thus now be assumed that only constants have zero derivatives.

   Under these conditions, if both $\Phi_1$ and $\Phi_2$ are such that $\partial_s\Phi_1$=$\partial_s\Phi_2$=$\Psi$ then $\partial_s$($\Phi_1$-$\Phi_2$)=0, so that $\Phi_1$=$\Phi_2$+c, for some constant c. Then $\Phi_2$(b,q$_\alpha$)-$\Phi_2$(a,q$_\alpha$) = $\Phi_1$(b,q$_\alpha$)-$\Phi_1$(a,q$_\alpha$) for such $\Phi_1$ and $\Phi_2$ and particular a and b. It is then possible to uniquely define
\begin{equation}\mbox{$\int^b_a\Psi$($\tau$,q$_\alpha$)d$\tau$ $\equiv$ $\Phi$(b,q$_\alpha$)-$\Phi$(a,q$_\alpha$) for any $\Phi$ such that $\partial_s\Phi$(s,q$_\alpha$)=$\Psi$(s,q$_\alpha$).}\label{eq:int4}\end{equation}

   This condition gives $\int^b_a\Psi$($\tau$,q$_\alpha$)d$\tau$ as a well-defined function of a, b, and q$_\alpha$, and also gives a relationship formally identical to the usual Fundamental Theorem of Calculus. For this reason it is reasonable to call $\int^b_a\Psi$($\tau$,q$_\alpha$)d$\tau$ the ``definite integral" of $\Psi$ from a to b. With this general definition it is easy to deduce the usual formal properties of integrals:

\begin{enumerate}\item $\partial_x\int^x_a\Psi$(s,q$_\alpha$)ds = $\partial_x$[$\Phi$(x,q$_\alpha$)-$\Phi$(a,q$_\alpha$)] = $\partial_x\Phi$(x,q$_\alpha$) = $\Psi$(x,q$_\alpha$).\item Suppose $\phi$(s)=$\tilde{\phi}$(u(s)). Then $\partial_s\phi$ = $\partial_u\tilde{\phi}\ast\partial_s$u, so $\int^\beta_\alpha\partial_u\tilde{\phi}\ast\partial_s$u ds = $\int^b_a\partial_s\phi$ds when u(a)=$\alpha$ and u(b)=$\beta$. Thus the Chain Rule leads, as usual, to the rule for changing the variables of integration.\item $\int^b_a\phi$d$\tau$+$\int^c_b\phi$d$\tau$=$\int^c_a\phi$d$\tau$. Thus integrals may be considered, as usual, to be given by sums. It also follows that $\int^b_a\phi$d$\tau$ = - $\int^a_b\phi$d$\tau$.\item $\int^b_a$($\alpha\ast\phi$+$\beta\ast\psi$)d$\tau$ = $\alpha\ast\int^b_a\phi$d$\tau$ + $\beta\ast\int^b_a\psi$d$\tau$ for constant $\alpha$ and $\beta$.\item $\partial_x$($\phi\ast\psi$) = ($\partial_x\phi$)$\ast\psi$ + $\phi\ast$($\partial_x\psi$) so $\int^b_a\phi\ast$($\partial_x\psi$)dx = $\int^b_a\partial_x$($\phi\ast\psi$)dx - $\int^b_a$($\partial_x\phi$)$\ast\psi$dx = $\phi\ast\psi|^b_a$ - $\int^b_a$($\partial_x\phi$)$\ast\psi$dx. This is the usual rule for integration by parts.\end{enumerate}

   The differentiation of polynomials follows the usual rules, so it follows that integration of polynomials will also follow the usual rules. It can thus be seen that the theory of integration given here gives a non-trivial extension of the usual theory.

   The tools just developed will now be applied to yield generalized versions of the logarithm and exponential functions.

   As every function is integrable, the integral $\int^b_a\tau^{-1}$d$\tau$ may be considered without regard to the behavior, finite or otherwise, of the integrand.

   Define now $\ln$(x)$\equiv\int^x_a\tau^{-1}$d$\tau$, with a as yet unspecified. Then $\ln$(x$\ast$y) = $\int^{xy}_a\tau^{-1}$d$\tau$ = $\int^x_a\tau^{-1}$d$\tau$ + $\int^{xy}_x\tau^{-1}$d$\tau$ = $\ln$(x) + $\int^{xy}_x\tau^{-1}$d$\tau$.

   Let u($\tau$) = x$^{-1}\ast\tau$. Then $\ln$(x$\ast$y) = $\ln$(x) + $\int^y_1$u$^{-1}$du = $\ln$(x) + $\int^y_a\tau^{-1}$d$\tau$ - $\int^a_1\tau^{-1}$d$\tau$ = $\ln$(x) + $\ln$(y) - $\int^a_1\tau^{-1}$d$\tau$. Now if a = 1 is chosen, then 
\begin{equation}\mbox{$\ln$(x$\ast$y) = $\ln$(x) + $\ln$(y).}\label{eq:log1}\end{equation}
   This choice is made, so that, in general:
\begin{equation}\mbox{$\ln$(x) $\equiv$ $\int^x_1\tau^{-1}$d$\tau$.}\label{eq:log2}\end{equation}

   Now $\partial_x\ln$(x)$|_{x=y}$ = $\frac{1}{y}$, so that $\ln$(w) = $\ln$(v) implies that w = v, and so $\ln$ is a one-to-one function and may be inverted. Let g be the inverse of $\ln$. Then, according to (\ref{eq:log1}), g must satisfy
\begin{equation}\mbox{g(x+y) = g(x)$\ast$g(y) with g(0)=1.}\label{eq:exp1}\end{equation}

   Now, for rational p, it follows that $\ln$(x$^p$) = p$\ast\ln$(x), so then 
\begin{equation}\mbox{x$^p$ = g($\ln$(x$^p$)) = g(p$\ast\ln$(x)).}\label{eq:exp2}\end{equation}

   This provides a general definition of x$^p$ regardless of whether or not p is
   rational. In fact, g defines irrational symbols, so that g definitely extends
   the known range of the symbolic system. Such irrationals correspond to
   infinite sequences of integers in their decimal expansions, so that the
   inclusion of irrationals in the symbolic system requires the admittance of
   the results of limits and, in particular, requires the inclusion of infinite
   symbols and of reals. If the symbol ``e" is \textit{defined} by the requirement
   that it be such that $\ln$(e) = 1,
   then e$^x$=g(x$\ast\ln$(e)) = g(x). Thus g clearly generalizes the exponential function.

\section{The Relationship to the Usual Calculus}

   An abstract theory of functions, differentiation, and integration has been
   given which relies on the constructibility of symbols with tailor-made
   commutation properties but does not rely on the use of limits nor assume that
   the quantities involved are finite. Moreover, differentiation obeys the usual
   rules when operating upon sums and products of functions and so the
   generalized theory is an extension of the usual Calculus. In fact, as a basis
   for comparison between the usual and the generalized Calculus, it may be
   helpful to think of implementing the given rules for the generalized theory
   in a symbolic math program. The computations performed would then be
   identical to those performed in a version of the program based upon the usual
   Calculus whenever exact results are required and recourse to numerical methods for approximating limits is not allowed.

   Conversely, if it is assumed that all quantities involved in computation are
   finite, then the generalized functions, differentiation, and integration
   given here reduce to those of the usual Calculus. It is thus possible to
   apply the formal methods of the Calculus in a domain where the finiteness of
   quantities is not presumed and to also, then, investigate explicitly where and how the finiteness of quantities affects the results.

%% file: rpichap6.tex

\chapter{COMPLETION OF THE SYMBOLIC SYSTEM}

\section{Functional Equations and Their Consequences}

   With the development of the Calculus it is now possible to derive some
   further results on the algebraic nature of the general symbolic system. This
   will be done by solving, by the methods of the calculus just developed,
   certain functional equations which are constructed based on the relationship
   between the formal system and an embedded lattice of meaningful symbols that results
   from it upon interpretation. These considerations will indicate that $\ast$
   may, in fact, be taken as the extension of $\wedge$ to the general symbolic
   system, and that complex numbers must necessarily be included in the general symbolic system.  

   Consider the meaningful symbols a, b, and c. In considering meaningful symbols, it follows that there is a function $\phi$ which maps these symbols into corresponding symbols in the symbolic system. Thus let $\phi$(a) = x, $\phi$(b) = y, and $\phi$(c) = z. Now $\phi$ must be such that, for some function F,  
\begin{equation}\mbox{$\phi$(a$\wedge$b) = F[$\phi$(a), $\phi$(b)] = F[x, y]}\label{eq:Homo}\end{equation}
The associative property of $\wedge$ then immediately results in 
\begin{equation}\mbox{F[x, F[y, z]] = F[F[x, y], z]}\label{eq:Assoc}\end{equation}
This will be referred to as the ASSOCIATIVITY EQUATION. 

   The Associativity equation is a functional equation for F, and the solution
   of functional equations is, in general, a difficult problem\footnote{ See
   \cite{Aczel1} and \cite{Aczel2}.} the solution of which generally requires
   special additional assumptions about the function to be determined. In the
   general symbolic system, however, this equation can be solved in all
   generality without restrictive assumptions about F\footnote{In this
   connection also consult \cite{Cox}.}. The solution process is intricate, however, and the details are left to the appendix. The solution is here simply given as
\begin{equation}\mbox{F[x, y] = $\Phi^{-1}$[$\alpha\ast\Phi$(x)$\ast\Phi$(y)]}\label{eq:Wedge}\end{equation}
where $\alpha$ is a fixed constant and the form of $\Phi$ is specified in the the appendix. The form of the equation (\ref{eq:Wedge}) indicates that the operation $\ast$ may definitely be taken to correspond to the operation $\wedge$. This is significant because the operation $\ast$ was merely motivated by the operation $\wedge$ without there being a definite connection between them assumed. 

   Further results about the symbolic system may be derived by considering the relationship between interpretation and operations between meaningful symbols. Consider, in particular, a Boolean Lattice of meaningful symbols, where a complement a$^\prime$ is given for every meaningful symbol a. Recall that (a$^\prime$)$^\prime$ = a. Now, for some function f, it must be that
\begin{equation}\mbox{$\phi$(a$^\prime$) = f($\phi$(a)).}\end{equation}
From this it follows that f satisfies 
\begin{equation}\mbox{f(f(x)) = x.}\end{equation}
This is another functional equation; it shall be referred to as the CATEGORICITY EQUATION.

   As was the case with the Associativity equation, the solution of the Categoricity equation may be found in the appendix. The solution is given, in general, by    
\begin{equation}\mbox{f(x) = [ c$^{r}$ - x$^{r}$ ]$^{\frac{1}{r}}$}\label{eq:Idemf}\end{equation}
where c and r are arbitrary fixed constants.

   So far, the consideration of finite constructions within the symbolic system
   has resulted in rationals and polynomials. The discussion of the Calculus has even extended the
   symbolic system to irrational numbers and reals. The equation
   (\ref{eq:Idemf}) must, however, be solvable for any and all constants c, r.
   This necessarily results in the admission of further kinds of symbols.
   Consider, in particular, r = 2, with c and x rationals such that x$^2$ $>$
   c$^2$ + 1. Then, in order that f(x) may be solved for, it is necessary that
   (-1)$^\frac{1}{2}$ be a symbol in the general symbolic system. Thus the
   general symbolic system must include complex numbers. This is quite an ironic
   result in the sense that the interpretability of a Boolean Lattice has
   resulted in the necessity of the symbolic system being complex, at least,
   whereas such complex quantities only entered physical theory with the advent
   of Quantum theory and are taken to reflect a non-Boolean logic. 

   The considerations of this section have shown a definite connection between
   the algebraic operations and algebraic types of the symbols of the general
   symbolic system on the one hand, and the corresponding operations in an
   embedded lattice of meaningful symbols on the other. With these results it is possible to proceed, in the next section, to a complete algebraic specification of the general symbolic system.

\section{Frobenius's Theorem and the Completion of the System}

   The specification of the general symbolic system is almost complete. It only remains to consider some consequences of the way in which finitely many symbols may be combined and encoded. This will have consequences for the algebraic form of the system.

   Consider the symbols that may be finitely generated by the algebraic operations and finitely many variables and complex numbers. The result will be a polynomial p of finite order, over the complex numbers, and in the given variables. Suppose attention is restricted to only the consideration of finite symbols. In this context it is a standard result that any such given finite order polynomial corresponds to a definite list of its distinct zeros, and conversely that a polynomial may be constructed which has precisely a given list of zeroes. Thus a correspondence may be given between finite selections of meaningful symbols and polynomials which represent them. It must certainly be possible to construct a restricted form of the formal system which is comprised only of symbols which are generated in this way. The general symbolic system must then be an extension of this system.

   The restricted finite form of the symbolic system would therefor have to be
   algebraic over the reals at least, and the general system must be an
   extension of this restricted system. In order that uniform rules as to the
   algebraic form of the symbols in the general symbolic system may be adopted,
   it is necessary that the general system have the same algebraic form as the
   finite restricted system. The general form of division rings of finite
   symbols algebraic over the reals has already been established in a theorem of
   Frobenius \footnote{ See pp. 327-329 of reference \cite{Frobenius}.}. This
   theorem indicates that such a division ring must be isomorphic to one of the
   real field, the complex numbers, or the real quaternions. 
   
   In addition to the
   information given by Frobenius's Theorem, it may also be concluded that
   neither the form of the reals nor complex numbers may be adopted for the
   general symbolic system because these are commutative fields, and the formal
   system must include non-commuting variables. The unequivocal conclusion is
   therefor that in the general symbolic system the symbols must have the
   algebraic form of the real quaternions, and in fact be real quaternions in
   the case that they are to be finitely generated. This simple algebraic result
   will later be shown to be of importance in accounting for the
   four-dimensionality of space-time in the derivation of the general
   relativistic formalism. This requirement will also distinguish the quantum
   theoretical derivation to follow in that while the usual theory operates over
   the complex field, there is no clear justification for or understanding of
   the significance of this particular choice, whereas such understanding is
   supplied here. 

   To sum up, it has been found that every symbol $\omega$ in the general symbolic system will have the form
\begin{equation}\mbox{$\omega$ = $\omega_0$ + $\omega_1\ast$i +$\omega_2\ast$j +$\omega_3\ast$k, where}\label{eq:quat1}\end{equation}
\begin{equation}\mbox{i$^2$=j$^2$=k$^2$= -1, i$\ast$j=-j$\ast$i=k, j$\ast$k=-k$\ast$j=i, and k$\ast$i=-i$\ast$k=j.}\label{eq:quat2}\end{equation}

   In general the $\omega_k$ will commute with one another, will satisfy the
   rules of a division ring, and will comprise the components of such
   four-tuples, but they will not generally be real numbers, for the finiteness
   of the $\omega_k$ is not presumed for the general symbolic system. It may
   also be observed that such $\omega$ will not be generally suitable as
   arguments for functions, as defined here, for only restricted classes of such symbols will commute with one another.

   One final note. Given the expression (\ref{eq:quat1}) for $\omega$ it is possible to define the CONJUGATE of $\omega$, denoted by $\bar{\omega}$, as 
\begin{equation}\mbox{$\bar{\omega}$ = $\omega_0$ - $\omega_1\ast$i -$\omega_2\ast$j -$\omega_3\ast$k.}\label{eq:Conj}\end{equation}
The introduction of the conjugate also allows the INNER PRODUCT, between two symbols a and b, to be defined by
\begin{equation}\mbox{(a, b) = a$\ast\bar{\mbox{b}}$.}\label{eq:InProd}\end{equation}
Then, in the usual way, the NORM may be defined by
\begin{equation}\mbox{$||$a$||$ = $\sqrt{\mbox{ (a, a)}}$.}\label{eq:Norm}\end{equation}

\section{Conclusions on the Algebraic System}

   The time has finally come to take stock of the net effect of all these
   abstract considerations about the general symbolic system. The following conclusions have been reached: 
\begin{enumerate}\item The general symbolic system is a division ring with the
algebraic form of the real quaternions.\item The formal system has a partial
order defined on part of it.\item Symbols have conjugates and norms, and an
inner product is defined.\item Symbols are not necessarily finite, and infinite
symbols must be included in the general symbolic system.\item A generalized
version of the Calculus has been defined on the symbolic system without
reference to finite quantities.\item The generalized Calculus gives an extension
of the usual limit-defined notions, so that all functions are both
differentiable and integrable.\end{enumerate}

   With these observations it will now be possible to proceed to the derivation of physical theories without the utilization of any experimental results whatsoever!

%% file: rpichap7.tex

\chapter{THE DERIVATION OF PHYSICAL THEORY}

\section{Introduction}

   The considerations of the previous chapters have been preliminary to this
   chapter, for to substantiate a claim that physical theory can be derived in a
   purely formal manner, without recourse to experiment, it is necessary to
   first develop a formal system on its own terms. Having done so, the task of
   constructing physical theory will be taken up in the following manner.

   It has already been argued that the very nature of experimentation is such
   that it must be possible to state its results and carry out its predictions in
   terms of finitely many finite symbols. It is therefor of particular interest
   to investigate, as a preliminary special case, the restriction of the general
   symbolic system that results if it is first assumed that the variables in the
   symbolic system are all finite. It will then be argued that the associated
   finite relations comprising physical laws will be expressible in a particular
   canonical form. The analysis of this form may be broken into two cases.

   First it will be shown what form of predictive theory results if it is
   assumed that the finite relations are to be determinate. This will yield a
   novel derivation and justification for General Relativity Theory\footnote{
   There are many presentations of Relativity. For an intuitive motivation and
   overview, see \cite{Einstein1}. For a more rigorous introduction in the
   geometric style, see \cite{Einstein2}. For an exposition given in the
   language of forms, see \cite{Misner}. A non-geometrical route is taken in
   \cite{Weinberg}. Finally, a concise mathematical presentation may be found in
   \cite{Dirac2}.} which is inevitable, on these assumptions, and which is free
   from experimental or merely plausible theoretical justification. It will also
   follow that Relativity is the unique form of finite deterministic theory.
   From Einstein's Field Equation, serving as the basis of General Relativity
   Theory, the dependent notion of Turing Computability will also arise.

   Finite laws that are statistical are discussed next. In order that these laws
   give information that is independent of Einstein's theory, it is necessary
   that the quantities they provide for prediction must not, at the very least,
   be determinable by Turing Machines. This will lead to a unique set of
   requirements for statistical predictions, and consequently to a propensity
   theory of probability which is similar, in its foundations, to Von Mises's
   Frequency Theory of Probability \footnote{ See references \cite{Mises1} and
   \cite{Mises2}.}. This theory will apply to the cases of both continuous and
   discrete representation of data, and so is perfectly general. 

   Combining these interpretive aspects of finite statistical theory with the
   previously determined form of finite relations will yield Quantum Theory as
   the unique embodiment of finite statistical theory\footnote{ For an axiomatic
   presentation of the quantum formalism see \cite{Jauch}. Note especially the
   comment on pg. 75 in regard to the finiteness of experimental data.}. The
   strength of this approach is shown in that it dismisses the controversies
   that have arisen in deciding the proper way to interpret the quantum formalism.

  Such discussion completely analyzes finitary theory but it is found, in fact,
  that these two canonical finite physical theories apparently result in a
  mathematical dichotomy. This difficulty may be understood to stem from one of
  two sources and lead to correspondingly definite means of resolution. One
  solution is to maintain the restriction to finite mathematical theories and
  resign one's self to forever deferring to experiment. Alternately, the
  assumption of finiteness could be dropped, and this, it will be argued, results
  in a unique non-finitary formalism. Interestingly enough, these two solutions,
  while embodying the contending philosophical approaches of pragmatic
  empiricism and of idealistic metaphysics, are, in a definite sense, not at all
  distinct.  

\section{The Form of Finite Relations}

  Now that the abstract algebraic properties of the general symbolic system have
  been specified it is possible to arrive at some concrete conclusions about the
  kinds of descriptions this system can provide. Starting from the formal
  assumption that all descriptive symbols are finite it will be found, to begin
  with, that such finite relations must be expressible in a particular canonical
  form. The necessity of this form will have implications for the construction of physical theory. 

   Consider any function in the general symbolic system. Such a function, and indeed any term in any equation, will, in general, have a value and arguments of the form of quaternions. Thus any equations can be expressed as a system of four functions set equal to zero, each function having values of the form of either real or complex numbers. For each such function, the corresponding arguments must be commuting symbols. 
   
   In the general case the constructibility of symbols which didn't commute with
   given arguments was of central importance in the derivation of the
   generalized Calculus. In the finite case, however, it has been observed that
   the operations of the usual Calculus, with their associated limit dependence,
   may be substituted and do not require that the arguments not commute with
   certain symbols. For this reason it is possible to take the arguments of all
   functions to be complex when it is assumed that all symbols are finite. 
   
   It
   should be pointed out that expressions in the finite case therefor, while
   yet in
   correspondence with expressions in terms of real quaternions, do not
   \textit{directly}
   express the non-commutative nature of the division ring of the general
   symbolic system. With this understanding any finite relations will, in fact, be expressed in terms of complex-valued functions of complex arguments.

   Consider any finite collection of distinct symbols each of which has been put into correspondence with, and therefor represented by, a distinct complex number q$_\alpha$. It is a standard result of complex variable theory that it is then possible to construct an analytic function $\phi$(z) which has zeroes at precisely the points z=q$_\alpha$. Now the condition $\phi$(z)=0 is equivalent to the condition $|\phi$(z)$|$=0, so that given a known collection of data represented by the complex numbers q$_\alpha$, the same collection is represented by a certain analytic function $\phi$(z) and is recovered by solving the equation $|\phi$(z)$|$=0. In this way it is seen that the analytic function $\phi$(z) associates the zeroes q$_\alpha$ and thus expresses a relation between them. It is thus the case that any quantity depending on the collection of data q$_\alpha$ will, in fact, be a function of $|\phi$(z)$|$. The general conclusion is reached that all statements and quantities upon which finite physical theory may be based are to be found in this form.   

   The considerations above were made under the assumption that the data q$_\alpha$ were already given and then $\phi$(z) was to be constructed to represent them. However, the converse procedure is completely justified, so that, starting with an analytic function selected according to some criterion, $|\phi$(z)$|$ will have to be the quantity upon which prediction depends and it will, in general, automatically engender an associated collection of data. It may be noted in passing that the consideration of analytic functions generally might also be used to extend the theory to the case of countably many data, as analytic functions with that many zeroes may be constructed. In the same way as above, the formal criteria which select $\phi$(z) also correspond to certain ``phenomena" so that the usual intuitive process of looking for quantities to represent what is observed may be reversed. This observation helps in understanding how a completely formal theory might reproduce the physical theories which have been motivated by experiments and observation in the past, although the difficulty remains of identifying such constructive and yet abstract criteria which $\phi$(z) is to satisfy. 

   It has been found that all finite predictive quantities must be expressible
   in a particular canonical form, and that this form also indicates a way to formally bypass experiment. The next section will move beyond these abstractions to explore an application of this idea and to indicate a particular success of this approach.

\section{Determinism and Relativity}

\subsection{Introduction}

   In the sections to follow it will first be demonstrated that a particular
   form of $|\phi$(z)$|$, called the interval, must be derivable which gives
   invertible relations between pairs of symbols. Next it will be shown how the
   interval may be put into a canonical form at a point by a change of
   coordinates, and that this form has implications for the corresponding
   physical theory. In particular, relations between events may be expressed in
   terms of the relationship between locally preferred coordinate systems, each
   of which give the interval the canonical form at a particular point,
   associated with each event. The sequence of events determined by the theory
   will then be defined by the allowable evolution of the preferred coordinate systems associated with these events. In order to determine this evolution it is next shown how the evolution of the preferred coordinate system may be expressed, via the geodesic equation, in any system of coordinates. Examination of these equations will indicate the quantities in the geodesic equation which will govern the evolution of the preferred coordinate system, and then it will be shown that there is a unique way that the behavior of the preferred coordinate systems may be determined in a manner independent of the coordinate system in which the definition is expressed. This will yield Einstein's Field Equations. Finally, it will be shown that Einstein's equations are, as they must be, equivalent to the formal requirement of the general symbolic system which started these considerations.  

\subsection{Derivation of the Interval}

  In this section it will be demonstrated that a particular form of $|\phi$(z)$|$ must be derivable which gives invertible relations between pairs of symbols. This will give a novel derivation and interpretation of General Relativity Theory without any further assumptions or recourse to experiment. This derivation will thus explain all relativistic phenomena and concepts, such as gravitation, the behavior of light signals and space-time, and the unification of the concepts of mass, momentum, and energy, as the natural implications of a formal necessity found in the general symbolic system.  

  In this section it will, in fact, be demonstrated that the necessity of
  allowing one-to-one relabeling in the general symbolic system requires that a
  particular form of finite relation, determined by $|\phi$(z)$|$, must be
  derivable which gives invertible relations between pairs of symbols.

  Recall the formal process of relabeling in the general symbolic system. In
  order to investigate a special restricted case ,require, in addition, that
  this procedure be one-to-one: Then x$\mapsto$f(x) is such that a = b exactly
  when f(a) = f(b). Thus such one-to-one relabeling maintains the distinctness
  of symbols: They may be said to be ``separated", so that if a$\not=$b then
  f(a)$\not=$f(b). This is a definite symmetric relationship between symbols
  which is maintained under invertible relabeling. As such manipulations are an
  inherent part of the general symbolic system, this precise relationship must find its expression for finite relations as well.

   Suppose $\phi$(z)=m$\ast$e$^{\imath\theta}$, where m$\geq$0, is to express
   this relationship for pairs of symbols a and b. Definite values for a and b
   must then give definite values for $\phi$(z). However, m=$|\phi$(z)$|$, upon
   which expression of this relationship depends, does not generally identify a
   unique function $\phi$(z): $\theta$ is as yet undetermined. Such a
   specification requires a \textit{conventional} value of $\theta$ be chosen,
   and the most convenient is $\theta$=0. Then $|\phi$(z)$|$=$\phi$(z)= m
   $\geq$0. In other words, the realization of the relationship maintained by
   invertible relabeling requires, for its finite expression, utilizing an
   analytic $\phi$-function which is real-valued and non-negative.

   The labels a and b referred to above are representatives of the general symbolic system and, as such, are given by real quaternions in the finite case. Each label can, therefor, be represented by 4-tuples of real numbers: x$^\mu$, where $\mu$=0,$\ldots$,3. Thus $\phi$(z) may be rewritten as $\delta$(x$^\mu_1$,x$^\nu_2$). Let dx$^\mu$=x$^\mu_2$-x$^\mu_1$. Then, with some abuse of notation, $\delta$(x$^\mu_1$,dx$^\mu$)$\equiv\delta$(x$^\mu_1$,x$^\mu_2$) may be defined. $\delta$, being analytic, may be expanded in a Taylor series for x$^\mu_1$ fixed and dx$^\mu$=O($\epsilon$): 
\begin{equation}\mbox{$\delta$(x$^\mu_1$, dx$^\mu$)=g(x$^\mu_1$)+g$_\alpha$(x$^\mu_1$)dx$^\alpha$+g$_{\alpha\beta}$(x$^\mu_1$)dx$^\alpha$dx$^\beta$+g$_{\alpha\beta\gamma}$(x$^\mu_1$)dx$^\alpha$dx$^\beta$dx$^\gamma$+O($\epsilon^4$).}\label{eq:Taylor1}\end{equation}

   Note here that Einstein's convention, where repeated indices are summed over, is utilized in (\ref{eq:Taylor1}). Also note that all g-functions are symmetric in their indices as $\delta$ is analytic. Now it may be seen, at O(1), that:
\begin{equation}\mbox{$\delta$(x$^\mu_1$, x$^\mu_2$)-$\delta$(x$^\mu_2$, x$^\mu_1$)= g(x$^\mu_1$)-g(x$^\mu_2$).}\label{eq:Taylor2}\end{equation}
   
   But $\delta$ is symmetric in its arguments, so that:
\begin{equation}\mbox{g(x$^\mu_1$)=g(x$^\mu_2$).}\label{eq:Taylor3}\end{equation}

   Thus g is locally constant, and this result may be continued so that g=0 may be stipulated globally. This amounts to choosing $\delta$(x$^\mu$, x$^\mu$)=0 rather than some other fixed number. Consider now $\delta$(x$^\mu_1$, -dx$^\mu$). As $\delta\geq$0, it follows that all odd-ordered terms must be zero, so that:
\begin{equation}\mbox{$\delta$(x$^\mu_1$, dx$^\mu$) = g$_{\alpha\beta}$(x$^\mu_1$)dx$^\alpha$dx$^\beta$+O($\epsilon^4$).}\label{eq:Taylor4}\end{equation}

   It may be recalled that $\delta$ was to fulfill the exact requirement of expressing the separateness of symbols and nothing more. Thus, because the second order term in (\ref{eq:Taylor4}) already suffices to make this distinction in general, the higher order terms must be dropped. The form of the function $\delta$ is thus uniquely given by
\begin{equation}\mbox{$\delta$(x$^\mu_1$, dx$^\mu$) = g$_{\alpha\beta}$(x$^\mu_1$)dx$^\alpha$dx$^\beta$.}\label{eq:Delta}\end{equation}

   Because $\delta\geq$0 and is real, it follows that $\delta$ may be rewritten as $\delta$=ds$^2$, where ds$\geq$0 is real-valued. The value ds corresponding to the labels x$^\mu_1$ and x$^\mu_2$ is called the ``interval" between them. This allows (\ref{eq:Delta}) to be put in the conventional form:
\begin{equation}\mbox{ds$^2$= g$_{\alpha\beta}$(x$^\mu_1$)dx$^\alpha$dx$^\beta$.}\label{eq:Delta2}\end{equation}

   This equation now defines, for given points x$^\mu_1$ and x$^\mu_2$, the interval ds once the functions g$_{\alpha\beta}$ may be stipulated. It must be remembered that conditions specifying the g$_{\alpha\beta}$ are required before this definition is complete. It should also be noted that, if ds$^2$ is to express only the separateness of symbols, then the g$_{\alpha\beta}$ must be determined by a self-consistent equation with no other data required. Otherwise, the g$_{\alpha\beta}$ will reflect both the separateness of symbols and something else.  

\subsection{Changes of Coordinates and the Canonical Form} 

   Now, in this section, it will be shown how the interval just derived may be,
   and must be able to be, put into a canonical form at each point by a change
   of coordinates. The preferred coordinate system which is thus associated with
   each point then serves as a definite means of speaking of a relationship
   between points: The canonical form at a point will identify a locally preferred coordinate system, so that events may be characterized in a definite way and it thus becomes possible to speak of definite relations between them. This will lead, in short order, to a logical derivation of the usual relativistic space-time picture of physical reality.

   The coordinates x$^\mu$ of labels, being subject to relabeling, cannot have a direct significance. It is possible, however, to indicate at this point how relabeling is effected and how this affects the terms of (\ref{eq:Delta2}).

    Suppose x$^\mu$=x$^\mu$($\bar{x}^\alpha$) gives a relabeling, or change of coordinates. It is only required that this substitution leave ds$^2$ invariant in (\ref{eq:Delta2}). Thus the change of coordinates need only be continuous, and substitution of
\begin{equation}\mbox{dx$^\mu$=$\frac{\partial\mbox{x}^\mu}{\partial\mbox{$\bar{x}$}^\alpha}$d$\bar{x}^\alpha$.}\end{equation}

into (\ref{eq:Delta2}) leads to 

\begin{equation}\mbox{$\bar{\mbox{g}}_{\mu\nu}$=g$_{\alpha\beta}\frac{\partial\mbox{x}^\alpha}{\partial\mbox{$\bar{x}$}^\mu}\frac{\partial\mbox{x}^\beta}{\partial\mbox{$\bar{x}$}^\nu}$.}\end{equation}

  Now because ds$^2$ is a real quadratic form it is, according to a standard theorem\footnote{ See pg. 207 of \cite{Shilov}.}, always possible, at any point x$^\mu$, to change coordinates continuously so as to obtain the canonical form 
\begin{equation}\mbox{ds}^2\mbox{=}\eta_{\alpha\beta}\mbox{d}\xi^\alpha\mbox{d}\xi^\beta\mbox{=}\eta_{\alpha\alpha}\mbox{(d$\xi^\alpha$)$^2$ where $|\eta_{\alpha\beta}|$=$\delta_{\alpha\beta}$.}\label{eq:canon}\end{equation}
 
   It is clear that a uniform definition of ds cannot be arrived at unless the signs of $\eta_{\alpha\alpha}$ are determined in the same way for all points x$^\mu$. Now each x$^\mu$ is to correspond to some event which an individual perceives at a particular instant, so that one of the variables x$^\mu$ must be specially identified, by a sign for $\eta_{\alpha\alpha}$ distinct from that of the others, to serve as an indicator of the instant in question. The arbitrary choice that x$^0$ be identified by $\eta_{00}$= 1, and thus also that $\eta_{m m}$= -1 for m=1, 2, and 3, is made. 

   Once the canonical form of ds$^2$ has been decided upon, the changes of
   coordinates that may lead to it at a particular point take on a definite
   significance. It is then possible to speak of definite relationships between
   different points based on the relationship between the preferred coordinate systems associated with these points.  

   This has the consequence, first of all, that the labels x$^\mu$ will now be formally identical to the usual space-time which comprises a pseudo-Riemannian manifold. In such a scheme there is an associated notion that the relation between events which ds specifies is given by a sort of signaling. In the standard empirical justification of Relativity, this signaling is taken to be a property of light. Here, no such recourse to experiment is implied, but it is instead asserted that the formal properties of the symbolic system itself require that such a relation between events obtain, and that, if it is desired that this relation be interpreted in in terms of signaling, there be ``something" which travels between events and acts as such a signal. This something, being always associated with the events which it acts as a signal for, can be thought of as being made up of whatever causes the perception of these events. It follows that if the identification of events is ultimately made visually, for example, then the signal between events must be realized by light.  

   The necessity of a uniform definition of ds also has the consequence that instants fall, from a finitely justifiable point of view, into a linear order as does the real-valued variable x$^0$, and thus introduces the qualitative temporal notions of ``earlier" and ``later" to the instants.  It is to be noticed that in this construction of the notion of time, the linear and continuous aspects of time are not assumed, based on experience, but are instead derived from the formal nature of the symbolic system as a whole.

   It must also be stressed that the three remaining non-temporal labels x$^m$ are not presumed in order that the usual notion of space may find a place within the formalism. Instead, this three-dimensional aspect of finite description rather justifies the notion of space, for the relations found in ``spatial perceptions" are never argued for on the basis of infinitely many examples. Because the notion of space may be derived simply from the assumption of a finite check of its relations, and always includes this assumption, it would be superfluous to assert any further source of the formal nature of spatial perceptions. Both space and time may thus be seen to be moulds into which experiences are poured when construed finitely.  

  Once the functions g$_{\alpha\beta}$ are determined the differential relation (\ref{eq:Delta2}) for ds$^2$ may, in most cases, be integrated along a path for which ds$\geq$0. Paths which give the least value may then uniquely define the interval between distant events. It will then be possible, by a chain of intermediate transformations, to relate such events by changes of variables.

    This process will not, however, \textit{necessarily} relate all pairs of points x$^\mu$. It may happen that the extension of a path along which ds$\geq$0 is blocked by the unsolvability of this condition. This will be the case where attempting to cross regions in which ds$\geq$0 cannot be satisfied or in which ds simply vanishes. Such pairs of events will then not be finitely related in this direct way, and Relativity is, in this sense, a ``local" theory. Such events cannot ``communicate" by (light) signaling, but the pairs of points x$^\mu$ will still be formal parts of the descriptive apparatus of the theory and serve to mathematically complete the manifold. A new notion, that of relatively isolated parts of ``reality", even parts which are never experienced, is thus introduced.  

  Another interesting remark may be made at this point. The functions specifying changes of coordinates have thus far been interpreted ``passively" as giving a mere renaming of the same events. It is unavoidable, however, that these functions will also relate pairs of events identified by the arguments and the values of these transformations. Because these events may both fall within a particular individual's experiences, the change of coordinates may be interpreted ``actively" as referring to transitions between that individual's experiences. The change of coordinates may thus show, at a particular instant, how that individual's experiences are related, or, if the transformation is between distinct instants for the individual, how ``earlier" events are related to ``later" ones.

   If these events are considered to be experienced by distinct individuals, then the change of coordinates specifies a ``change of reference frame." Once the interval is determined for all events, it follows that it will inter-relate the experiences of all distinct observers. The equal admissibility of all continuous changes of variables then gives all observers, which may be related in this way by changes in reference frame, an equal observational status.

\subsection{The Geodesic Equation}

   The canonical form discussed in the last section identifies a locally
   preferred coordinate system that is associated with each point in space-time.
   With these coordinate systems characterizing each point it now has become
   possible, in principle at least, to speak in some definite way about the
   relationship between events. Progress on this count will be made in this
   section, where the form of the equations describing the evolution of the locally preferred coordinate systems will be determined. It will then be apparent which quantities in these equations must be determinative in any law for this evolution. 

   It has been found that the finite-relatedness, generated by invertible
   relabeling, of pairs of events may be expressed in terms of the real quadratic form ds$^2$, and that ds$^2$ depends, in turn, on the functions g$_{\alpha\beta}$. The first step in the determination of the g$_{\alpha\beta}$ will now be undertaken.  

   For ds$>$0 it follows from (\ref{eq:canon}) that:
\begin{equation}\mbox{ 1 = $\eta_{\alpha\alpha}$}\left(\frac{\mbox{d}\xi^\alpha}{\mbox{ds}}\right)^2.\label{eq:geod1}\end{equation}

   From this it follows that 
\begin{equation} 0\mbox{ = }\eta_{\alpha\alpha}\left(\frac{\mbox{d}\xi^\alpha}{\mbox{ds}}\right)\left(\frac{\mbox{d$^2$}\xi^\alpha}{\mbox{ds$^2$}}\right).\label{eq:geod2}\end{equation}

   In order that (\ref{eq:geod2}) hold generally it is necessary that
\begin{equation}\frac{\mbox{d$^2$}\xi^\alpha}{\mbox{ds$^2$}}\mbox{ = 0.}\label{eq:geod3}\end{equation}

   For any parameter $\omega$ and new coordinate system $\xi^\alpha$=$\xi^\alpha$(x$^\mu$) the equation (\ref{eq:geod3}) takes the form
\begin{equation}\frac{\mbox{d$^2$x}^\lambda}{\mbox{d$\omega^2$}}\mbox{ + }\Gamma^\lambda_{\mu\nu}\frac{\mbox{dx$^\mu$}}{\mbox{d$\omega$}}\frac{\mbox{dx$^\nu$}}{\mbox{d$\omega$}}\mbox{ =0}\label{eq:geod4}\end{equation}

\begin{equation}\mbox{ where $\Gamma^\lambda_{\mu\nu}\equiv\frac{\partial\mbox{x}^\lambda}{\partial\xi^\alpha}\frac{\partial^2\xi^\alpha}{\partial\mbox{x}^\mu\partial\mbox{x}^\nu}$.}\label{eq:geod5}\end{equation}

   (\ref{eq:geod4}) are usually known as the ``geodesic equations", while the $\Gamma^\lambda_{\mu\nu}$ are known as the ``Christoffel" symbols of the first kind. These equations indicate the form taken by the conditions (\ref{eq:geod3}) in any coordinate system. Note that any differences between points x$^\mu$ must be determined by $\Gamma^\lambda_{\mu\nu}$ at the points, but that it is always possible to choose a coordinate system so that $\Gamma^\lambda_{\mu\nu}$ vanishes at a given point. Thus the $\Gamma^\lambda_{\mu\nu}$ determine the relationship between points under relabeling, but not in a manner independent of the associated change in coordinates.

\subsection{The Metric Field Equation}

   It will now be shown how the metric field g$_{\alpha\beta}$ must be determined in a manner independent of the coordinates. With the determination of g$_{\alpha\beta}$ the interval will also be defined.

   Because both g$_{\alpha\beta}$ and $\Gamma^\lambda_{\mu\nu}$ determine ds, it must be the case that they are related to each other. In fact, it may be shown that they satisfy the relations
\begin{equation}\frac{\partial\mbox{g}_{\mu\nu}}{\partial\mbox{x}^\lambda}\mbox{ = }\Gamma^\rho_{\lambda\mu}\mbox{g}_{\rho\nu}+\Gamma^\rho_{\lambda\nu}\mbox{g}_{\rho\mu}\mbox{  ,   and}\label{eq:gamma1}\end{equation}
  
\begin{equation}\Gamma^\sigma_{\lambda\mu}\mbox{ = }\frac{1}{2}\mbox{g}^{\nu\sigma}\left[\frac{\partial\mbox{g}_{\mu\nu}}{\partial\mbox{x}^\lambda}\mbox{ + }\frac{\partial\mbox{g}_{\lambda\nu}}{\partial\mbox{x}^\mu}\mbox{ - }\frac{\partial\mbox{g}_{\mu\lambda}}{\partial\mbox{x}^\nu}\right]\label{eq:gamma2}\end{equation}

\begin{equation}\mbox{where g}^{\nu\sigma}\mbox{g}_{\rho\nu}\mbox{ = }\delta^\sigma_\rho\mbox{ is the delta function.}\label{eq:gamma3}\end{equation}

   Note that g$^{\nu\sigma}$ may be defined because the coordinate transformations considered are invertible. Because all $\Gamma^\rho_{\lambda\nu}$ may be made to vanish at a point, it follows from (\ref{eq:gamma1}) that the same may be done for all $\frac{\partial\mbox{g}_{\mu\nu}}{\partial\mbox{x}^\lambda}$. Then the equations (\ref{eq:gamma1}) and (\ref{eq:gamma2}) give no further information. A non-vanishing relationship may, however, be derived. This relationship will be expressed in a coordinate system in which $\Gamma^\sigma_{\lambda\mu}$ vanishes, and may be derived by differentiating (\ref{eq:gamma2}) with respect to x$^\beta$. Then
\begin{equation}\frac{\partial\Gamma^\sigma_{\lambda\mu}}{\partial\mbox{x}^\beta}\mbox{ = }\frac{1}{2}\mbox{g}^{\nu\sigma}\left[\frac{\partial^2\mbox{g}_{\mu\nu}}{\partial\mbox{x}^\beta\partial\mbox{x}^\lambda}\mbox{ + }\frac{\partial^2\mbox{g}_{\lambda\nu}}{\partial\mbox{x}^\beta\partial\mbox{x}^\mu}\mbox{ - }\frac{\partial^2\mbox{g}_{\mu\lambda}}{\partial\mbox{x}^\beta\partial\mbox{x}^\nu}\right].\label{eq:gamma4}\end{equation}

   It should be noted that this relationship is linear in the second derivatives
   of g$_{\mu\nu}$ with coefficients depending upon g$_{\mu\nu}$. Such a
   differential equation is amenable to integration by parts, so that an
   equivalent condition may be derived from a variational principle. This will
   have the advantage of guaranteeing that the equation thus derived will
   determine g$_{\mu\nu}$ in a way that is independent of the coordinate system
   chosen. It is, in fact, necessary that such an equation may be found if
   invertible relabeling is to be realized in the finite system.

   Consider the variational principle 
\begin{equation}\mbox{ I = }\int \mbox{R}\sqrt{-\mbox{g}}\,\mbox{d$^4$x}\mbox{,  where }\delta\mbox{I = 0.}\label{eq:var1}\end{equation}

   Here g is the Jacobian of the transformation from the locally canonical coordinate system, which is given by the determinant of g$_{\alpha\beta}$, and R is a scalar function of g$_{\alpha\beta}$ and its first and second derivatives. 

   Now equation (\ref{eq:gamma4}) is a relation in symbols with four indices, so it will be advantageous to think of R as
\begin{equation}\mbox{R = g}^{\mu\nu}\mbox{R}_{\mu\nu}\mbox{ = g}^{\mu\nu}\mbox{R}^\lambda_{\mbox{ }\mu\lambda\nu}.\end{equation}

   If R is to be a scalar, then R$^\lambda_{\mbox{ }\mu\omega\nu}$ must be a
   tensor, and if $\delta$I = 0 is to yield a relation of the form of
   equation(\ref{eq:gamma4}), then this tensor must be linear in the second
   derivatives of g$_{\alpha\beta}$. Adventitiously, it's the case that there
   is, up to constant multiples, exactly one tensor with these
   properties\footnote{Proof of this assertion may be found on pp. 131-134 of
   \cite{Weinberg}.}. It is the Riemann-Christoffel Curvature Tensor. Then R must be the Riemann Curvature Scalar.  

   Applying this information to (\ref{eq:var1}) yields 

\begin{equation}\mbox{R}^{\mu\nu}\mbox{ - }\frac{1}{2}\mbox{g}^{\mu\nu}\mbox{R = 0.}\label{eq:Field}\end{equation} 

   This is Einstein's Field Equation (in free space). It is, according to the previous analysis, the unique equation defining a self-determined g$_{\alpha\beta}$ field which may merely express a finite local relation within the general symbolic system, this relation being determined solely by the requirement that relabeling be carried out in a one-to-one fashion. 

\subsection{The Significance of the Field Equation}

  The equation (\ref{eq:Field}) must be satisfied by the metric in order to result in a self-consistent finite theory which embodies simply the notion of relabeling. It should be possible, according to the reversible way in which the theory has been constructed, to make the equation (\ref{eq:Field}) the sole basis of the entire theory. This is, as will be indicated, in fact possible. This approach may also be extended to provide the basis of all theories which are merely consistent with relabeling.  

   The theory presented so far relating to the interpretation of equation
   (\ref{eq:Field}) is somewhat unsatisfactory in that a manifold on which
   (\ref{eq:Field}) may be solved anywhere does not in any way indicate any
   finite number of points which may be distinguished from the others. There has
   to be some way in which a particular finite collection of points may be
   singled out so that equation (\ref{eq:Field}) describes their behavior. The
   very construction of these equations indicates that such a distinction must
   be provided by a topological invariant. The curvature, in itself, is not
   eligible to make such a finitely defined distinction, so the only possibility
   is that points may be distinguished by there being singularities in the metric.

   This representation of the interaction of a finite number of distinguished
   points was achieved by Einstein and Infeld \footnote{ See reference
   \cite{Infeld}.}. Their conclusions were that the equations (\ref{eq:Field})
   were, in themselves, sufficient to completely determine the behavior of the
   singularities representing distinguished points. Moreover, it was found that
   the condition that the singularities of the metric field be restricted to the
   distinguished points resulted in these points moving according to the
   geodesic equation. Thus the Field Equation implies, in this sense, the
   geodesic equation and consequently that the intervals between points be
   invariant under one-to-one relabeling. 

   The preceding analysis also indicates how to determine all finite relations
   which are \textit{consistent} with relabeling, for such relations must be determined by solutions of an extension of equation (\ref{eq:Field}). Extensions of equation (\ref{eq:Field}) will have to be of the form
\begin{equation}\mbox{R}^{\mu\nu}\mbox{ - }\frac{1}{2}\mbox{g}^{\mu\nu}\mbox{R = kT}^{\mu\nu}.\label{eq:Field2}\end{equation} 

   An identity of Bianchi\footnote{ For this and other mathematical details in this section, see \cite{Dirac2}.} then indicates, as a consequence of (\ref{eq:Field2}), that 
\begin{equation}\mbox{T}^{\mu\nu}_{\mbox{      :}\nu}\mbox{ =0 , where``:" indicates covariant differentiation.}\label{eq:Conserve}\end{equation}

   This is a conservation equation, and T$^{\mu\nu}$ will give what will be interpreted as an Energy-Momentum Tensor. In fact this interpretation of T$^{\mu\nu}$ can be justified fully. The phenomena characterized by (\ref{eq:Field2}) will, because of the maintenance of separation which the construction of (\ref{eq:Field2}) uniquely reflects, be localizable, and localizable phenomena may be described by corresponding proper velocities  
\begin{equation}\mbox{v$^\mu$ = $\frac{\mbox{dx}^\mu}{\mbox{ds}}$.}\label{eq:vel}\end{equation} 

   Now a general second order tensor may be expressed as a sum of terms which are the product of a scalar and two vectors. Thus, for a given scalar field $\rho$,
\begin{equation}\mbox{T}^{\mu\nu}\mbox{ = }\rho\mbox{v}^\mu\mbox{v}^\nu\label{eq:source}\end{equation}
gives the most general expression for a source contribution for (\ref{eq:Field2}).

   Applying (\ref{eq:Conserve}) to this source results in 
\begin{equation}\mbox{v$^\mu$($\rho$v$^\nu$)$_{\mbox{:}\nu}$ + $\rho$v$^\nu$(v$^\mu$)$_{\mbox{:}\nu}$ =0.}\label{eq:energy1}\end{equation}

   Multiplying (\ref{eq:energy1}) by v$_\mu$ yields
\begin{equation}\mbox{v$_\mu$v$^\mu$($\rho$v$^\nu$)$_{\mbox{:}\nu}$ + $\rho$v$^\nu$(v$_\mu$v$^\mu_{\mbox{:}\nu}$)=0.}\label{eq:energy2}\end{equation}

   Now g$_{\mu\nu}$v$^\mu$v$^\nu$ = 1 implies that 0 = (g$_{\mu\nu}$v$^\mu$v$^\nu$)$_{\mbox{:}\sigma}$=2g$_{\mu\nu}$v$^\mu$v$^\nu_{\mbox{:}\sigma}$, and this yields
\begin{equation}\mbox{v$_\nu$v$^\nu_{\mbox{:}\sigma}$ = 0.}\label{eq:sub1}\end{equation}

   With this the last term of (\ref{eq:energy2}) drops out, leaving
\begin{equation}\mbox{($\rho$v$^\nu$)$_{\mbox{:}\nu}$=0.}\label{eq:conserve2}\end{equation}

   This is the equation of conservation of Energy-Momentum, so that imposing the field equation has the automatic consequence of requiring the conservation of energy-momentum.

   Returning attention to (\ref{eq:energy1}), it is seen that the first term vanishes, so that, for $\rho\not=$0, this leaves
\begin{equation}\mbox{v$^\nu$v$^\mu_{\mbox{:}\nu}$=0.}\label{eq:geo1}\end{equation}

   Consider the geodesic equation when v$^\mu$=v$^\mu$(x$^\alpha$(s)), so that multiple paths may be explored. It may be confirmed that in this case the geodesic equation is given by (\ref{eq:geo1}). Thus imposing the field equation also requires that phenomena follow the geodesic equation.  

   Thus, in both the free space case, given by (\ref{eq:Field}), and in the
   presence of ``matter", as is the case for (\ref{eq:Field2}), the Field
   Equation is, in itself, a complete basis for the entire Relativity Theory.
   This is of particular importance in that it finally justifies, in a restricted
   case, the faith that a purely formal symbolic theory of experience would not necessarily
   degenerate to mere tautologies, but rather could lead to a constructive
   theory which accounts for some of the structure in the diversity
   of experience. Relativity does, after all, describe gravitation and therefor
   leads to an agglomeration and organization of the various labels which
   correspond to ``matter". This leads to the conclusion that, for a given
   finite level of description which is inter-related in a one-to-one manner,
   the general symbolic system has thus far succeeded in formally generating
   structure and the corresponding metalabeling.

\section{On Turing Computability}

   Einstein's Field Equation leads, as demonstrated in the last section, to the satisfaction of the geodesic equation for the description of traceable phenomena in space-time. An important limiting case for the solution of the geodesic equation is that of arbitrarily slow processes. In such cases the geodesic equation reduces\footnote{ See, again, reference \cite{Dirac2}.} to the Newtonian form 
\begin{equation}\frac{\mbox{dv}^m}{\mbox{dx}^0}\mbox{ = }({\mbox{g}_{00}}^\frac{1}{2})_{\mbox{, m}}\mbox{  for m = 1,2,3.}\label{eq:newton}\end{equation}

   Here ``," indicates partial differentiation with respect to the variable with the indicated index. The form of these equations indicates that the particle motions described are those of motion according to Newton's laws in a conservative potential. Such potentials may describe, among other things, elastic collisions. Thus the notion of a system of particles that interact by elastic collisions has a strictly theoretical derivation. 

   From this intuitively appealing special case of phenomena that may be
   generated from Einstein's Field Equation, important conclusions follow. In
   fact, such elastic collisions define the ideal behavior of billiard balls,
   and, as a purely conceptual matter, it has been demonstrated \footnote{ See
   reference \cite{Fredkin}.} that such a ``physical system" may perform all of
   the computational steps necessary in order to realize a Turing Machine. Thus
   Relativity, as a special case, contains all of Recursive Function Theory. If
   Church's Thesis, that all computations may be carried out on a Turing
   Machine, is accepted then this demonstrates that that Relativity entails, in
   fact, all conceivable symbolic computations. Moreover, as all Turing
   Machines may be encoded as integers\footnote{Kleene gave such a standard
   form. See \cite{Kleene}.}, it would follow that any symbolic computation would be specified by a particular integer, and the effects of Relativity, in restricted special cases, could be given by this same number.

   It should also be observed that both Set and Class Theory are given in the context of a formal logic. Thus all derivations of statements in Set and Class Theory follow steps that may be directed by Turing Machines. It follows that the General Symbolic System cannot be any less powerful than either Set or Class Theory, as it, in fact, includes them.

\section{Statistical Prediction and Probability}

   Relativity Theory, its derivation, and some of its consequences have now been
   discussed. It has been found that Relativity Theory may be derived on a strictly formal basis and includes, as a special case, all of Recursive Function theory and Set and Class Theory as well. Given the breadth of these results it may well be wondered whether or not there can be any other finite theory which hasn't already been encompassed in Relativity Theory. The investigation of this possibility will be taken up next.

   Relativity Theory has been shown to be the canonical finitary formalism in
   the case when invertible operations are performed on individual symbols. It
   will be recalled, however, that the general symbolic system, besides such
   individual symbols, also admits of metalabels or, in other terminology,
   ``categories" as fundamental concepts. If the general symbolic system were
   exhausted by Relativity, then this would amount to saying that no formal laws
   may be stated in terms of categories. It is to be expected, and it will be
   shown in the following, therefor, that
   there are some finitary laws which take metalabels as such as their
   arguments. These relations will go beyond Relativity Theory. 
   
   Any finite relations which are to go beyond the predictions which Relativity
   already provides must, at least, not be computable by Turing Machines. Thus the considerations to follow will begin by considering what role Turing Machines may play in the theory of finite relations.

   Consider a general function $\rho$ which is to represent a finite relation. As indicated in the last section, any Turing computable relations may be represented by an integer m which encodes an appropriate Turing Machine. Thus the function $\rho$ must have the dependence    
\begin{equation}\rho=\rho\mbox{(}\phi\mbox{, m)}\end{equation} 
where $\phi$ is itself an undetermined function which gives a finite relation.
$\phi$ will therefor be, in general, a complex analytic function. Such functions
correspond, as has been noted, to at most countably many zeros, and so $\phi$
can give a relation between at most countably many data. The domain, and thus
the range, of $\phi$ also may thus, without loss in generality, be taken to be
the Natural numbers. The distinct natural numbers making up the range of $\phi$
will be referred to as ``attributes". These attributes naturally classify the action of $\phi$ in the sense that $\phi$(n) = $\phi$(m) indicates that n and m have the same attributes. 

   It may thus be said that a finite relation $\rho$ corresponds to a particular
   at most countable variety of integral attributes which "occur" in some sequence,
   together with the specification of a Turing Machine. The sequence itself
   corresponds to some finite relation which may be specified by an analytic
   function $\phi$.

   Now, for a fixed variety of attributes, any such function $\rho$($\phi$, m)
   may be recoded as 
\begin{equation}\mbox{$\rho$($\phi$, m) = $\omega$($\tilde{\phi}_m$)  where}\label{eq:rho1}\end{equation} 
\begin{equation}\mbox{$\tilde{\phi}_m$(n)$\equiv\phi$(T$_{\mbox{m}}$(n))  }\label{eq:rho2}\end{equation}

   If there is to be such a thing as a finite relation $\rho$ which is independent of Relativity, and thus also of Turing Computability, then it must be the case that 
\begin{equation}\rho\mbox{(}\phi\mbox{, m) = }\rho\mbox{(}\phi\mbox{, s)  for all naturals m,s .}\end{equation}
Such $\rho$ will be said to be ``T-independent". Some conclusions about T-independent functions will now be derived.

   Consider those Turing Machines T$_m$ which permute initial segments of the
   sequence corresponding to $\phi$. If $\rho$ is to be invariant under the action of all such Turing Machines then it must, first of all, be invariant under the action of all T$_m$ which permute the same initial segment of the sequence of attributes of $\phi$. Then the dependence of $\rho$ on this initial segment must be expressible in terms of the relative frequencies of each attribute within this segment. If this is to apply, for a general finite function $\rho$, for all initial segments, then, first of all, the relative frequencies of each attribute must approach limits. $\rho$ will then have to be a function of these limiting relative frequencies. In particular, $\rho$ may be the limiting relative frequency of one of the attributes of $\phi$. The limiting value, p(a), of the relative frequency of an attribute ``a" for a T-independent function will be called its ``probability".

   Now suppose T$_m$ yields an increasing function of n. Then, in general,
   $\tilde{\phi}_m$(n) = $\phi$(T$_m$(n)) will have the effect of generating a
   new sequence in which certain of the original sequence will have been omitted
   based solely on the place, in terms of n, they occupied. Such ``place
   selections" must not, therefor, alter the limiting relative frequency of an
   attribute if $\rho$ is to give a finite relationship independent of what
   Turing Machines may yield. Such sequences, being ``insensitive" to place
   selections, will be said to be ``Random". It is of particular significance
   that ''random" sequences \textit{must} be convergent as any particular segments
   of such sequences which might prevent convergence of the associated relative
   frequencies may be removed by place selections.  
   
   It should be pointed out that, for
   a given sequence to be random, it can in no sense be ``known" term-wise, for
   the sequence is only to be acknowledged in the role which the theory assigns
   to it, and the theory considers a random sequence and any other sequence 
   derived from it by place selections to be interchangeable. With this
   observation it becomes clear that while the argument of a probabilistic function
   may be described as a random sequence, it may more simply be described as a
   \textit{proper} category, where a proper category is a metalabel which \textit{cannot} be defined in
   terms of its corresponding assigns. To be clear, the unknowability of the random sequence entails the impossibility of uniquely determining $\phi$.

   From the above considerations it may be concluded that the quantities which T-independent finite functions may give as description beyond those obtainable from Turing machines may be thought of as being limiting relative frequencies of sequences of attributes. Furthermore, the sequences of attributes which these limits apply to must be random, that is, they must be insensitive to place selection. These two properties correspond, in an obvious way, to the axiomatic basis for Von Mises's frequency theory of probability, although the theory presented here has a strictly conceptual basis and is not asserted to have an empirical basis as is that of Von Mises. It should also be stressed that this theory is distinct from the usual measure-theoretic formulation of probability\footnote{ See \cite{Kolmogorov}.} or the classical theory of probability\footnote{ For a comprehensive survey of distinct theories of probability see \cite{Howson}. \cite{Cox} determines features which all theories of probability ought to share.}. 

   Recall that the question at issue here is the possibility of there being a
   finite theory that is independent of Turing computability. With this in mind
   it is clear that something must be said about the acceptability of infinite
   sequences necessarily playing a role in the theory. In this connection it
   must be recalled that it is the limiting relative frequency itself, and the
   randomness of the sequence, which are demanded by the theory, and not any
   \textit{particular} infinite sequence of occurrences of attributes nor any calculations
   involving an infinite quantity; The calculation of this limiting frequency,
   which depends solely on $\rho$, need not require any such limiting processes
   to be carried out.
   
   In fact, a finite T-independent theory treats probability as a
   ``propensity". This propensity, as defined by $\rho$ and not requiring the
   elaboration of a sequence of attributes, operates
   somewhat like a force field in classical physics: Just as a force causes a
   definite dynamical evolution, so a probability causes an evolution of the
   relative frequency of an attribute towards a definite limit. This action is
   unlike that of a classical force, however, in so far as the propensity
   governs the collective behavior of the whole sequence and may therefor be
   considered to act non-locally\footnote{ Quantum theory may, in fact, be
   formulated as a strictly causal though non-local theory as was done by
   deBroglie and Bohm. See \cite{Holland}.}. In this sense probability may be
   thought of as a non-local relation. It is only after viewing probability in
   this finitely-oriented way that a finite sequence of attributes may be
   explicitly introduced which is supposed to be governed by a law of
   probability. Then it may be argued, based on this, that as the sequence is
   extended under the given probability conditions its relative frequencies will
   approach particular limits. There can, in fact, be no cause to doubt that the
   sequence is random as the sequence is, after all, an artifact of the
   probabilistic theory itself. 

   One last remark on statistical theory in general. It will be noticed that the above theory, being based on the notion of T-independence, treats of discrete attributes. This is not, however, an essential restriction, for this probability theory may be extended, in the usual way, to probabilities over continuous attributes. Randomness may still play a role in the continuous theory as continuous distributions may be approximated by discrete ones on which the above notions of Randomness make sense. The discussion to follow will therefor retain the discrete perspective for the sake of ease of presentation and derivation.

\section{Quantum Theory}

   The possibility of extending the theory of finite relations beyond what is given in Relativity was investigated in the last section. These considerations led to the conclusion that if such were possible, then the corresponding theory would have to be a probabilistic one: The foundations of the necessary form of such a theory were in fact given. Of particular interest was the conclusion that a finite theory of probability is necessarily a ``realistic" one in which probability is a certain propensity for relative frequencies to approach particular limits. These were important but also very general conclusions. The goal of this section is to continue on to derive a more precise and quantitative theory. The natural development of the foregoing theory will lead uniquely to Quantum Theory, thus giving it a definite theoretical significance quite apart from any appeal to experimental justification.

   The probability of an attribute has been defined to be its limiting relative
   frequency in a random sequence. Suppose that certain attributes a$_\alpha$
   are grouped together and all considered instances of the attribute a.
   Calculation of p(a) must then be such that the appearance of a$_\alpha$ in
   the random sequence adds one to the frequency of occurrence of the a's, and
   thus contributes to the relative frequency also. Consider an initial segment of length n. Suppose a appears n$_a$ times, while each a$_\alpha$ appears n$_\alpha$ times. Then n$_a$ = $\sum_\alpha$ n$_\alpha$. Because all terms are positive, limits may be interchanged to derive
\begin{equation}\mbox{p(a) = }\lim_{n\rightarrow\infty}\frac{\mbox{n$_a$}}{\mbox{n}}\mbox{ = }\sum_\alpha\lim_{n\rightarrow\infty}\frac{\mbox{n}_\alpha}{\mbox{n}}\mbox{ = }\sum_\alpha\mbox{p(a$_\alpha$)}\label{eq:prob1}\end{equation}

   As a special case of this suppose all attributes $\alpha$ are considered to be instances of the attribute u. Then
\begin{equation}\sum_\alpha\mbox{p($\alpha$) = p(u) = 1.}\label{eq:prob1a}\end{equation}

   Suppose an attribute a is also considered to be a case of an attribute b, though not necessarily the other way around. Suppose also that p(b) $\not=$0. Consider, again, an initial segment of length n. Let n$_{\mbox{a$|$b}}$ be the number of attributes a found among the b's that were found in the initial segment. Also let n$_{\mbox{a}}$ be the number of a's found and n$_{\mbox{b}}$ the number of b's found. Note that then n$_{\mbox{a$|$b}}$ = n$_{\mbox{a}}$, and if n$_{\mbox{b}}\rightarrow\infty$ then it follows that n$\rightarrow\infty$.  Then
\begin{equation}\mbox{p(a$|$b) }\equiv\lim_{{\mbox{n}_{\mbox{b}}}\rightarrow\infty}\frac{{\mbox{n}}_{\mbox{a$|$b}}}{{\mbox{n}}_{\mbox{b}}}\mbox{ = }\lim_{{\mbox{n}}\rightarrow\infty}{\left(\frac{{\mbox{n}_{\mbox{a}}}}{{\mbox{n}}}\right){\left(\lim_{{\mbox{n}}\rightarrow\infty}\frac{\mbox{n}_{\mbox{b}}}{\mbox{n}}\right)}^{-1}}\mbox{ = p(a)[p(b)]$^{-1}$.}\end{equation}
so that it follows that
\begin{equation}\mbox{p(a) = p(a$|$b)$\ast$p(b).}\label{eq:prob2}\end{equation}

  It should be stressed that in order to apply this rule to factor a probability
  it must be the case that the occurrence of a implies the occurrence of b.

   The considerations above have indicated some relations, holding between the probabilities determined by a finite relation, that must be satisfied, but didn't tell anything of how these probabilities were, themselves, to be determined. This will be addressed next.

   In the determination of an appropriate function to represent a finite
   relation there will generally be certain data which is considered to be
   dependent on other data and primarily varies and other data which is
   presumed, and not considered to vary. Such is the case, in standard physical
   theories, with dependent variables on the one hand, and with independent
   variables, parameters, and ``physical constants" on the other. Such a
   segregation of the data is not proven to be absolute, but, rather, is a
   matter of convenience in stating the relations which are utilized in the theory. For this reason, a means of indicating a segregation of data will also  be adopted here.

   It has already been shown above that finite relations must take a particular form. Probabilities, being determined by finite relations, must also take this form, and so must be given as functions of the magnitude, or, equivalently, of the square of the magnitude, of some complex analytic function.
 The particular way in which a probability is to depend upon the square of the magnitude of an analytic function must be well-defined and the same for all probability calculations but is otherwise unrestricted and so may be determined simply as a matter of convenience. It turns out to be convenient to choose the convention that probability be given, in fact, by the square of the magnitude of some complex analytic function. This choice results in the probability function being a measure\footnote{ See pp. 70-72 of \cite{Everett}.} and also avails the theory of a geometric analysis on a Hilbert Space, the given magnitude easily being constructed by an inner product\footnote{ For the relevant functional analysis see, e.g., \cite{Bachman}.}. Combining the above observations results in the following notation and results:

   In the context of certain presumed data which are symbolized by ``i", the
   probability, p(a), of a particular attribute a is given by the square of the
   magnitude of a certain analytic function, referred to from now on as the corresponding ``probability amplitude"\footnote{ For an exposition of Quantum Theory on this basis, see \cite{Feynman1}.}, determined by the attributes under consideration and evaluated at a. This amplitude is denoted, as was done by Dirac\footnote{ See \cite{Dirac3}.}, by $<$a$|$i$>$. Thus
\begin{equation}\mbox{ p(a) = }{|<\mbox{a}|\mbox{i}>|}^2\label{eq:quant1}\end{equation}

   Now suppose that p(a)=$|<$a$|$i$>|^2$, p(b)=$|<$b$|$i$>|^2$, and p(a$|$b)=$|<$a$|$b$>|^2$. Applying equation(\ref{eq:prob2}) results in 
\begin{equation}<\mbox{a}|\mbox{i}>\mbox{ = }{\mbox{e}}^{\imath\theta}\ast<\mbox{a}|\mbox{b}>\ast<\mbox{b}|\mbox{i}>\mbox{.}\label{eq:quant2}\end{equation}

Here it is to be noted that $\theta$, if it need be determined, must yet be
determined by other considerations. In fact, it is a standard insight of Quantum
Theory that such exponential factors allow for, and demand, the conservation of
such quantities as energy, charge, angular momentum, etc.\footnote{ See vol. III
of \cite{Feynman2}.}. Also, it should be reiterated that this result applies, as
does (\ref{eq:prob2}), only in the case that the occurrence of a implies the
occurrence of b. Conversely, whenever an amplitude may be factored into a product of amplitudes it will have to be the case that any corresponding probabilities will be for attributes which are related in this way.

  The only other relation from probability theory is equation(\ref{eq:prob1}). It applies in the case when a variety of attributes a$_\alpha$ are to be considered to be instances of another attribute a. This leads to the relation
\begin{equation}{|<\mbox{a}|\mbox{i}>|}^2\mbox{ = }\sum_\alpha |<\mbox{a}_\alpha|\mbox{i}>|^2\mbox{.}\label{eq:quant3}\end{equation}

   This rule is evidently not very helpful in determining $<$a$|$i$>$ itself,
   although it does lead to the conclusion, following the case in
   (\ref{eq:prob1a}), that $|<$u$|$i$>|^2$ =1. In other words, the ``total"
   amplitude must be normalized. Thus the probability rules offer little help,
   but also little hindrance, in the determination of the amplitudes that will
   play a role in the theory. The only possible approach that remains in
   determining a constructive theory is thus to analyze amplitudes according to the simple fact that they are given by analytic functions.

   A defining characteristic of analytic functions is that they may be expressed
   as power series sums, and such series may be manipulated term-wise, and
   analyzed in this way. Consider, then, an amplitude and a power series
   expansion of it. Each of the summands, in order that it may be interpreted in
   the theory, must be somehow related to an amplitude. Each term may, in fact, be formally analyzed according to the rule in equation (\ref{eq:quant2}), so that, in general, an amplitude $<$a$|$i$>$ may be analyzed as
\begin{equation}<\mbox{a}|\mbox{i}>\mbox{ = }\sum_\beta{\mbox{e}}^{\imath{\theta_\beta}}\ast<\mbox{a}|\beta>\ast<\beta|\mbox{i}>\mbox{.}\label{eq:quant4}\end{equation}

   It must be remembered that the $\beta$ are introduced here in a strictly
   formal sense, for if the occurrence of a were to imply the occurrence of any such $\beta$ then application of equation (\ref{eq:quant2}) would yield 
\begin{equation}<\mbox{a}|\mbox{i}>\mbox{ = }{\mbox{e}}^{\imath\delta}\ast<\mbox{a}|\beta>\ast<\beta|\mbox{i}>\mbox{.}\end{equation}
and thus the sum in equation (\ref{eq:quant4}) would have to either degenerate into this single term or else lead to contradictions if more than one distinct such $\beta$ were to be found. On the other hand, if, starting from a situation described by the amplitude $<$a$|$i$>$ given by (\ref{eq:quant4}), an event described by a particular $\beta$ is considered to occur, then the amplitude for a would be of the form $<$a$|\beta>$ and the transition between these situations could, to needlessly indulge in a ``physical" description of what may be considered a logical process, be said to be a case of a ``collapse" of the ``wave function" $<$a$|$i$>$. It should be remembered that the theory is derived here under the assumption that there is no separate realm of experience which may affect the developments within the theory.

  If the expansion (\ref{eq:quant4}) is not to refer to any necessary
  intermediate process $\beta$ in the description of the attribute a, then there
  must be a great deal of freedom in the choice of such formally chosen
  ``virtual" attributes. The selection of any particular expansion must,
  therefor, be justified on the basis of calculational advantages. The benefit
  of this analysis consists in two effects. First of all, each summand may be
  simple and so much so, in fact, that the values of such amplitudes may be
  \textit{assumed} in the construction of a calculational theory. Such is, in
  fact, done in the case of the so-called ''bare" mass and ''bare" charge of the
  electron in Quantum Theory. It may also happen, as a second benefit, that the
  selection of a particular expansion may result in rapid convergence of the sum
  (\ref{eq:quant4}) so that it might even be terminated after a few terms. 

   At this point it is clear that the above considerations have completely
   reproduced Feynman's formalization of Quantum Theory, including a
   justification of the central role played by the Feynman diagram approximation
   technique\footnote{ See \cite{Feynman1} for the quantum formalism, and
   consult \cite{Feynman2} for a discussion of the Feynman diagram technique.}
   but with the advantage of showing that corresponding graph-theoretical
   concepts naturally arise in the algebraic analysis of $\phi$. It should be noted that this approach, in basing all calculations on a single history, is unsuitable for addressing the intrinsic compatibility of quantum theory with ``multiple observers". However, because this approach is avowedly applied as an approximation, it is also not necessary to do so, for any effects associated with this kind of situation may always be relegated to the uncalculated terms in the approximation. 

   Another formal consequence of the analytic nature of the amplitudes is of
   paramount importance when it is desired that exact calculations may, in
   principle, be made. Because expansions of analytic functions, such as in
   (\ref{eq:quant4}), may be manipulated term-wise in the case of many
   mathematical operations, such operations will be linear in the amplitudes.
   Thus the theory given here may also, for those that prefer to do so, be
   translated into the coordinate-free operator formalism which Von Neumann gave
   to Quantum Theory\footnote{ Again, see \cite{Neumann1}.}, with one important
   difference: \textit{All} transformations of state must now be represented by
   application of a (Hermitian) linear operator as there can no longer be
   special procedures for representing ``measurements". This then leads uniquely
   to a quantum theory which is formally identical to that of Everett's Many
   Worlds approach\footnote{ See pp. 3-149 of \cite{Everett} for an exposition
   of this theory.} although it must be stressed that in the context of this
   thesis there is no implication of, nor can there be, histories and/or universes which are anything but \textit{formally} separate. As all changes of state are of the same formal nature, it would be unacceptable, besides merely troubling or unbelievable, for any to be especially felt as a ``split of the universe" or a ``branch in one's history".

   Everett's derivation of his theory assumed that a Schr\"{o}dinger equation was
   provided for the time evolution of the state, but the theory may be derived
   without this assumption. As Everett has shown, there is, corresponding to a
   sequence of attributes which defines a history, a corresponding sequence of
   applications of corresponding operators. Each history corresponds, exactly as
   in Feynman's sum over histories approach to Quantum Mechanics\footnote{See
   \cite{Feynman3} for an exposition of this theory.}, to a summand in the superposition (\ref{eq:quant4}) which formally represents an amplitude. In this way the superposition may be thought of as a net of histories. 
   
   Begin with some operator H which is to be a ``Hamiltonian" which gives the
   evolution of what is to be considered, by definition of this choice, an
   ``isolated" system. In that case all other operators must represent, by
   definition, ``interactions". In examining any history one may then refer to
   ``change" whenever H alternates with some other operator. One may then
   construct a corresponding notion of time for each individual history where
   distinct points in time are identified by such transitions while interactions
   shared with other histories identify related moments. 
   
   Everett has
   shown\footnote{ Refer, once more, to pp. 70-72 of \cite{Everett}.} that there
   is a measure on histories which agrees with the probability quantum mechanics
   assigns to them. Thus, it may be observed here that a direction of a history which progresses towards
   the appropriate limiting frequencies may be identified with the future along
   that history. This then accounts for the ordinal nature of time. If it is
   desired that time also possess a measure, then this may be achieved by
   selecting \textit{some} operator, T say, distinct from H and identifying any successive
   interactions involving T with unit intervals in time along that history.
   The rate of change will then correspond to the likelihood of such
   interactions. 

   While all formal histories which correspond to terms in (\ref{eq:quant4})
   play a role, they weigh in according to the measure assigned to them - in
   other words, according to their probability: This is the quantal basis for
   the variational form of the laws of physics. It is thus possible to dispense
   with relatively unlikely histories and, conversely, any initial conditions
   which may play a role in theory must be sufficiently likely. In this sense initial
   conditions no longer enter into theory as independent data. This makes sense
   as, after all, the distant past, in not being accessible to personal
   experience, has always been \textit{reconstructed} in
   light of
   current understanding or \textit{theory}. 
   
   The whole character of the theory is determined, in fact, by H, because
   identifying H is a matter of identifying when, by contrast, something may
   happen. If an operator has a particular symmetry, then the history
   corresponding to it will appear with a corresponding degeneracy, or repeated
   appearance, in the superposition (\ref{eq:quant4}) so that symmetries
   correspond to prevalent terms and thus to significant histories. This
   explains why the search for the laws governing quantum-scale interactions has
   been successfully guided by considerations of symmetry. Although this account of quantum theory is by no means a usual one, the formal rigor of all mathematical steps has already been born out in the references cited and the work done above, with additional confirmation to be found in the history of the development of quantum theory. 

   The above analysis is self-contained and fully determinate excepting that it
   remains to explain how the Hamiltonian operator arises as, surely, not just
   any one will do! Such an explanation will, however, proceed from the
   following discussion which links the above ideas with formally thermodynamic considerations\footnote{ For an exposition of the required Thermodynamic theory see, for example, vol. I of \cite{Kubo}.}. 

   First of all, all component histories will, by definition, converge in their futures to the respective probable states corresponding to the analytic state function $\phi$ which finitely defines the system. This being true individually, it also follows that all histories proceeding from a single state $\phi$, and thus forming a system, will also converge towards \textit{mutual} equilibrium. This limiting condition is that which is guaranteed by the so-called Zeroth Law of Thermodynamics, wherein a temperature function is introduced to express the conditions of relative equilibrium holding between any collection of systems. That the progress towards the limit identifies a direction in which time ``flows" in a particular history is the essential content of the Second Law of Thermodynamics.  

   The First Law of Thermodynamics is that of Energy Conservation. It has,
   however, already been noted, in the discussion following equation
   (\ref{eq:quant2}), that energy is conserved in quantum theory. It is also well-known that the boundedness, and general smallness, of the ground state energy of quantal systems leads to the Third Law of Thermodynamics. It may be therefor concluded from the above analysis that all of the laws of Thermodynamics follow within this quantum formalism.  

   Quantal systems being therefor thermodynamic systems, it follows that they
   may be analyzed thermodynamically. Since the theory presented here is
   strictly formal, it necessarily follows that the ensembles used in the
   analysis \textit{must} be (virtual) Gibbs Ensembles. As is always the case in
   the thermodynamic analysis of a system, the properties of the thermodynamic
   system are completely defined by the (Grand) Partition Function which
   corresponds to the Gibbs Ensemble. Such ``bulk" properties as the volume and
   pressure exerted by the system are, in fact, expressible, in a standard way,
   in terms of the partition function\footnote{Refer, again, to \cite{Kubo}.}. Now note: \textit{Both} the partition function and $\phi$ \textit{completely} determine precisely the behavior of the system so that these functions are interchangeable! Thus $\phi$ will be taken to \textit{be} the partition function. 

   Identifying $\phi$ to be the partition function has immediate consequences. First of all, as $\phi$ is necessarily a complex analytic function, it naturally follows that this is also the proper way of thinking about the partition function. In this case one no longer introduces the complex ``fugacity"\footnote{ Refer to \cite{Kubo} for an explanation of this concept.} into the partition function as a formal trick but, instead, the question rather becomes how one starts with a complex-valued partition function with complex arguments and ends up only concerned with a real-valued partition function with real-valued arguments. In this connection it should be recalled that $\phi$ is constructed so as to \textit{encode} certain corresponding complex zeroes. The nature of the events which may occur in the given thermodynamic system is therefor determined by the location of these zeroes: There must be certain criteria as to the situation of these zeroes the satisfaction of which results in singular phenomena. 

   At this point it is appropriate to introduce, as a formally necessary
   development of the finite case rather than as a merely plausible addition, a
   classic result of Yang and Lee\footnote{ See \cite{Yang}.} which states that
   any ambiguity in the value of certain ``bulk" variables, such as density, and
   the corresponding changes of ``phase" of the system, such as from liquid to
   solid, must occur precisely at points \textit{on the real axis} which are not
   separated, by any neighborhood in the complex plane, from the zeros of $\phi$ (the
   partition function). This shows, to begin with, that the variables which are
   to represent points of transition between phases must be real: This justifies
   requiring all variables, in the end, to be real-valued. 
   
   According to this result it also follows that, as the positioning of the zeroes
   of $\phi$ corresponds to condensation phenomena, these zeroes also
   define ``material structure" in concrete and tactile terms such as pressure, volume, and
   temperature. Such relatively stable material structure is a candidate for
   treatment as an isolated system. The symmetries of such systems then will
   identify corresponding Hamiltonians. 
   
   This self-contained Quantum-Theoretical
   analysis, like that of the Relativistic analysis before it, reduces the
   problem of the variety of concrete experience to that of the problem of the
   ``existence" of \textit{any} phenomena whatsoever. In other words,
   \textit{if} $\phi$ has any zeroes at all then this automatically engenders a
   corresponding physical picture, operating according to the quantal laws,
   comprised of interacting phases of matter, such as solids, liquids, and gases,
   each occupying certain volumes and ``influencing" one another barometrically
   and thermally or in terms of other recognizable thermodynamic variables
   (Such as, for example, the chemical potential).
     
   In reviewing the considerations above it is clear that the attempt to
   construct a finite theory beyond that which is determined by Relativity has
   resulted in an essentially unequivocal derivation of the mathematical
   formalism of Quantum Theory. It has done so without appeal to experiment and
   without accepting the notion of an external realm of experience to which the
   formalism need refer. In doing this it has been demonstrated that the great
   variety of physical interpretations of the quantum formalism are, presumably,
   unnecessary, for from a finite strictly mathematical standpoint a unique
   propensity interpretation of probability and of Quantum Theory arises.  

\section{Conclusions on the Derived Theories}

   In the preceding it has been found possible to derive both Relativity and
   Quantum Theory on a strictly conceptual basis. That these derivations may
   have been performed without appeal to experiment obviously undercuts the
   notion that these theories must be thought of as describing or being
   associated with an external physical reality separate from language.
   
   Again, no appeal to experiment has
   been made, but, rather, the principle aspects of the phenomena which these
   theories address have been shown to follow from strictly formal principles;
   It should be understood that, instead of experiment justifying theory, it has
   been found that the scope of thinkable finitary theory has shown \textit{why} such
   such phenomena presumably \textit{must} occur.
   In particular, Relativity Theory has been argued to uniquely reflect relations
   which maintain one-to-one relabeling within the finite general symbolic
   system. Quantum Theory was subsequently argued to reflect every and anything
   else that may be called a finite relation which is not justified on a
   Relativistic basis. In particular, it has been indicated that Quantum Theory,
   instead of treating of relationships between individual symbols, 
   yields rather all finite relations which govern proper categories or, in other
   words, are based upon metalabeling relationships as such. 
   
   As it has been argued that consistency in the utilization of individual
   symbols, together with the conveyance of metalabeling
   relationships between symbols, exhausts all that may be required of language,
   it is to be expected that Quantum Theory, in virtue of the nature of its
   construction, should, together with Relativity Theory, provide a complete and
   consistent description, if such is possible, of all finite relations within
   the general symbolic system. This possibility is precisely what is
   investigated in the next section. 

\section{Einstein's Alternatives and Finite Reality}

    Relativity and Quantum Theory should together, if such is possible, comprise
    a complete finite theory. It will be found, however, that they are not
    merely distinct but also, in the context of a finite restriction of the
    general symbolic system, incompatible. This incompatibility will be shown by
    an argument similar to one that Einstein gave\footnote{ See pp. 168-173 of
    \cite{Born}.}. The possible ways of resolving this incompatibility will be
    investigated and this will lead, on the basis of the general symbolic
    system, to plausible far-reaching conclusions about physical theory. 

   The general symbolic system now has, in Quantum Theory and Relativity, two
   distinct formal procedures which may be applied to the data which comprise
   any given ``physical situation". The question now arises as to which parts of
   the data each of the theories is to apply. Is it possible, in other words, to
   recognize distinctly ``classical" data to which Relativity alone applies?
   
   Relativistic phenomena are characterized by the energy-momentum tensor, which
   is, it may be recalled, the right-hand side of equation (\ref{eq:Field2}).
   Conversely, equation (\ref{eq:Field2}) indicates that this tensor is a
   function of the space-time itself, so that a flow of energy must correspond
   to changes in the space-time manifold. As quantum processes are not
   energetically isolated it follows, therefor, that they too must be dependent,
   at least in part, on the space-time variables. This influence may also be seen
   to go the other way, as it is well known that, in accordance with the Third
   Law of Thermodynamics, Quantum Theory results, in a number of cases, in
   predictions that agree with those of Classical Physics and, in fact with,
   Relativity. This is also the content of Bohr's Correspondence Principle.
   
   Because of this limiting behavior there can be no finitary conceptual
   distinction in general between a result produced by quantal laws and a result
   which is the outcome of local deterministic processes. In short, as is
   usually taken as a physical necessity, Quantum and Relativistic processes
   ``interact": They may not be considered to be disjoint physical theories
   which don't share common variables upon which they depend. Thus
   quantum-theoretical amplitudes must generally depend, in part, on space-time
   variables.

   Let a be an attribute which indicates something about events at a particular
   space-time point P$_1$. Also let $<$a$|$i$>$ be the amplitude associated with
   this attribute. Recall that, in a finite probabilistic theory, this amplitude
   must be considered to be a kind of propensity, changes of which therefor have an
   immediate space-time significance. 

   Now consider some attributes $\beta$ which aren't implied by a so that equation (\ref{eq:quant4}) may be applied. Then
\begin{equation}<\mbox{a}|\mbox{i}>\mbox{ = }\sum_\beta{\mbox{e}}^{\imath{\theta_\beta}}\ast<\mbox{a}|\beta>\ast<\beta|\mbox{i}>\end{equation}
holds. Note that $\beta$ could represent almost anything.

   Consider an attribute $\beta$ which indicates something about events at a
   space-time point P$_2$. Further suppose that the interval from P$_1$ to P$_2$
   is positive. Now the occurrence of $\beta$, according to
   (\ref{eq:quant4}), requires the formal transition
   $<$a$|$i$>\mapsto<$a$|\beta>$. If thought about in physical terms, such a
   change in the amplitude is an immediate ``reality", so that a change at P$_2$
   entails an instantaneous change at P$_1$. But, even if viewed in a strictly
   formal sense, this is obviously a violation of the restriction to local interactions in Relativity Theory. Thus, while Relativity and Quantum Theory cannot be kept apart, they are nevertheless incompatible.

   There are exactly two assumptions that have been made in the construction of
   these finitary physical theories; That there is no necessity for a separate
   realm of experience apart from language itself, and that all descriptive
   quantities in the theory must be finite. At least one of these assumptions
   must be dropped.  

   If the first assumption is dropped, then, as noted before, the permanent role
   of experiment entails that there can never be a final understanding or
   ultimate physical theory. If the second assumption is dropped then it will be
   necessary to have a mathematics that can naturally handle infinite
   quantities, as was attempted in the physical programme known as
   ``renormalization"\footnote{ See \cite{Manoukian}.},  and interpret such
   quantities in a sensible fashion. In order to maintain the derivation of the
   theory of the general symbolic system without recourse to experiment the
   second option will next be explored.

\section{On Generalizing Finite Theories}

   It may have seemed at the time that consideration of infinite quantities and
   the development of a Calculus which doesn't depend upon finitely defined
   limits was an extravagance, but it is clear now that there was a definite
   justification for these steps. In fact, had the Calculus been restricted to
   the usual finitely defined operations it would have been impossible to
   identify the cause of the inconsistency, in a non-empirical setting, with
   simply the finiteness of the variables, for the operations of the Calculus also went into the derivation of the theory and might have been connected with the difficulties.

   Having already laid the groundwork, the extension of the previous theory to
   the infinite realm is relatively simple. It has already been noted that
   the operations of the generalized Calculus extend those of the usual Calculus
   to the infinite case, so the Calculus may still be applied in the derivations
   of exact theory. Suppose now that some equation is to hold in the potentially
   infinite case. This same equation must still apply even if all of its
   arguments are finite, but in that case the given equation would have to be
   part of the theory for the finite case. It thus follows that the general
   theory for the manipulation of symbols is formally identical to, in fact it
   is determined by the same equations (\ref{eq:Field2}), (\ref{eq:quant1}),
   (\ref{eq:quant2}), and (\ref{eq:quant4}) as govern the finitely-based theory which has already been
   derived above. The \textit{only} difference between the finite and non-finite
   cases is, therefor, merely the potentially non-finite nature of the variables
   employed in the theory.
   
   While an accurate conclusion, it is still not very informative to merely
   imply that the variables in the general symbolic system are (generally)
   non-finite analogues of the finite space-time coordinates and probabilities
   which characterize the usual Relativistic and Quantum-theoretical formalisms.
   In order to more definitely relate the general symbolic system to its finite
   restriction it is therefor desirable to introduce the methods of Non-Standard
   Analysis. These methods allow for the rigorous definition and algebraic
   manipulation of non-finite quantities in accordance with the rules of a
   division ring, as well as for a canonical representation of all such
   quantities as a unique sum of finite and non-finite parts.
   
   Non-Standard Analysis, as previously noted, has been rigorously developed
   elsewhere, so it is neither necessary nor desirable to go into much detail in
   explaining or justifying these methods. It should be helpful, however, if an
   intuitively plausible exposition is provided which, moreover, shows that the
   definition of the extension of the finite real number system employed in
   non-standard analysis may reasonably be thought of as arising from a
   generalization, to a non-finite case, of the notions already introduced in
   the analysis of the finite case. 
   
   The most general kind of relation in the general symbolic system is that
   which allows for the non-invertible manipulation of symbols. As was
   previously indicated, such relations were, in the finite case, determined by
   the formal rules of Quantum Theory. These rules still apply and will
   determine the relations which define the non-standard extension of the reals.
   
   Recall the Quantum-theoretical rule (\ref{eq:prob1}) and its special case
   (\ref{eq:prob1a}): These rules, considered respectively, require the finite
   additivity of probability and that probabilities should be real numbers in
   the unit interval. Considered without regard to the notion of randomness,
   the development of the theory of finite relations, in accordance with these
   rules, leads to the construction of a finitely additive measure on sequences.
   Given such a measure, m, and any two sequences, \{a$_n$\} and \{b$_n$\}, the
   measure may be used to quantify the degree of term-wise agreement between any
   two such sequences: m\{n$|$ a$_n$ = b$_n$\} gives this measure. 
   
   In the finite case the desired T-independence of this measure requires
   randomness of such sequences. It may be recalled that the relative
   frequencies of random sequences necessarily converge to a limit, which is the
   probability, and that any two random sequences which correspond to the same
   measure are interchangable and correspond to the same probability. In this
   sense sequences are, as a natural development within the finite general
   symbolic system, ``identified" according to their having the same limit, so
   that limits play the role of measure in this case. 
   
   The identification of sequences with common limits suggests the construction
   of equivalence classes $\langle$ a$_n$ $\rangle$ defined by $\langle$ a$_n$
   $\rangle$ $\equiv$ \{\{b$_n$\} $|$ lim$_{n\rightarrow\infty}$ (a$_n$ -
   b$_n$)=0 \}. The equivalence between sequences which is thus defined with
   respect to limits is a very liberal one, there being very many sequences
   which converge to a given limit. In other words, probability, thought of as a
   measure, distinguishes relatively few sequences because it neglects the mode
   of convergence of such sequences.
   
   Algebraic operations may be defined upon sequences, though if such
   operations are to be relevant to, and reliably reflected in, the
   probabilities associated with such sequences, then these operations must be
   defined term-wise. Therefor the definitions \{a$_n$\}$\ast$\{b$_n$\} $\equiv$
   \{a$_n$$\ast$b$_n$\} and \{a$_n$\}+\{b$_n$\} $\equiv$ \{a$_n$+b$_n$\} are
   made. With these definitions it may be confirmed that the equivalence classes
   corresponding to such sequences also satisfy these equations. 
   
   The above construction of limit-based equivalence classes is reminiscent of
   the construction of the real number system as a collection of equivalence
   classes of Cauchy sequences of rational numbers. In this derivation algebraic
   operations on the equivalence classes are like-wise well-defined in terms of
   the corresponding term-wise operations on the elements of representative
   sequences from each of the classes. In order to generalize the real numbers
   to the non-finite case it therefor is natural to attempt a generalization of
   this procedure by replacing the ``limit measure" with the least restrictive
   measure conceivable. 
   
   As indicated above, a measure categorizes a sequence according to the value
   it assigns it. In order to make the greatest number of sequences distinct it
   is necessary to allow the measure to take on the least range of values. This
   identifies the measure to be employed in the extension of the reals as the
   discrete measure, where this measure only takes on either of the two values
   0 or 1 for any subset of the integers. In order that this measure lead
   to a number system which extends the reals it is necessary that it assign
   the value 0 to all finite subsets of the integers. This corresponds to the
   indifference of the ``limit measure" to any finite portion of a sequence.
   This requirement also leads to the conclusion that infinite subsets of the
   integers have discrete measure 1.
   
   Given such a discrete measure m defined on the integers, a new definition of
   equivalence classes of sequences may be specified in analogy with that
   already determined by limits. In particular: $\langle$ a$_n$ $\rangle$ $\equiv$
   \{\{b$_n$\} $|$ m\{ n $|$ a$_n$=b$_n$\}=1 \}. When this is done it is easily
   confirmed that, as before, the addition and multiplication of equivalence
   classes is well-defined. It is also clear that the real numbers may be
   embedded in this scheme with the mapping a$\mapsto\langle$a$\rangle$, which
   takes the real number a to the equivalence class of sequences which contains
   a sequence \textit{all} of the terms of which are identically a. The above
   construction therefor extends the real number system. This extension is
   referred to as the hyper-real number system.
   
   Unlike before, however, the discrete measure takes into account the
   \textit{whole} sequence, and not just its limit, so that it categorizes
   convergent sequences also according to their modes of convergence. To see
   this in detail it suffices to extend the notion of order to the new
   equivalence classes: Take $\langle$a$_n$$\rangle<\langle$b$_n$$\rangle$
   exactly wherever m\{n $|$ a$_n<$b$_n$\}=1. Then it may be seen that,
   according to this definition,
   $\langle\frac{1}{n^2}\rangle<\langle\frac{1}{n}\rangle$ and
   $\langle$n$\rangle<\langle$n$^2\rangle$ are examples of inequalities which
   hold in the hyper-real number system. As may be expected, these inequalities
   are between what may be referred to, in a precise way, as infinitesimal and
   infinite numbers, respectively. Indeed, a \textit{finite} number x may be
   defined to be one which is bounded by \textit{some} positive real a so that
   -a$<$x$<$a. Numbers which are not finite are termed \textit{infinite}.
   \textit{Infinitesimals} are taken to be numbers bounded by \textit{all}
   positive real numbers. 
   
   The foregoing makes it clear that the hyper-reals constitute a
   \textit{proper} extension of the real number system. Beyond this, the
   formal relationship between the hyper-reals and the embedded finite real
   number system may readily be made clear\footnote{Rigorous proof of the
   following and other results of Non-Standard Analysis may be found in
   \cite{Cutland}. See, in particular, pp. 1-105 there for an introduction to
   non-standard methods along the lines of the presentation here}. Consider a
   finite hyper-real x. Let a be the least upper bound of the real numbers less
   than x. Then $\epsilon$=x-a is infinitesimal. The decomposition x =
   a+$\epsilon$ is unique. Thus all hyper-reals may be expressed as a unique sum
   of finite and non-finite numbers. 
   
   With this result it is possible to have an at least intuitive understanding
   of the hyper-real number system and how it relates to the previously
   developed finitary theory. It now remains to apply this general understanding
   to see how working in the context of the hyper-real number system might
   resolve the previously mentioned incompatibility between the finite
   Relativistic and Quantum-theoretical formalisms. 
   
   Consider any combination of Relativistic and Quantum-Theoretic data in the
   non-finitary general symbolic system. Express this data as canonical sums of
   finite and non-finite numbers. The standard (finite) parts of the data may
   then be identified with the data upon which the finitary theory holds. As
   previously discovered, such restricted data cannot form the basis of a
   satisfactory physical theory in the general symbolic system. 
   
   Such data, while giving a finite \textit{level} of description, do not,
   however, give the ``whole picture" in the non-finitary case. The residual
   parts of the data, due to their infinitesimal or infinite nature, while
   either empirically entirely inaccessible or irrelevant in principle in the
   context of a finitary theory, are, nevertheless, critical in resolving the
   apparent incompatibility of Quantum Theory and Relativity. Two numbers which
   differ, even if only by an infinitesimal amount, are \textit{different}, and
   the differences to be found in the residual parts of data are presumably the
   \textit{only} means of \textit{formally} rejecting the arguments of the last
   section. 
   
   Put more concretely, the non-local correlations which Quantum Theory formally
   demands but finitary Relativity forbids are expressed by
   quantities neglected in the finitary theory. It may be said, in this sense,
   that the elaboration of finitary theory, while not adopting any axioms, has
   nevertheless resulted in a \textit{level} of description. Changes in such
   neglected quantities amount to signals which may travel at infinite speed and
   yet are not incompatible with the finitary theory. This is apparently the
   unique way, String Theory not withstanding, in which the Quantum Theoretical and
   Relativistic formalisms may be unified in the general symbolic system. This
   unification also suggests a resolution between the corresponding conflict
   between a preference for a discrete or a continuous mode of description of
   experience: The non-finitary general theory comprises a third kind of
   description which chooses neither over the other, and in this context neither
   probabilistic nor deterministic theory takes precedence over the other or is
   wholely adequate. 

   As a final observation, the role which Non-Standard Analysis plays in
   presumably reconciling Relativity and Quantum Theory as parts of the general
   symbolic system suggests that an identification may be made between the notion
   of the externality of the ``physical reality" which finite empiricism forever
   appeals to and the presumed conceptual inadequacy of \textit{any} finite
   structures within the general symbolic system. In particular, there
   \textit{is} a similarity between these two approaches in that finitely based
   empirical theories are subject to arbitrarily many experimental revisions of
   their data, while the finite structures of the general symbolic system are
   themselves subject to infinite elaboration. If this identification is made
   then, it is proposed, these two approaches may be thought of as being
   indistinct. While it would be inconsistent with the very notion of a
   non-axiomatic theory to demand such an identification, nevertheless it is
   hoped that such an identification might make rejecting serious consideration
   of such systems less likely.

%% file: rpichap8.tex

\chapter{DISCUSSION AND CONCLUSIONS}

   The object of this thesis has been the development of a written discrete combinatorial symbolic system based on formal rules for the manipulation of symbols which apply without regard to the particular meaning of the symbols involved. Motivation for the particular rules to be adopted was sought through examination of some contemporary scientific theories. Beginning with discussions of Linguistics, Metamathematics, and Physics it was found that, although each proceeded, in part, from a formal axiomatic basis, there were still a number of difficulties associated with the incorporation of an additional intuited meaningful aspect of experience. The discussion of Metamathematics, however, led to the suggestion that the notions of the truth and the formal provability of any particular assertion were, in part, determined by the formally unrestricted \textit{choice} of an interpretation. Consequently it was suggested that perhaps it is unnecessary and undesirable to make such an apparently arbitrary choice. Perhaps, instead, the formalism could be shown to yield its own interpretation and thus avoid the possibility of the formal rules conflicting with intuition.  

   In order to further motivate and justify this approach a schematic overview
   of physics in which, beyond the idealistic presentation of the introduction, a practical description of the activities of physicists and their implicit philosophies, as well as a discussion of the history of the development of Physics itself, was given. This presentation clarified the nature of the conflict between formalism and intuition and indicated the way in which this difficulty has been addressed, though by necessity impermanently, through the adoption of either statistical or non-statistical theories. As one is always free to adopt theories of either character, it was next attempted to take a step back to a stage prior to theoretical analysis in order to seek a basis for a formal theory.  

   The first stage of the scientific procedure, prior to that of theory construction, is that of observation and description. At this point it was argued, aiming at plausibility rather than proof, that a number of the features of experience which seem to be merely intuited may nevertheless be shown to derive from a merely formal basis. It was argued, in fact, that while any particular grouping, or categorization, of symbols( such a process being a necessary prerequisite for any formal theory ) cannot be defended a priori, nevertheless the mere \textit{faculty} of being able to adopt \textit{some} such description leads to the utilization of realistic structures within language. It was therefor decided that, even on an intuitive level, the formalism should be based not on the expression of ``truths" or ``facts", but rather constructed in order to express contingency, in which case a theory may be judged completely satisfactory if its elements are convincingly inter-related and the structure of the language is rich enough to be descriptive.

   Such a formal theory was then presented. It was shown that in such a theory
   the freedom of labeling consequent to the absence of facts leads to groupings of symbols, here called metalabelings, and to a further facility for relabeling which, almost paradoxically, enables one to \textit{conventionally} reserve symbols to be used in particular ways. Such considerations were then applied to the development of the so-called ``general symbolic system". This system takes the algebraic \textit{form} of the real quaternions, the system inheriting a partial order based on metalabeling, though it also naturally admits of infinite and infinitesimal quantities. In order to demonstrate the feasibility of such a system it was necessary to next develop a generalization of the Calculus which is not based on $\epsilon$-$\delta$ arguments. With this it was possible to complete the development of the formal system, although it remained to demonstrate its relevance by showing its utility in application to Physics. 

   Physical theory was developed in accordance with the requirements of
   empiricism which dictate that physical description may be given in terms of
   finitely many finite variables. Such a starting point was invoked in basing
   all theory on the location of the zeroes of a complex analytic state function
   $\phi$. Theory was then developed along two lines. First it was shown that
   Relativity Theory, including the space-time structure to which it applies,
   resulted if it was desired that all operations in the theory should express
   invertible relations. As was also indicated, it turns out that Turing
   computability is a notion which may be thought of as being subsidiary to
   Relativity Theory. 

   Relativity presumably being the complete expression of determinism, extension of the
   theory led to probabilistic developments. It was then argued that the
   propensity-based probability theory arrived at, expressing relations between
   proper categories, was, in fact, Quantum Theory. As was to be expected, this
   derivation also serves to justify the assertion that Quantum Theory is an
   intrinsically probabilistic theory. The strictly formal basis of this
   derivation bypassed the notorious ``measurement problem" of quantum
   theory and resulted in a formalism and descriptive apparatus formally similar
   to that of Feynman's Sum Over Histories and Everett's Many Worlds
   interpretations of quantum mechanics. The resulting perspective is unique,
   however, in indicating that all features of the dynamical and thermodynamical
   laws and description, including those materialistic phenomenological aspects of
   individual experience described by pressure, volume, and temperature, may be
   thought of as simply being consequences of basing theory on the finitary
   state function $\phi$. 

   The finite application of the general symbolic system had, at this point,
   reasonably justified its relevance in
   comparison with the usual empirical epistemology. Investigation of its
   internal coherence led, however, to the interesting conclusion that retaining
   a formally finite basis for theory \textit{required} that the formalism be
   supplemented by something outside itself, such as is done in experiment, and
   yet does not result in a \textit{conceptually} coherent description of
   experience. Conversely, retaining a strictly formal perspective apparently
   requires rejecting finitism and therefor utilizing symbols which
   \textit{cannot} be empirically verified if it is assumed that measurements
   are represented by finite collections of finite numbers. Extension of the
   formal theory in such an infinitary approach turns out to be surprisingly
   easy and is perhaps the only way to achieve a desired unification of the
   Relativistic and Quantum-theoretical formalisms.

%% file: rpiapp.tex

\appendix  
\addcontentsline{toc}{chapter}{APPENDICES}
\chapter{THE SOLUTION OF FUNCTIONAL EQUATIONS}
\section{The Associativity Equation}

   The equation to be solved is 
\begin{equation}\mbox{F[x,F[y,z]] = F[F[x,y],z].}\label{eq:a1}\end{equation}
   From (\ref{eq:a1}) may be derived 
\begin{equation}\mbox{$\partial_1$F[x,F[y,z]] =
$\partial_1$F[F[x,y],z]$\ast\partial_1$F[x,y]}\label{eq:a2}\end{equation} 
and
\begin{equation}\mbox{$\partial_2$F[x,F[y,z]]$\ast\partial_1$F[y,z] =
$\partial_1$F[F[x,y],z]$\ast\partial_2$F[x,y].}\label{eq:a3}\end{equation} 
   Note that the order of the factors is unimportant because both terms in each
   product are functions of the same variables and therefor commute. For F not
   constant in any of x, y, or z it follows that $\partial_1$F[x,y] $\not=$0 and
   $\partial_2$F[x,y] $\not=$0. Now (\ref{eq:a2}) and (\ref{eq:a3}) may be
   combined in the form 
\begin{equation}\mbox{$\partial_1$F$^{-1}$[x,y]$\ast\partial_2$F[x,y] =
$\partial_1$F$^{-1}$[x,F[y,z]]$\ast\partial_2$F[x,F[y,z]]$\ast\partial_1$F[y,z].}\label{eq:a4}\end{equation}

   This expression may be simplified by letting
\begin{equation}\mbox{G[x,y] = $\partial_1$F$^{-1}$[x,y]$\ast\partial_2$F[x,y].}\label{eq:a5}\end{equation}
   Then follows
\begin{equation}\mbox{G[x,y] = G[x,F[y,z]]$\ast\partial_1$F[y,z]}\label{eq:a6}\end{equation}
and
\begin{equation}\mbox{G[x,y]$\ast$G[y,z] = G[x,F[y,z]]$\ast\partial_2$F[y,z].}\label{eq:a7}\end{equation} 
   Now
\begin{equation}\mbox{$\partial_z$G[x,y] = 0 =$\partial_2$G[x,F[y,z]]$\ast\partial_2$F[y,z]$\ast\partial_1$F[y,z] + G[x,F[y,z]]$\ast\partial_{21}$F[y,z]}\label{eq:a8}\end{equation}
and
\begin{equation}\mbox{$\partial_y$[G[x,y]$\ast$G[y,z]] = $\partial_2$G[x,F[y,z]]$\ast\partial_1$F[y,z]$\ast\partial_2$F[y,z] + G[x,F[y,z]]$\ast\partial_{12}$F[y,z].}\label{eq:a9}\end{equation}
   Subtracting (\ref{eq:a8}) from (\ref{eq:a9}) results in 
\begin{equation}\mbox{$\partial_y$[G[x,y]$\ast$G[y,z]] = G[x,F[y,z]]$\ast$[$\partial_{12}$F[y,z] - $\partial_{21}$F[y,z]].}\label{eq:a10}\end{equation}
   Now $\partial_{12}$F = $\partial_{21}$F so that
\begin{equation}\mbox{$\partial_y$[G[x,y]$\ast$G[y,z]] = 0.}\label{eq:a11}\end{equation}
   This may be solved by 
\begin{equation}\mbox{G[x,y] = $\beta\ast$H(x)$\ast$H$^{-1}$(y)}\label{eq:a12}\end{equation}
where H is any function such that H(y) $\not=$ 0 and $\beta$ is a constant.
   Now from this and (\ref{eq:a6}) it follows that
\begin{equation}\mbox{$\partial_1$F[y,z] = H(F[y,z])$\ast$H$^{-1}$(y).}\label{eq:a13}\end{equation}
   Similarly, (\ref{eq:a7}) and (\ref{eq:a12}) yield 
\begin{equation}\mbox{$\partial_2$F[y,z] = $\beta\ast$H(F[y,z])$\ast$H$^{-1}$(z).}\label{eq:a14}\end{equation} 
   Now define the function $\Phi$ by the requirement that 
\begin{equation}\mbox{$\ln$($\Phi$($\omega$)) = $\int^\omega_1$H$^{-1}$($\tau$)d$\tau$.}\label{eq:a15}\end{equation}
   Then 
\begin{equation}\mbox{$\ln$($\Phi$(F[y,z])) = $\int^{\mbox{F[y,z]}}_1$H$^{-1}$($\tau$)d$\tau$   and   $\ln$($\Phi$(y)) = $\int^{\mbox{y}}_1$H$^{-1}$($\tau$)d$\tau$.}\label{eq:a16}\end{equation} 
Thus $\partial_{y}\ln$($\Phi$(y)) = H$^{-1}$(y)  and, by (\ref{eq:a13}), it follows that
\begin{equation}\mbox{$\partial_{y}\ln$($\Phi$(F[y,z])) = H$^{-1}$(F[y,z])$\ast\partial_1$F[y,z] = H$^{-1}$(y).}\label{eq:a17}\end{equation}
Thus
\begin{equation}\mbox{$\ln\left(\frac{\Phi\mbox{(F[y,z])}}{\Phi\mbox{(y)}}\right)$ = $\ln$(K(z)) for some function K.}\label{eq:a18}\end{equation}
   Now (\ref{eq:a14}) indicates that 
\begin{equation}\mbox{$\partial_{z}\ln$($\Phi$(F[y,z])) = H$^{-1}$(F[y,z])$\ast\partial_2$F[y,z] = $\beta\ast$H$^{-1}$(z)}\label{eq:a19}\end{equation}
so that 
\begin{equation}\mbox{$\partial_{z}\ln$($\Phi$(F[y,z])) = $\beta\ast\partial_{z}\ln$($\Phi$(z)) = $\partial_{z}\ln$($\Phi^\beta$(z)).}\label{eq:a20}\end{equation}
   Thus
\begin{equation}\mbox{$\partial_{z}\ln\left(\frac{\Phi\mbox{(F[y,z])}}{\Phi^\beta\mbox{(z)}}\right)$ = 0.}\label{eq:a21}\end{equation} 
   Now considering (\ref{eq:a18}) also this leads to
\begin{equation}\mbox{$\partial_z\ln$(K(z)) = $\partial_z\ln$($\Phi$(F[y,z])) = $\beta\ast\partial_z\ln$($\Phi$(z)).}\label{eq:a22}\end{equation}
Thus, for a constant c, and by (\ref{eq:a18}), it follows that 
\begin{equation}\mbox{$\ln$(K(z)) = $\ln$($\Phi^\beta$(z)) - $\ln$(c) = $\ln$($\Phi$(F[y,z])) - $\ln$($\Phi$(y)).}\label{eq:a23}\end{equation}
This last equation is equivalent to 
\begin{equation}\mbox{c$\ast\Phi$(F[y,z]) = $\Phi$(y)$\ast\Phi^\beta$(z).}\label{eq:a24}\end{equation}
   Note that c must be non-zero. Now it remains to determine $\beta$. Starting from (\ref{eq:a1}) it is evident that 
\begin{equation}\mbox{c$\ast\Phi$(F[x,F[y,z]]) = c$\ast\Phi$(F[F[x,y],z]).}\label{eq:a25}\end{equation}
Applying (\ref{eq:a24}) to (\ref{eq:a25}) twice results in 
\begin{equation}\mbox{$\Phi$(x)$\ast\mbox{[ c$^{-1}\ast\Phi$(y)$\ast\Phi^\beta$(z)]}^\beta$ = c$^{-1}\ast\Phi$(x)$\ast\Phi^\beta$(y)$\ast\Phi^\beta$(z).}\label{eq:a26}\end{equation} 
and this is equivalent to 
\begin{equation}\mbox{c$^{\beta-1}$ = $\Phi^{\beta(\beta-1)}$(z).}\label{eq:a27}\end{equation}
Thus it must be that $\beta$ = 1 if $\Phi$ is not to be constant.
   The solution of (\ref{eq:a1}) is therefor 
\begin{equation}\mbox{$\Phi$(F[x,y]) = c$^{-1}\ast\Phi$(x)$\ast\Phi$(y).}\label{eq:a28}\end{equation} 
  
\section{The Categoricity Equation}

   The equation to be solved is
\begin{equation}\mbox{$\phi$(a$^\prime$) = f($\phi$(a)).}\label{eq:b1}\end{equation}
   This correspondence must hold in all cases, including when attention is restricted to a Boolean Lattice. This is the case under which (\ref{eq:b1}) will be solved.
  
   Let h = $\Phi\phi$. Then, by (\ref{eq:a28}), it follows that 
\begin{equation}\mbox{c$\ast$h(a$\wedge$b) = h(a)$\ast$h(b).}\label{eq:b2}\end{equation}
 
   Let h(a$^\prime$) = g(h(a)). From (\ref{eq:b1}) it follows that
\begin{equation}\mbox{$\phi$(a$^\prime$) = [$\Phi^{-1}$g$\Phi$]$\phi$(a).}\label{eq:b3}\end{equation} 
Thus f = $\Phi^{-1}$g$\Phi$ so that finding f is reduced to finding g.

   Now, according to (\ref{eq:b1}),  (\ref{eq:b2}), and DeMorgan's Law, it follows that
\begin{equation}\mbox{c$\ast$g(h(a$\vee$b)) = g(h(a))$\ast$g(h(b))}\label{eq:b4}\end{equation}
so that
\begin{equation}\mbox{h(b) = g(g(h(b))) = g$\left(\frac{\mbox{c$\ast$g(h(a$\vee$b))}}{\mbox{g(h(a))}}\right)$.}\label{eq:b5}\end{equation}
   Similarly, with the help of (\ref{eq:b5}), it follows that
\begin{equation}\mbox{c$\ast$h(a$^\prime\wedge$b) = g(h(a))$\ast$g$\left(\frac{\mbox{c$\ast$g(h(a$\vee$b))}}{\mbox{g(h(a))}}\right)$.}\label{eq:b6}\end{equation}
   Now (\ref{eq:b2}) may be used to evaluate this same quantity in another way;
\begin{equation}\mbox{c$\ast$h(a$^\prime\wedge$b) = h(b)$\ast$g$\left(\frac{\mbox{c$\ast$h(a$\wedge$b)}}{\mbox{g(h(a))}}\right)$.}\label{eq:b7}\end{equation}
   Then (\ref{eq:b6}) and (\ref{eq:b7}) combine to yield
\begin{equation}\mbox{g(h(a))$\ast$g$\left(\frac{\mbox{c$\ast$g(h(a$\vee$b))}}{\mbox{g(h(a))}}\right)$ = h(b)$\ast$g$\left(\frac{\mbox{c$\ast$h(a$\wedge$b)}}{\mbox{g(h(a))}}\right)$.}\label{eq:b8}\end{equation}
   Consider a special case where a = i$\wedge$j and b = i$\vee$j. Then a$\wedge$b = a and a$\vee$b = b. Applying this to (\ref{eq:b8}) yields
\begin{equation}\mbox{g(h(a))$\ast$g$\left(\frac{\mbox{c$\ast$g(h(b))}}{\mbox{g(h(a))}}\right)$ = h(b)$\ast$g$\left(\frac{\mbox{c$\ast$h(a)}}{\mbox{h(b)}}\right)$.}\label{eq:b9}\end{equation}
   Let y = g(h(a)), z = h(b), u = $\frac{\mbox{c$\ast$g(y)}}{\mbox{z}}$, and v = $\frac{\mbox{c$\ast$g(z)}}{\mbox{y}}$. Then (\ref{eq:b9}) becomes simply
\begin{equation}\mbox{z$\ast$g(u) = y$\ast$g(v).}\label{eq:b10}\end{equation}
   Operating with $\partial_{y}$ on (\ref{eq:b10}) results in 
\begin{equation}\mbox{c$\ast$g$^\prime$(u)$\ast$g$^\prime$(y) = g(v) - v$\ast$g$^\prime$(v).}\label{eq:b11}\end{equation}
   Similarly, taking $\partial_{z}$ of (\ref{eq:b10}) leads to
\begin{equation}\mbox{c$\ast$g$^\prime$(v)$\ast$g$^\prime$(z) = g(u) - u$\ast$g$^\prime$(u).}\label{eq:b12}\end{equation}
   Operating on (\ref{eq:b12}) with $\partial_{y}$ now gives
\begin{equation}\mbox{u$\ast$y$\ast$g$^{\prime\prime}$(u)$\ast$g$^\prime$(y) = v$\ast$z$\ast$g$^{\prime\prime}$(v)$\ast$g$^\prime$(z).}\label{eq:b13}\end{equation}
By using (\ref{eq:b10}) this may be reduced to 
\begin{equation}\mbox{u$\ast$g$^{\prime\prime}$(u)$\ast$g(u)$\ast$g$^\prime$(y) = v$\ast$g$^{\prime\prime}$(v)$\ast$g(v)$\ast$g$^\prime$(z).}\label{eq:b14}\end{equation}
   Equations (\ref{eq:b11}) and (\ref{eq:b12}) may be solved for g$^\prime$(y) and g$^\prime$(z) and the results substituted into (\ref{eq:b14}). This results in
\begin{equation}\frac{\mbox{u$\ast$g$^{\prime\prime}$(u)$\ast$g(u)}}{\mbox{g$^\prime$(u)$\ast$(u$\ast$g$^\prime$(u) - g(u))}}\mbox{ = }\frac{\mbox{v$\ast$g$^{\prime\prime}$(v)$\ast$g(v)}}{\mbox{g$^\prime$(v)$\ast$(v$\ast$g$^\prime$(v) - g(v))}}\mbox{ = $\lambda$ ( a constant ).}\label{eq:b15}\end{equation}
   Evidently, then, the variables u and v may be separated in this differential equation. The first integral of the resulting equation is given by
\begin{equation}\mbox{g$^{-\lambda}$(x)$\ast$g$^\prime$(x) = $\alpha\ast$x$^{-\lambda}$}\label{eq:b16}\end{equation}
where $\alpha$ is a constant of integration. 
 
   The solution of (\ref{eq:b16}) falls into two cases corresponding to $\lambda$ = 1 and $\lambda$ $\not=$ 1.

   For $\lambda$ = 1 the solution of (\ref{eq:b16}) is of the form g(x) = A$\ast$x$^\alpha$. Then
\begin{equation}\mbox{x = g(g(x)) = g(A$\ast$x$^\alpha$) = A$^{\alpha\mbox{+}1}\ast$x$^{\alpha^2}$}\end{equation}
   This amounts to
\begin{equation}\mbox{A$^{\mbox{-(}\alpha\mbox{+}1\mbox{)}}$ = x$^{\mbox{(}\alpha\mbox{+}1\mbox{)(}\alpha\mbox{-}1\mbox{)}}$.}\label{eq:b17}\end{equation}
This implies that $\alpha$ = $\pm$1. For $\alpha$ = 1, A = $\pm$1. For $\alpha$ = - 1, A need merely be non-zero. Thus the $\lambda$ = 1 case solutions are of the form
\begin{equation}\mbox{g(x) = $\pm$x, or g(x) = A$\ast$x$^{-1}$.}\label{eq:b18}\end{equation}

   According to the previous section of the appendix, it was found that $\ast$
   formally corresponds to $\wedge$. It is necessary, then, that $\ast$-products
   may be factored so that an interpretation may be uniquely determined. It may
   therefor be seen that all of the solutions for g(x), for the $\lambda$ = 1
   case, must be rejected. In particular, g(x) = x obviously doesn't distinguish
   x from g(x) in the first place. g(x) = -x must be rejected because products
   of negatives don't retain specific signs for the factors, so that it would be
   generally impossible to determine which factors represent g(x) for some x.
   Finally, g(x) = A$\ast$x$^{-1}$ must also be rejected. In this case it is
   because the operation of inversion with respect to $\ast$ has already been
   reserved, by design, for the task of indicating erasures. It would be
   impossible to necessarily determine, then, whether a factor x$^{-1}$ would
   indicate that x is to be erased at some point or that it indicates that x is
   not a meaningful symbol. There therefor only remains one possible class of solution.

   For $\lambda\not=$ 1 the solution of (\ref{eq:b16}) is defined by 
\begin{equation}\mbox{g$^{\mbox{r}}$(x) = $\alpha\ast$x$^{\mbox{r}}$ + B, where r = (1-$\lambda$).}\label{eq:b19}\end{equation}
Applying (\ref{eq:b19}) twice to (\ref{eq:b10}) results in
\begin{equation}\mbox{($\alpha^2\ast$c$^{\mbox{r}}$ - B)$\ast$(y$^{\mbox{r}}$ - z$^{\mbox{r}}$) = 0, so B = $\alpha^2\ast$c$^{\mbox{r}}$.}\label{eq:b20}\end{equation}

   In light of (\ref{eq:b20}), it follows that substituting (\ref{eq:b19}) into x = g(g(x)) results in
\begin{equation}\mbox{(1 - $\alpha^2$)$\ast$x$^{\mbox{r}}$ = $\alpha^2\ast$($\alpha$ + 1)$\ast$c$^{\mbox{r}}$, so $\alpha$ = $\pm$1.}\label{eq:b21}\end{equation}
The only solution for the case $\lambda\not=$1 which, at least, has not already been found in the previous case is then   
\begin{equation}\mbox{g$^{\mbox{r}}$(x) = c$^{\mbox{r}}$ - x$^{\mbox{r}}$.}\label{eq:b22}\end{equation}
This general solution of this equation requires that complex numbers be considered, and thus such numbers must be admitted to the general symbolic system. Such a number system may be used to distinguish the factor x from g(x). Thus the solvability of (\ref{eq:b1}) necessitates the inclusion of complex numbers in the symbolic system, and, conversely, gives a definite rationale for using such symbols.

%% file: rpibib.tex
 
\addtocontents{toc}{\parindent0pt\vskip12pt BIBLIOGRAPHY}

\

%% file: rpithes.bbl
\begin{thebibliography}{99}
\bibitem{Pinker1} Pinker, Steven \textit{The Language Instinct}, W. Morrow and Company, 1994.
\bibitem{Papadimitriou}  Papadimitriou, Christos and Lewis, Harry \textit{ Elements of the Theory of Computation}, Prentice-Hall, 1981.
\bibitem{Pinker2} Pinker, Steven \textit{How the Mind Works}, Norton, 1997.
\bibitem{Hofstadter} Hofstadter, Douglas R. \textit{G\"{o}del, Escher, Bach: An Eternal Golden Braid}, Vintage Books, 1980.
\bibitem{Kleene} Kleene, Steven C. \textit{Introduction to Metamathematics}, D. Van Nostrand, 1952. 
\bibitem{Cohen} Cohen, Paul J. \textit{Set Theory and the Continuum Hypothesis}, W. A. Benjamin, 1966.
\bibitem{Robinson} Robinson, Abraham \textit{Non-Standard Analysis}, North-Holland Publishing Company, 1966.
\bibitem{Gentzen} Szabo, M.E., ed \textit{The Collected Papers of Gerhard Gentzen}, North-Holland Publishing Company, 1969. 
\bibitem{Espagnat} Espagnat, Bernard d' \textit{In Search of Reality}, Springer-Verlag, 1983. 
\bibitem{Newton} Newton, Isaac, Sir \textit{Philosophiae Naturalis Principia Mathematica}, Harvard University Press, 1972.
\bibitem{Everett} DeWitt, Bryce S. and Graham, Niell,  eds. \textit{The Many-Worlds Interpretation of Quantum Mechanics. A Fundamental Exposition by Hugh Everett III, with papers by J.A. Wheeler (and others)}, Princeton University Press, 1973.
\bibitem{Neumann1} Neumann, John von \textit{Mathematical Foundations of Quantum Mechanics}, Princeton University Press, 1955.
\bibitem{Bell} Bell, John S. \textit{Speakable and Unspeakable in Quantum Mechanics: Collected Papers on Quantum Philosophy/ J.S. Bell}, Cambidge University Press, 1987.
\bibitem{Wheeler} Wheeler, J.A. and Zurek, W. \textit{Quantum Theory and Measurement}, Princeton University Press, 1983.
\bibitem{Baer} Baer, Reinhold \textit{ Linear Algebra and Projective Geometry}, Academic Press, 1952.

\bibitem{Varadarajan} Varadarajan, V.S. \textit{Geometry of Quantum Theory}, Springer-Verlag, 1985. 
\bibitem{Bohr} Bohr, Niels. \textit{ Essays 1958-1962 on Atomic Physics and Human Knowledge}, Interscience Publishers, 1963.
\bibitem{Hermes} Hermes, Hans \textit{ Enumerability, Decidability, Computability. An Introduction to the Theory of Recursive Functions}, Academic Press, 1965.
\bibitem{Neumann2} Neumann, John von \textit{Theory of Self-Reproducing Automata}, University of Illinois Press, 1966.
\bibitem{Neumann3} Neumann, John von \textit{ Probabilistic Logics and the Synthesis of Reliable Organisms from Unreliable Components}, Collected Works, vol. 5, pp. 329-378, 1963.
\bibitem{Pears} Pears, David \textit{Ludwig Wittgenstein}, The Viking Press,
1970.
\bibitem{Wittgenstein1} Wittgenstein, Ludwig \textit{Philosophical
Investigations}, Macmillan, 1953.
\bibitem{Wittgenstein2} Wittgenstein, Ludwig \textit{Tractatus
Logico-Philosophicus}, Routledge Humanities Press International, 1988.
\bibitem{Schilpp} Schilpp, Paul Arthur, ed. \textit{Albert Einstein: Philosopher Scientist}, Library of Living Philosophers, Inc., Vol. 7, 1949.
\bibitem{Fine} Fine, Arthur \textit{The Shaky Game: Einstein, Realism and the Quantum Theory}, University of Chicago Press, 1996. 
\bibitem{Popper} Popper, Karl. \textit{ Post-Script to the Logic of Scientific Discovery}, Rowman and Littlefield, 1983. 
\bibitem{Poincare} Poincar\'{e}, Henri \textit{The Foundations of Science}, The Science Press, 1921. 
\bibitem{Birkhoff} Birkhoff, Garrett. \textit{ Lattice Theory}, AMS Colloquium Publications, vol. 25, 1967.
\bibitem{Hilbert} Hilbert, David. \textit{ The Foundations of Geometry}, Open Court, 1971.
\bibitem{Dirac} Dirac, Paul A. M. \textit{ On Quantum Algebra}, Procedings of the Cambridge Philosophical Society, vol. 23, pp. 412-418, July 17, 1926.
\bibitem{Aczel1}  Aczel, J. \textit{Lectures on Functional Equations and Their Applications}, Academic Press, 1966.
\bibitem{Aczel2}  Aczel, J. and Dhombres, J. \textit{Functional Equations Containing Several Variables}, Cambidge University Press, 1989.
\bibitem{Frobenius} Herstein, I.N. \textit{ Topics in Algebra}, Blaisdell Publishing Company, 1964. 
\bibitem{Einstein1} Einstein, Albert \textit{ Relativity: the Special and General Theory}, Hartsdale House, 1947. 
\bibitem{Einstein2} Einstein, Albert \textit{ The Meaning of Relativity}, Princeton University Press, 1956. 
\bibitem{Misner} Misner, Charles and Thorne, Kip S. and Wheeler, John A. \textit{Gravitation}, W.H. Freeman, 1973. 
\bibitem{Weinberg} Weinberg, Steven \textit{ Gravitation and Cosmology: Principles and Applications of the General Theory of Relativity}, Wiley, 1972.
\bibitem{Dirac2} Dirac, Paul A. M. \textit{ General Theory of Relativity}, John Wiley and Sons, 1975. 
\bibitem{Mises1} Mises, Richard von. \textit{ Probability, Statistics and Truth}, Dover, 1981.
\bibitem{Mises2} Mises, Richard von \textit{Mathematical Theory of Probability and Statistics}, Academic Press, 1964.  
\bibitem{Jauch} Jauch, Josef M. \textit{Foundations of Quantum Mechanics}, Addison-Wesley, 1968. 
\bibitem{Shilov} Shilov, Georgi E. \textit{ Linear Algebra}, Dover, 1977. 
\bibitem{Infeld} Einstein, Albert and Infeld, Leopold. \textit{ On the Motion of Particles in General Relativity Theory}, Canadian Journal of Mathematics, vol. 3, pp. 209-241, 1949.
\bibitem{Fredkin} Fredkin, Edward and Toffoli, Tomasso. \textit{ Conservative Logic}, International Journal of Theoretical Physics, vol. 21, pp. 219-253, 1982. 
\bibitem{Kolmogorov} Kolmogorov, Andrei N. \textit{ Foundations of the Theory of Probability}, Chelsea Publishing Company, 1956.
\bibitem{Howson}  Howson, Colin and Urbach, Peter \textit{ Scientific Reasoning: the Bayesian Approach}, Chicago: Open Court, 1993.
\bibitem{Cox}Cox, Richard T. \textit{The Algebra of Probable Inference}, Johns Hopkins Press, 1961.
\bibitem{Holland} Holland, Peter R. \textit{The Quantum Theory of Motion: an Account of the de Broglie-Bohm Causal Interpretation of Quantum Mechanics}, Cambidge University Press, 1993.
\bibitem{Bachman} Bachman, George and Narici, Lawrence \textit{Functional Analysis}, Academic Press, 1966.
\bibitem{Feynman1} Feynman, Richard \textit{ Theory of Fundamental Processes}, N. A. Benjamin, 1961.
\bibitem{Dirac3} Dirac, Paul A. M. \textit{ The Principles of Quantum Mechanics}, Oxford Clarendon Press, 1930.
\bibitem{Feynman2} Feynman, Richard and Leighton,Robert and Sands,Matthew \textit{ Feynman Lectures on Physics}, Addison-Wesley, 1963.
\bibitem{Feynman3} Feynman, Richard and Hibbs, A.R. \textit{Quantum Mechanics and Path Integrals}, McGraw Hill, 1965.
\bibitem{Kubo} Kubo, R. and Toda, M. and Saito, N. \textit{Statistical Physics}, Springer-Verlag, 1991. 
\bibitem{Yang} Yang, C.N. and Lee, T.D. \textit{Statistical Theory of Equations of State and Phase Transitions I: Theory of Condensation}, Physical Review, vol. 87, number 3, pp. 404-409, 1952.
\bibitem{Born} Born, Max \textit{The Born-Einstein Letters - Correspondence Between Albert Einstein and Max and Hedwig Born from 1916 to 1955 with comments by Max Born.}, Walker and Company, 1971.
\bibitem{Manoukian} Manoukian, Edward B. \textit{ Renormalization}, Academic Press, 1983. 
\bibitem{Cutland} Cutland, Nigel, ed. \textit{Nonstandard Analysis and its
Applications}, Cambridge University Press, 1988.
\bibitem{Davis} Davis, Martin \textit{Applied Nonstandard Analysis}, Wiley, 1977.
\end{thebibliography}
